\documentclass[11pt,a4paper,oneside]{article}
\usepackage{graphicx,amssymb,amsmath,color}
\usepackage[colorlinks=true,urlcolor=blue,anchorcolor=blue,citecolor=red,filecolor=blue
,linkcolor=blue,menucolor=blue,linktocpage=true,pdfproducer=medialab,pdfa=true
]{hyperref}
\usepackage{cite}
\usepackage{relsize}
\usepackage{enumerate}
\usepackage[margin=2 cm]{geometry}
\usepackage{slashed}
\usepackage{multirow}
\usepackage[normalem]{ulem}
\usepackage[dvipsnames, x11names]{xcolor}
\usepackage[square,comma,sort&compress,numbers]{natbib}
\usepackage{soul}
\usepackage{bigints}
\usepackage{relsize,exscale}
\usepackage{stmaryrd}

\usepackage{cleveref}
\usepackage{cancel}
\usepackage{comment}

\usepackage[utf8]{inputenc}
\usepackage[T1]{fontenc}

\def\beq{\begin{equation}}
\def\eeq{\end{equation}}
\def\bea{\begin{eqnarray}}
\def\eea{\end{eqnarray}}

\def\bit{\begin{itemize}}
\def\eit{\end{itemize}}

\def\baa{\begin{array}}
\def\eaa{\end{array}}

\def\simgt{\mathrel{\lower2.5pt\vbox{\lineskip=0pt\baselineskip=0pt
           \hbox{$>$}\hbox{$\sim$}}}}
\def\simlt{\mathrel{\lower2.5pt\vbox{\lineskip=0pt\baselineskip=0pt
           \hbox{$<$}\hbox{$\sim$}}}}

\def\bfc{\begin{Fig.}\begin{center}}
\def\efc{\end{center}\end{Fig.}}
\def\nn{\nonumber\\}

\definecolor{chromeyellow}{rgb}{1.0, 0.65, 0.0}
\definecolor{darkcoral}{rgb}{0.8, 0.36, 0.27}
\definecolor{cadmiumgreen}{rgb}{0.0, 0.42, 0.24}

\newcommand{\CL}{{\tt ${\mathcal C}$osmo${\mathcal L}$attice}}

\begin{document}

\begin{minipage}{16.5cm}
\vspace{0cm}
    \begin{flushright}
DESY-25-161
\end{flushright}
\end{minipage}

\begin{flushright}
\hspace{3cm} 

\end{flushright}
\vspace{.6cm}
\begin{center}

\hspace{-0.4cm}{\huge \bf 
Domain walls in the scaling regime: \\
\vspace{.5cm}
\Large
Equal Time Correlator and Gravitational Waves
}

\vspace{1cm}{}
\end{center}

\begin{center}
 Simone Blasi$^{\dagger\,1}$, Alberto Mariotti$^{\# \, 2}$, A\"aron Rase$^{\star \, 2}$ and Miguel Vanvlasselaer$^{\ddagger \, 2,3}$ \\
\vskip0.4cm

{\it $^1$ Deutsches Elektronen-Synchrotron DESY, Notkestr.~85, 22607 Hamburg, Germany} \par
\vskip0.2cm
{\it $^2$ Theoretische Natuurkunde and IIHE/ELEM, Vrije Universiteit Brussel,
\& The International Solvay Institutes, Pleinlaan 2, B-1050 Brussels, Belgium }
\vskip0.2cm
{\it $^3$Departament de Física Quàntica i Astrofísica and Institut de Ciències del Cosmos (ICC), Universitat de Barcelona, Martí i Franquès 1, ES-08028, Barcelona, Spain.}
\end{center}

\bigskip \bigskip \bigskip

\begin{abstract}
\noindent
Domain walls are topological defects that may have formed in the early Universe through the spontaneous breakdown of discrete symmetries, and can be a strong source of gravitational waves (GWs). We perform 3D lattice field theory simulations with \CL, considering grid sizes $N = 1250$, $2048$ and $4096$, to study the dynamics of the domain wall network and its GW signatures. We first analyze how the network approaches the scaling regime with a constant $\mathcal{O}(1)$ number of domain walls per Hubble volume, including setups with a large initial number of domains as expected in realistic scenarios, and find that scaling is always reached in a few Hubble times after the network formation. To better understand the properties of the scaling regime, we then numerically extract the Equal Time Correlator (ETC) of the energy--momentum tensor of the network, thus determining its characteristic shape for the case of domain walls, and verifying explicitly its functional dependence as predicted by scaling arguments. The ETC can be further extended to the Unequal Time Correlator (UTC) controlling the GW emission by making assumptions on the coherence of the source. By comparison with the actual GW spectrum evaluated by \CL, we are then able to infer the degree of coherence of the domain wall network. Finally, by performing numerical simulations in different background cosmologies, \emph{e.g.} radiation domination and kination, we find evidence for a universal ETC at subhorizon scales and hence a universal shape of the GW spectrum in the UV, while the expansion history of the Universe may instead be determined by the IR features of the GW spectrum.
\\
\vspace{1cm}

\end{abstract}

\vfill
\noindent\line(1,0){188}
{\footnotesize{ \\ 
\text{$^\dagger$~simone.blasi@desy.de}\\
\text{$^\#$alberto.mariotti@vub.be}\\
\text{$^\star$~aaron.rase@vub.be}\\
\text{$^\ddagger$~miguel.vanvlasselaer@ub.edu}}}

\newpage

\hrule
\tableofcontents
\vskip.8cm
\hrule

\section{Introduction}

Symmetries and their breaking represent a central topic in high energy physics. In a cosmological context, this is typically associated to phase transitions as symmetries are typically restored in the early Universe for a large enough reheating temperature, and undergo spontaneous breaking while the Universe expands and cools down. Phase transitions are then often predicted in particle physics models, including compelling theories beyond the Standard Model (BSM).

If the symmetry-group breaking pattern, say $G\rightarrow H$, leads to a vacuum manifold, $\mathcal{V}=G/H$, with a non-trivial homotopy structure, the phase transition leads to the formation of topological defects via the Kibble mechanism\,\cite{Kibble:1976sj}
(see Ref.\,\cite{Vilenkin:2000jqa} for an extensive review on topological defects). In this paper we will focus on domain walls (DWs), two-dimensional defects that are formed when the vacuum manifold $\mathcal{V}$ consists of disconnected parts, namely it has a non-trivial $\pi_0(\mathcal{V})$ homotopy group
\,\cite{Kibble:1976sj,Vilenkin:2000jqa,Zeldovich:1974uw}.
The prototypical example for DW formation is the spontaneous breaking of a $\mathbb{Z}_{N}$ symmetry to nothing\,\footnote{The $\mathbb{Z}_N$ symmetry may or may not emerge as the remnant of a global $U(1)$ symmetry, as for axion DWs\,\cite{Sikivie:1982qv,Vilenkin:1982ks}.
In this case the network of defects is more complex and includes cosmic strings.}, even though more complex patterns can lead to the same outcome. For this reason, DWs arise in several BSM theories entailing the breakdown of a discrete symmetry.

From a phenomenological point of view, DWs can play a key role in a large variety of processes in the early Universe. These include the production of primordial black holes\,\cite{Ferrer:2018uiu,
Gelmini:2022nim,Gouttenoire:2023gbn,Gouttenoire:2023ftk,Ferreira:2024eru,Dunsky:2024zdo}, the generation of the baryon asymmetry\,\cite{
Brandenberger:1994mq,
Brandenberger:1998kn,Brandenberger:2005bx,Schroder:2024gsi,
Daido:2015gqa, Sassi:2023cqp,Sassi:2024cyb,Mariotti:2024eoh,Vanvlasselaer:2024vmi,Azzola:2024pzq}, the production of dark matter \,\cite{Harigaya:2018ooc,Babichev:2021uvl,Gelmini:2022nim,Gorghetto:2022ikz, DEramo:2024lsk}, inducing observable cosmic birefringence \,\cite{Gasparotto:2022uqo, Ferreira:2023jbu,Blasi:2024xvj}, as well as seeding bubble nucleation during first order phase transitions\,\cite{Blasi:2022woz,Blasi:2023rqi,Agrawal:2023cgp,Li:2023yzq,Sassi:2025dyj}. 

More importantly for the purpose of our work, DWs are known to be a strong source of a gravitational wave (GW) background in the early Universe, see Ref.\,\cite{Saikawa:2017hiv} for a review. In fact, the corresponding signal is searched for at GW experiments such as Pulsar Timing Arrays, where DWs have been shown to possibly explain the observed GW background\,\cite{NANOGrav:2023hvm,Gouttenoire:2023ftk,Blasi:2023sej, Kitajima:2023cek}. GWs from DWs have been recently searched by the LIGO--Virgo--KAGRA (LVK) collaboration as well\,\cite{LIGOScientific:2025kry} (see also\,\cite{Jiang:2022svq,Miritescu:2023ruv}), and are considered for the prospects of next--generation detectors such as the Einstein Telescope (ET)\,\cite{ET:2025xjr}. In general, DWs can be a source of GWs in a wide range of frequencies and thus relevant for LISA \cite{LISACosmologyWorkingGroup:2022jok} and other future experiments.

The dynamics of a DW network in the early Universe is understood as follows. At the time of the symmetry breaking phase transition, a dense network of DWs is formed via the Kibble mechanism, with a correlation length that is parametrically smaller than the Hubble horizon. After friction from the thermal plasma becomes negligible, the network is expected to reach an attractor solution (\emph{i.e.}, independent of the initial conditions) known as the \emph{scaling} regime, where the number of DWs per Hubble volume is of order unity 
and the average velocity of the DWs is mildly relativistic,
as confirmed in several numerical studies
\cite{Press:1989yh,Garagounis:2002kt,Oliveira:2004he,Avelino:2005kn,Leite:2011sc, Leite:2012vn, Avelino:2005pe,Kawasaki:2011vv, Hiramatsu:2013qaa,Kitajima:2023kzu, Ferreira:2024eru, Dankovsky:2024zvs, Notari:2025kqq, Cyr:2025nzf,Hindmarsh:1996xv,Hindmarsh:2002bq}.
During this regime, the energy density of a DW network decreases more slowly than the critical (background) energy density. Thus, if the network persists for too long, it will eventually dominate the energy budget of the Universe -- a scenario inconsistent with observations. This well-known domain wall problem \cite{Zeldovich:1974uw,Sikivie:1982qv} can be resolved, for example, by introducing a small explicit breaking (bias) of the underlying discrete symmetry \cite{Sikivie:1982qv,Gelmini:1988sf}, causing the network to collapse before domination occurs. 

As already mentioned, GWs are continuously generated during the scaling regime, with the overall emission dominated by the latest times near the decay of the network, when the DW energy density relative to the background is maximal. Several studies have investigated the GW spectrum numerically \cite{Hiramatsu:2013qaa, Kitajima:2023kzu, Li:2023yzq, Ferreira:2023jbu, Dankovsky:2024zvs, Notari:2025kqq, Cyr:2025nzf, Dankovsky:2024ipq, Gruber:2024pqh}, including the phase of collapse
\cite{Kitajima:2023cek, Ferreira:2024eru, Babichev:2025stm, Notari:2025kqq, Cyr:2025nzf}, and consistently found a broken power-law behaviour, albeit with varying numerical parameters. 

In this paper, we inspect several aspects of the DW network dynamics by performing lattice field theory  simulations with \CL 
~\cite{Figueroa:2020rrl, Figueroa:2021yhd}. In particular, we study in detail how the DW network approaches the scaling regime by considering a different set of initial conditions, including setups with a large initial number of DWs per Hubble volume. We show that, neglecting particle friction, the scaling regime is always reached in a few Hubble times, and that this occurs in a time scale that is in agreement with the Velocity--dependent One--Scale (VOS) model
\,\cite{Avelino:2005kn,Avelino:2019wqd,Martins:2016ois}.

We then investigate the scaling regime and the associated GW spectrum by
extracting for the first time
the Equal Time Correlator (ETC) of the (TT projected) energy momentum tensor of the DW network\,\footnote{
Notice that a similar approach was pursued in an early study \cite{Kawasaki:2011vv}, however the results were plagued by a bug in the numerical code.
}. The ETC provides a clean diagnostic of the approach to scaling and allows us to identify the relevant physical scales in the problem. In particular, we will compare the results of our numerical simulations with the form of the ETC as expected from general arguments characterizing the scaling regime, namely its self--similarity and the presence of essentially a single scale given at each time by the Hubble horizon. In fact, these arguments have been already applied to a scaling network of (possibly non--topological) defects in Refs.\,\cite{Figueroa:2012kw,Figueroa:2020lvo}. We will show how the case of scaling DWs differs from these previous studies in a way that is consistent with the generation of a peaked (rather than scale invariant) GW spectrum in radiation domination.

By making specific assumptions regarding the temporal coherence of the DW network as a GW source, one can extend the ETC to the Unequal Time Correlator (UTC)\,\cite{Caprini:2007xq,Caprini:2009fx,Fenu:2009qf, CamargoNevesdaCunha:2022mvg} (see also \,\cite{Albrecht:1995bg, Durrer:2001cg, Bevis:2010gj,Figueroa:2020lvo} for a numerical determination of the UTC for defects)
and predict the corresponding GW spectrum.
Within this framework, we will then be able to infer the degree of coherence of the network by matching the GW spectrum predicted by different extensions of the ETC with the one obtained directly from \CL. We find that the GW spectrum is best reproduced when the temporal coherence of the DW network is assumed to decrease at high momenta -- a behavior also noted for other defects, in particular cosmic strings\,\cite{Bevis:2010gj, CamargoNevesdaCunha:2022mvg}.

Finally, we repeat our analysis for simulations of a DW network conducted within different background cosmologies, \emph{i.e.} with a different Equation of State\,\footnote{We will never consider inflationary epochs in our work.}. We find evidence for a universal ETC in the scaling regime consistent with a power law $\sim (k \tau)^{-q}$ for subhorizon modes, with $k$ and $\tau$ the comoving momentum and conformal time, respectively, and $q \approx 2.8$ independently of the Equation of State. Together with simple considerations linking the ETC to the UTC,
this implies a universal shape of the GW spectrum from scaling DWs for subhorizon modes. This represents another difference between DWs and the scaling source considered in Ref.\,\cite{Figueroa:2020lvo}, for which the UV spectrum was found to depend on the cosmology. 

During our analysis,
we will also compare the GW spectrum obtained from our simulations of scaling DWs with existing results in the literature for radiation domination
\,\cite{Hiramatsu:2013qaa,Kitajima:2023kzu, Li:2023yzq, Ferreira:2023jbu, Dankovsky:2024zvs, Notari:2025kqq, Cyr:2025nzf},
and provide new results for the other cosmologies that we consider. Our results confirm a broken power law with a universal UV spectral index $f^{-n}$ with $n \approx 1.3$, while the IR spectrum and the overall GW amplitude will be affected depending on the background cosmology during (and possibly after) the time of the DW network annihilation.

The remainder of this paper is organized as follows. In Section \ref{sec:num_setup}, we briefly review the fundamentals of the DW dynamics and describe our numerical setup. Section \ref{sec:approach_scaling} discusses the approach to the scaling regime of the DW network. In Section \ref{sec:GW_from_scaling}, we numerically compute the GW spectrum and compare our results with previous numerical studies. Section \ref{sec: UTC} presents the computation of the ETC and explores its connection to both the GW spectrum and the coherence properties of the DW network. In Section \ref{sec: other cosmologies}, we investigate other cosmological backgrounds beyond radiation domination and confirm the universality of the ETC shape.
Section \ref{sec:pheno} explores some phenomenological implications of our results, while Section \ref{sec:conclusion} summarizes our conclusions. To facilitate the navigation of this paper easier, we gather our main results in Table \ref{tab:summary}.

\section{From theory to simulation: domain wall fundamentals}
\label{sec:num_setup}

Before presenting the numerical details of the simulations, we begin with a brief introduction to the physics of domain walls. We then extend this framework to the lattice and outline the numerical methods employed to simulate the domain wall network.
For the simulations we use the public code \CL ~to which we refer \cite{Figueroa:2020rrl} 
for further details.

\subsection{Theoretical framework of domain walls}

Through the spontaneous breaking of a discrete symmetry, DWs appear as topological defects dividing regions of space associated with separate but energetically equivalent vacua. These walls contain a high energy density as their profile interpolates between the different minima, thereby forcing the scalar field to sit atop of the potential at the center of the wall. Owing to this concentrated energy density, DWs can be a strong source of GWs. 

A typical and simple model to describe DWs arises from the following Lagrangian,
\begin{equation}
\label{exotic cosmological:potential}
    \mathcal{L}= \frac{1}{2} g^{\mu\nu}\partial_\mu\phi \partial_\nu\phi - V(\phi)\,,\quad \text{with} \quad V(\phi) = \frac{\lambda}{4}(\phi^2 - \eta^2)^2\,,
\end{equation}
where $\phi$ is a real scalar field, $\lambda$ is the self-coupling constant, $\eta$ is the vacuum expectation value (VEV) and we assume a flat expanding Universe,
\begin{equation}
    ds^2 = a^2(\tau)\left(d\tau^2 - dx^2\right)\,,
\end{equation}
where $a(\tau)$ denotes the scale factor, $\tau$ is the conformal time, and $x$ is the comoving coordinate. The potential $V(\phi)$ possesses two degenerate minima at $\phi = \pm \eta$, leading to the spontaneous breaking of the discrete $\mathbb{Z}_2$ symmetry and the formation of DWs interpolating between these vacua, with the mass of the scalar field in one of the two equivalent minima given by $m = \sqrt{2\lambda}\eta$. 

To be specific, DWs are produced in the early Universe via the \emph{Kibble mechanism} \cite{T_W_B_Kibble_1976}. At sufficiently high temperatures, typically for $T \gg \eta$, thermal corrections to the scalar potential restore the symmetry, leaving the scalar field with a vanishing vacuum expectation value. As the Universe cools below a critical temperature, the field begins to roll toward one of the degenerate minima. Consequently, causally disconnected regions of space will settle into different vacua of the theory. In fact, the scalar field is correlated only over distances comparable to the \emph{correlation length}, which is determined by the inverse Ginzburg temperature, $1/T_G \sim 1/m$, around the time of the phase transition. 
The interface separating regions associated with distinct vacua corresponds to a DW.

The equation of motion (EoM) for the scalar field, obtained from the Lagrangian in Eq. \eqref{exotic cosmological:potential}, is given by
\begin{equation}
\label{eq:EoM1}
   \frac{\partial^2\phi}{\partial\tau^2} 
   +2\mathcal{H}\frac{\partial\phi}{\partial\tau} 
   -\nabla^2\phi 
   + a^2\lambda \phi\left(\phi^2-\eta^2\right) = 0\,,
\end{equation}
with $\mathcal{H} = a'/a$ the conformal Hubble rate (where a prime denotes a derivative with respect to $\tau$), related to the physical Hubble $H$ by $\mathcal{H} = aH$. Notice that, as the focus of our work is to study DWs in the scaling regime, we will neglect particle friction altogether from our analysis of the DW dynamics, even though this can be relevant in certain ranges of temperature depending on the model, see \emph{e.g.}\,\cite{Blasi:2022ayo}.

The static DW solution to the EoM is a scalar field configuration interpolating between the two vacua $\phi = \pm \eta$, corresponding to the profile
\begin{equation}
     \phi(x) = \eta\  \text{tanh} \left(\frac{ax}{\delta_\text{dw}}\right)\,,
\end{equation}
where the wall thickness is $\delta_{\rm dw} \equiv  2/m$. The DWs are further characterized by their tension, \emph{i.e.} the energy per unit area, which in this model is given by $\sigma = (2/3)m\eta^2$. 

\subsection{To the lattice}

Let us now turn to the numerical setup that we will consider to simulate the dynamics of domain walls in the early Universe. First, we note that the EoM for the scalar field depends only on the ratio $m/H$ (or equivalently $m/\mathcal{H}$). To make this dependence explicit, we rescale the field and coordinates as
\begin{equation}
    \phi = \tilde{\phi}\,\eta\,, 
    \quad \tau = \frac{\tilde{\tau}}{m}\,, 
    \quad x = \frac{\tilde{x}}{m}\,,
\end{equation}
which transforms the EoM in Eq.\eqref{eq:EoM1} into
\begin{equation}
\label{eq:eoms}
    \frac{\partial^2\tilde{\phi}}{\partial\tilde{\tau}^2}
    + 2\left(\frac{\mathcal{H}}{m}\right)\frac{\partial\tilde{\phi}}{\partial\tilde{\tau}} 
    - \tilde{\nabla}^2 \tilde{\phi}
    + \frac{a^2}{2} \tilde{\phi} \left( \tilde{\phi}^2 - 1 \right) = 0\,.
\end{equation}
In all numerical simulations, we adopt this rescaling and set $\eta = m = 1$ by choosing the coupling $2\lambda = 1$ for the remainder of the paper (the dynamics of the system is primarily controlled by the ratio $m/H$, and the precise value of $\lambda$ is not crucial).

In our numerical setup, the lattice box with periodic boundaries represents a flat expanding Universe. We consider an expansion that is driven by an external fluid (e.g. radiation), so that the scale factor is determined by the Equation of State (EoS) parameter $\omega = P/\rho$, where $P$ and $\rho$ denote the pressure and energy density of that said fluid. For radiation, we have $\omega=1/3$, and the scale factor evolves as
\begin{equation}\label{eq: a}
    a(\tau) = a(\tau_i)\left(1+\frac{\mathcal{H}_i}{p}(\tau-\tau_i)\right)^p\,, 
    \quad p = \frac{2}{3(1+\omega)-2}\,,
\end{equation}
with $\tau_i$ an initial time. 
We define $\mathcal{H}_i\equiv \mathcal{H}(\tau_i)$ and, throughout the remainder of this work, set $a(\tau_i) = 1$.

\subsubsection{Initial fluctuations}\label{sec: initial fluctuations}

In order to mimic the Kibble mechanism arising from the cooling of the Universe, one would typically need to consider a temperature-dependent scalar potential. In this work, however, we neglect temperature corrections and the intrinsic dynamics of the phase transition, and instead start directly with the symmetry-breaking potential given in Eq.\eqref{exotic cosmological:potential}. Initially, the field is distributed around the maximum of the potential at $\phi \sim 0$, with small fluctuations. As the scalar field rolls down toward the symmetry-breaking vacua, domain walls will be naturally formed. The resulting number density of these walls is determined by the characteristics of the initial fluctuations.

We hence decompose the scalar field in a homogeneous component $\bar{\phi}$ and fluctuations $\delta \phi$, such that
\bea
\phi = \bar{\phi} + \delta \phi \,.
\eea 
We set the initial homogeneous amplitude of the scalar field at $\bar\phi = 0$. The fluctuations $\delta\phi$ are drawn from a distribution expressed in terms of a power spectrum $\mathcal{P}_{\delta\phi}(k)$,
\begin{equation}     \langle\delta\phi(\mathbf{k})\delta\phi(\mathbf{q})\rangle \equiv (2\pi)^3\mathcal{P}_{\delta\phi}(k)\delta(\mathbf{k}+\mathbf{q})\,.
\end{equation}
For the initial field configuration, we consider small Gaussian fluctuations $\delta\phi(\mathbf{k})$ in Fourier space  drawn from a white noise spectrum, and thus we impose $\mathcal{P_{\delta\phi}}$ to be constant. We normalize the power spectrum by demanding that the standard deviation of the scalar field in real space is equal to $0.1\eta$, \emph{i.e.}
\begin{equation}
     \sqrt{\langle\delta\phi^2(\mathbf{x})\rangle} = \sqrt{\int_0^{k_\text{cut}}  \frac{k^3}{2\pi^2}\mathcal{P}_{\delta\phi}(k)\ \mathrm{d}(\ln k)} = 0.1\eta\,,
\end{equation}
which results in 
\begin{equation}
    \mathcal{P}_{\delta\phi} = \frac{6\pi^2\langle\delta\phi^2(\mathbf{x})\rangle}{k_\text{cut}^3}\,.
\end{equation}
Here we applied a cutoff momentum $k_\text{cut}$ to avoid a divergent spectrum. Adjusting this cutoff changes the correlation length of the scalar field, which determines the initial number of domain walls. Indeed, increasing $k_{\rm cut}$ shortens the correlation length, meaning that the field is correlated over smaller distances and, as a result, more domain walls form within a given volume\footnote{
We notice here that alternative initial conditions (namely vacuum, thermal and scale invariant) were used in e.g. \,\cite{Hiramatsu:2013qaa, Kitajima:2023kzu, Ferreira:2023jbu, Li:2023yzq, Dankovsky:2024zvs, Cyr:2025nzf, Babichev:2025stm}. While this paper was in completion, Ref.\,\cite{Dankovsky:2025pjg} appeared
where they compare scenarios with different initial fluctuations.
}.

\subsubsection{Limitation of the lattice}

The dynamical range of the simulation is limited by the capability of resolving the DW width while having at least one Hubble patch at the end of the simulation. Indeed, within our comoving simulation box two effects occur over time. First, each Hubble volume enclosed in the box expands and can eventually become comparable to, or even exceed, the size of the simulation box. Second, the comoving DW width, given by $\delta_{\rm dw}/a(\tau)$, decreases as the scale factor grows, so that the DW can become smaller than the lattice spacing and therefore under-resolved. Given the comoving size of the simulation box $L$ and the total number of grid points $N^3$, we demand 1) that along a spatial direction, $L$ must be at least $\alpha$ times the comoving Hubble distance $\mathcal{H}^{-1}$ and 2) the DW width must be resolved by at least $\beta$ grid points. Picking reasonable values for $\alpha, \beta \geq 1$ imposes the following bounds:
\begin{equation}\label{eq: limitations}
  \alpha \mathcal{H}^{-1} < L\,, \quad \text{and}\quad \Delta x \equiv L/N < \frac{\delta_\text{dw}/a}{\beta}\,,
\end{equation}
 where $\Delta x$ is the grid spacing. Satisfying both of these conditions sets a maximum timescale up to which the simulation can run for. Using the evolution of the scale factor as given by Eq.\eqref{eq: a}, the final simulation time is determined by
\begin{equation}\label{eq: tend}
    \tau_\text{end} = \tau_i + \frac{p}{\mathcal{H}_i}\left[\left(\frac{2N}{\alpha\beta}\frac{\mathcal{H}_i}{m}\right)^{\frac{1}{p+1}}-1\right]\,,
\end{equation}
where we used $\delta_\text{dw} = 2/m$. By plugging this back in Eq.\,\eqref{eq: limitations}, we can deduce the box size $L$ that allows us to reach that time scale, namely
\begin{equation}\label{eq: box size}
    L = \frac{\alpha}{\mathcal{H}_i}\left(\frac{2N}{\alpha\beta}\frac{\mathcal{H}_i}{m}\right)^{\frac{1}{p+1}}\,.
\end{equation}

\bigskip
Having established the lattice setup, we now turn to discuss the results of the simulations themselves. In the following section, we investigate the approach to the scaling regime of the domain wall network by varying the initial field fluctuations and analyzing how rapidly scaling is achieved.

\section{Domain wall evolution: from formation to scaling}
\label{sec:approach_scaling}

In this section, we investigate the dynamics of the domain wall network in the expanding Universe.
After formation, the network evolves rapidly to an attractor solution with $\mathcal{O}(1)$ domain walls per Hubble volume, \emph{i.e.} the so-called \emph{scaling} regime \cite{Press:1989yh,Garagounis:2002kt, Oliveira:2004he,Avelino:2005pe,Kawasaki:2011vv}.
We will study how fast this attractor solution is reached within our numerical simulations considering a different set of initial conditions.

\subsection{Basics of the domain wall network}

\paragraph{Energy density and area parameter.}
The energy density in the domain wall network is determined by the total area covered 
by the walls and by the domain wall tension.
Following the discussion in \cite{Hiramatsu:2013qaa}, it is computed as the energy per unit comoving volume,
\begin{equation}
\rho_\text{dw} = \frac{\sigma A}{a(\tau) V},
\end{equation}
where $\sigma$ is the wall tension, $A$ the total comoving area of the walls, $a(\tau)$ the scale factor, and $V$ the comoving volume.
The total comoving area  $A$ can be extracted from the numerical simulations with the methods that we will explain later.
From the total area occupied by the domain walls, 
one can define the area parameter $\mathcal{A}$, which is a dimensionless quantity measuring the amount of domain walls per Hubble volume,
\begin{equation}\label{eq:method 1}
    \mathcal{A} = \frac{A}{2 \mathcal{H} V}\,.
\end{equation}
With the previous definition, the energy density in domain walls can be simply expressed as 
\begin{equation}\label{eq:rho_dw}
    \rho_\text{dw} = 2\mathcal{A} \sigma H,
\end{equation}
where we used $\mathcal{H} = a H$.
Note that $\mathcal{A}$ is generically a function of time encoding the time evolution of the domain wall network energy density.

\paragraph{The domain wall network at formation.}
The domain walls are formed at the phase transition through the Kibble mechanism. As we expect the initial size of the correlated patches to be roughly given by the inverse Ginzburg temperature $1/T_{G} \sim 1/m$,  the value of $\mathcal{A}$ is controlled by the ratio of the Hubble radius to the correlation length:
\bea\label{eq: A initial}
\mathcal{A}_{\rm initial} \sim \frac{m}{H(T\sim m)} \sim \frac{M_{\rm Pl}}{m} \gg 1\, \qquad \text{(initial number of DWs per Hubble volume)} \,,
\eea 
where we assumed a radiation dominated Universe, \emph{i.e.} $H \sim T^2/M_\text{Pl}$, with $M_{\rm Pl}$ the reduced Planck mass.
The number of DWs per Hubble will however rapidly drop, 
as we shall see from the numerical simulations.

\paragraph{The scaling regime.}
After formation, the domain wall network is rapidly attracted to a regime where the area of the DWs scales like the Hubble radius, and the number of DWs per Hubble volume is constant and $\mathcal{O}(1)$.
This attractor solution is set by the interplay between the expansion of the Universe and the recombination of the different portions of the domain wall network upon entering the horizon.  
In this regime, the only relevant scale controlling the dynamics of the domain wall network is the Hubble rate.
As a result, the network is characterized by a single typical length scale, $\ell$, that follows the relation $\ell \sim H^{-1}$. This length scale represents not only the average wall length, but also the mean curvature radius and the typical separation between neighboring walls.
From this description, one can directly infer a simple scaling behaviour for the energy density associated to the domain wall network:
 \bea \label{eq: rho dw wiggle}
 \rho_{\rm dw} \sim \frac{\sigma}{\ell} \sim
 \sigma H.
 \eea 
Comparing this with the definition of the area parameter in Eq.\,\eqref{eq:rho_dw}, 
we conclude that during scaling the area
parameter $\mathcal{A}$ is a constant,
corresponding to the number of DWs per Hubble volume.
Note that the factor of $2$ in Eq.\,\eqref{eq:rho_dw} is conventional and comes from comparison with \cite{Kawasaki:2011vv}, where the scaling behaviour was expressed in terms of cosmic time $t$ in a radiation dominated Universe, for which $H =1/(2t)$. 

The approach to scaling has been investigated previously by several groups. One of the first numerical studies was carried out in \cite{Press:1989yh}, where the authors introduced the so-called PRS algorithm. In this method, the \emph{comoving} DW width, rather than the physical width as considered in our case, was kept fixed by modifying the scalar field’s EoM such that the overall wall dynamics remained largely unaffected. The DW network was found to exhibit a scaling behaviour with a logarithmic correction. However, later simulations \cite{Garagounis:2002kt} which were also using the PRS algorithm, did not reproduce this correction. Subsequent studies \cite{Oliveira:2004he, Avelino:2005pe} showed that the network approaches scaling slowly, with strong deviations from the linear scaling behaviour at early times. Advances in computational power later allowed for a cleaner demonstration of the scaling solution and a quantitative calibration of numerical results against the analytical VOS model \cite{Leite:2011sc}, which we will discuss briefly in Section~\ref{sec: VOS}. More recent simulations \cite{Kawasaki:2011vv, Hiramatsu:2013qaa}, employing a fixed \emph{physical} width, have confirmed the existence of a scaling regime, and the latest studies \cite{Kitajima:2023kzu, Ferreira:2024eru, Dankovsky:2024zvs, Notari:2025kqq, Cyr:2025nzf} indicate that the area parameter $\mathcal{A}$ eventually remains constant and $\mathcal{O}(1)$ across different initial conditions, although its precise value varies. Analytical approaches using the thin-wall limit further support the existence of the scaling regime \cite{Hindmarsh:1996xv, Hindmarsh:2002bq}.

Note that the scaling regime is such that the 
fraction of energy density of the DW network with respect to the critical density of the Universe grows with time, specifically
$\Omega_{\rm dw} \sim  \sigma/(H M_{\rm pl}^2)$.
This would lead to the DW energy density dominating the energy content of the Universe at sufficiently late times after formation, leading to an inconsistent cosmology.
This is the cosmological DW problem\,\cite{Zeldovich:1974uw,Sikivie:1982qv}.
It is usually solved by introducing a small deformation of the theory, consisting of a small energy bias between otherwise degenerate minima that eventually makes the DW network collapse before it dominates the energy content of the Universe.
In this work, we focus only on the characterization of the scaling regime and its GW signatures, so that we will not introduce a bias term in our numerical simulations. This would be however crucial for simulations of the DW network collapse, see \cite{Kitajima:2023cek, Ferreira:2024eru, Babichev:2025stm, Notari:2025kqq, Cyr:2025nzf}.

\subsection{Approach to scaling: numerical results}\label{sec: scaling numerical results}

In this subsection we study the approach to the scaling regime for the domain wall network.
In particular, we would like to determine how fast the network reaches a configuration with $\mathcal{O}(1)$ domain walls per Hubble volume\,\footnote{See \emph{e.g.} Refs.\,\cite{Gorghetto:2018myk,Correia:2024cpk} for similar studies in the context of cosmic string networks.}.

To demonstrate the (in)dependence of the rapid onset of the scaling regime on the initial conditions and the initial number of domain walls, we performed several simulations with different setups. First, the scalar field is distributed randomly according to a white noise spectrum, as outlined in Section \ref{sec: initial fluctuations}. A natural choice for the momentum cutoff is $k_{\rm cut} = m$, since in the initial stage the size of the uncorrelated patches is expected to be given by the inverse Ginzburg temperature $\sim 1/m$. However, in order to mimic the larger initial domain wall density suggested by Eq.\,\eqref{eq: A initial} in realistic scenarios, we also consider enhanced cutoffs with $k_{\rm cut}/m = 2$ and $5$.
Additionally, we vary the initial Hubble scale by considering a range of mass-to-Hubble ratios $m/H_i$ from $1$ up to $500$, as larger values of $m/H_i$ effectively increase the number of domain walls contained within a single Hubble patch. 

In order to derive the area parameter $\mathcal{A}$
of the DW network from
the simulation data, we should introduce an algorithm that can identify the position of the domain walls
and estimate their surface.
This is performed according to the prescription of \cite{Press:1989yh}. The idea is to loop over all grid points and look for \emph{links}, \emph{i.e.} neighboring grid points for which the sign of the scalar field $\phi$ is changing. When such a link is found, the wall area is incremented by an amount $(\Delta x)^2$ divided by a weighting factor. One then has:
\begin{equation}
\label{eq:A_total}
    A = (\Delta x)^2 \sum_{\text{links}}\delta\frac{|\nabla \phi|}{\left|\frac{\partial\phi}{\partial x}\right| + \left|\frac{\partial\phi}{\partial y}\right| + \left|\frac{\partial\phi}{\partial z}\right|}\,,
\end{equation}
where the summation goes over all links and $\delta = 1$ if the link crosses a wall and $0$ otherwise. 
This provides an estimate for the total area covered by the domain walls, which is then used to obtain the area parameter with the relation \eqref{eq:method 1}
(later on we will introduce another equivalent but more intuitive method to compute the are parameter).

\begin{figure}
    \centering
    \includegraphics[width=0.6\linewidth]{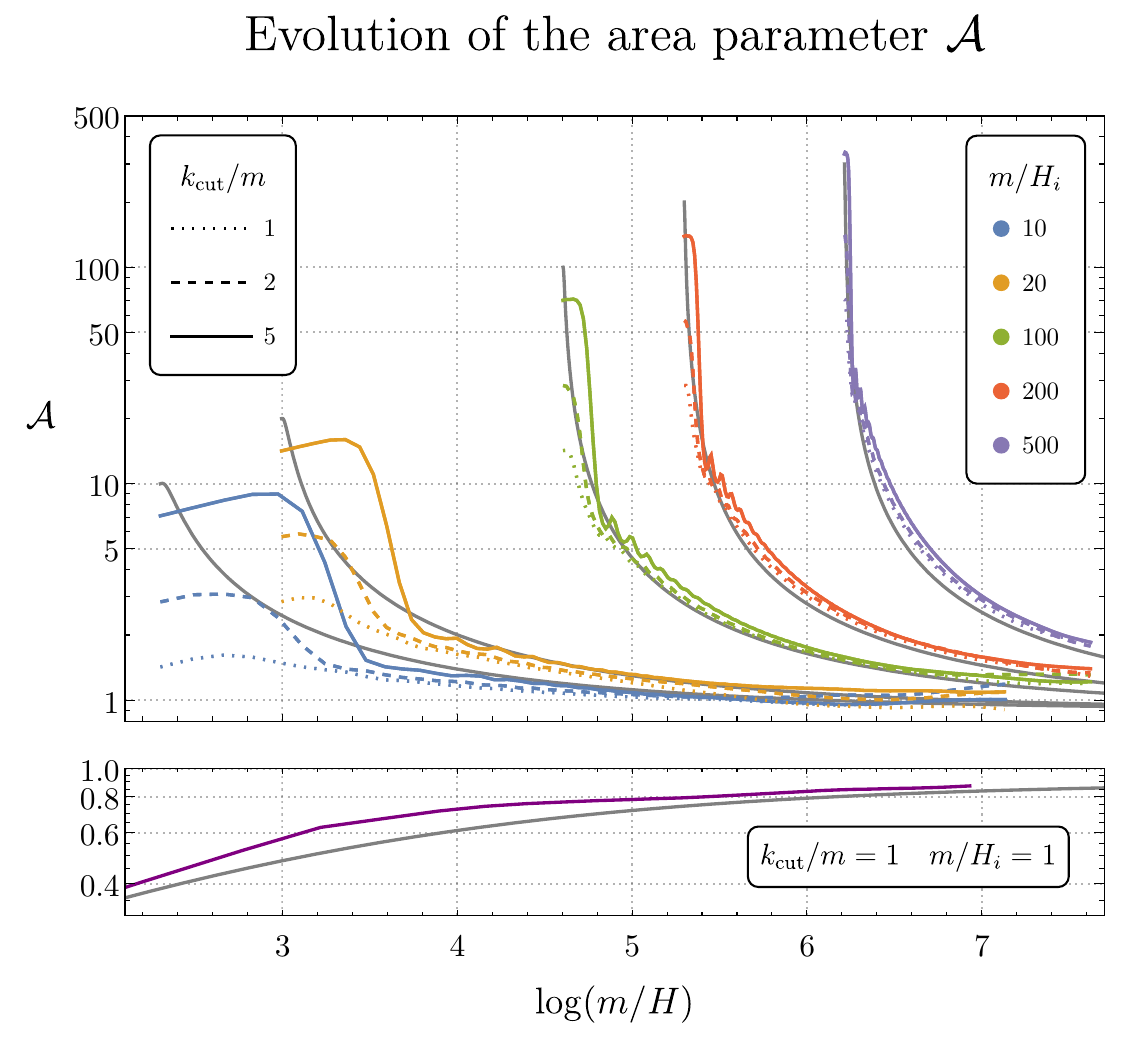}
    \caption{\textbf{Top:} Evolution of $\mathcal{A}$ for different initial conditions. Here, each color denotes a different initial ratio $m/H_i$. The dotted, dashed and solid lines correspond to different  initializations of the fluctuations with $k_{\rm cut}/m = 1,2,5$. Each curve displays the average of three simulations, with $N^3 = 1250^3$ for $m/H_i = 10$, $20$ and $N^3 = 2048^3$ otherwise. For each simulation, the box size $L$ is chosen such that we are left with one Hubble volume by the end of the simulation and we resolve the DW width by 2 grid points, which corresponds to choosing $\alpha = 1$ and $\beta = 2$ in Eq.\eqref{eq: box size}. Finally, the gray lines display the result for $\mathcal{A}$ via the VOS model as described in Section \ref{sec: VOS}, where in particular we compare to the cases with $k_{\rm cut}/m = 5$. \textbf{Bottom:} Same as above, where the purple curve represents the average of 5 simulations for $N^3 = 2048^3$, $k_{\rm cut}=m$ and $H_i = m$. In this case, and contrarily to the Top plot, the initial DW network is underdense. Here, the box size $L$ was chosen such that $\alpha = \beta = 2$ in Eq.\eqref{eq: box size}. The gray line displays the area parameter as given by the VOS model. }
    \label{fig:approach_scaling}
\end{figure}

\begin{figure}
    \centering
    \includegraphics[scale = 0.32]{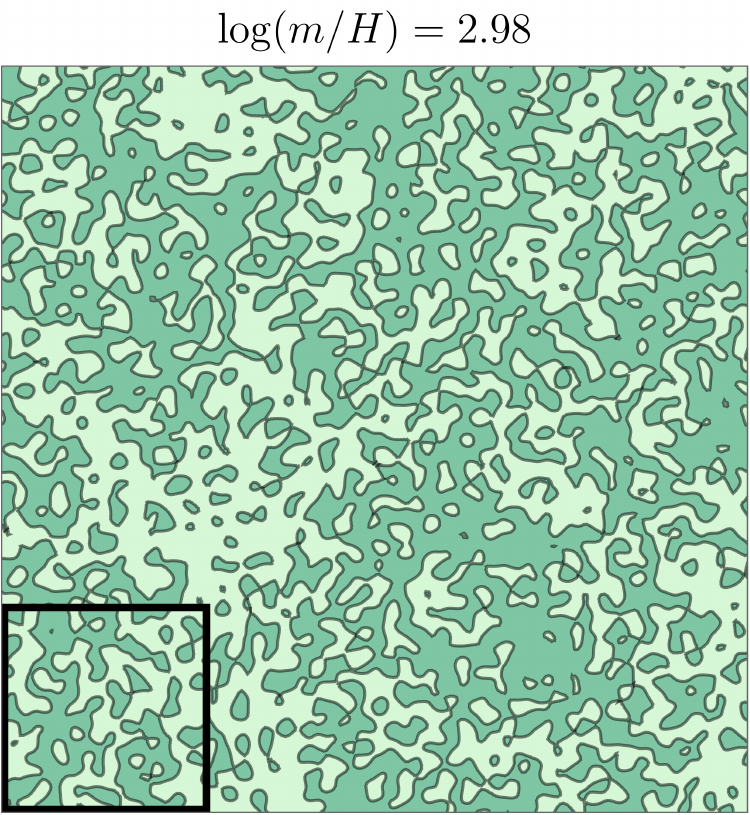}
    \includegraphics[scale = 0.32]{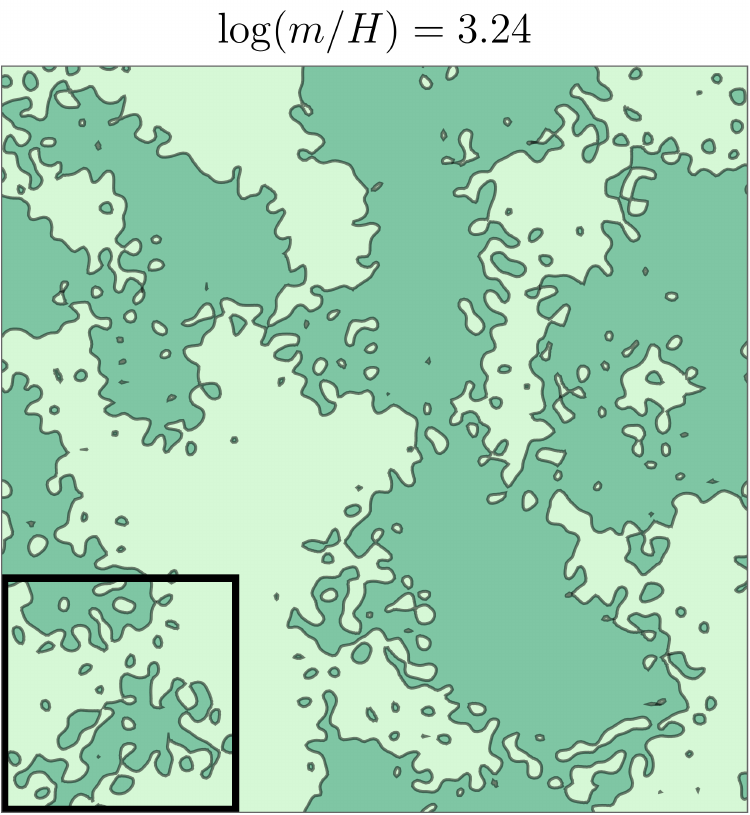}
    \includegraphics[scale = 0.32]{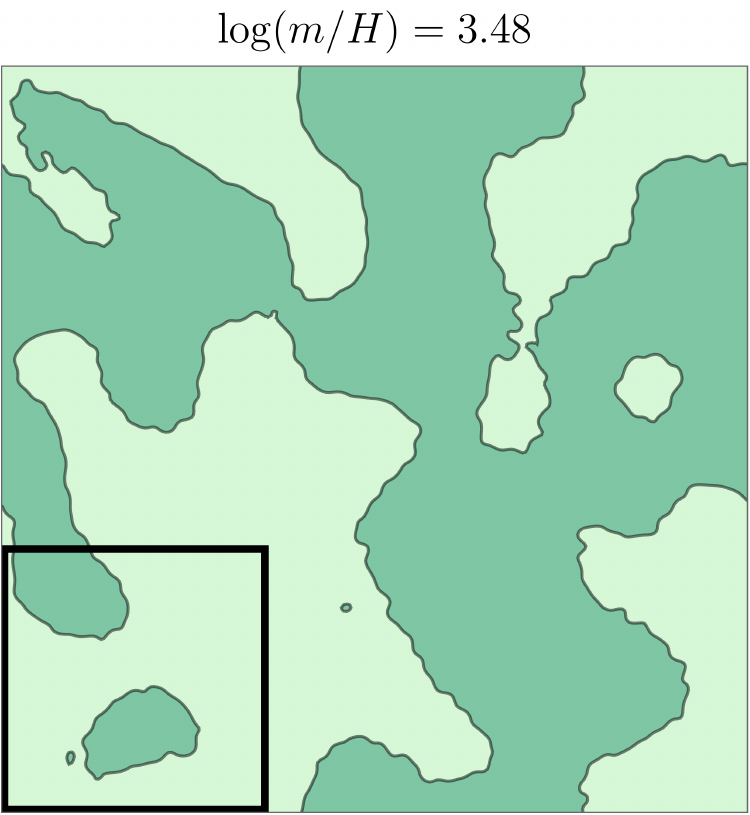}
     \includegraphics[scale = 0.32]{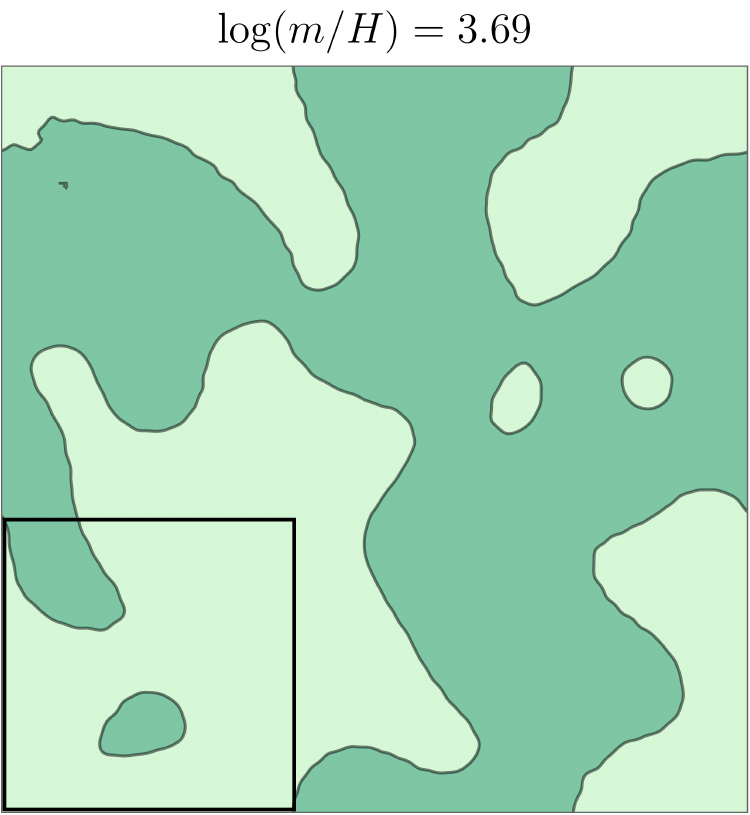}
    \caption{Slices of a 3D simulation of a DW network at different times specified by $m/H$ with initial conditions such that $m/H_i = 10$ and $k_{\rm cut}/m = 5$ for $N^3 = 256^3$. The black square indicates the size of one Hubble patch.}
    \label{fig:XiSim}
\end{figure}

In Fig. \ref{fig:approach_scaling}, we show the evolution of the parameter $\mathcal{A}$ considering a radiation dominated Universe, \emph{i.e.} $p = 1$, using the method above. It is observed that within a single $e$-fold, the number of domain walls decreases significantly, and the network eventually reaches the scaling regime, with $\mathcal{O}(1)$ domain walls per Hubble patch. As expected, a higher density of domain walls leads to increased interactions between them, which accelerates the approach to scaling. For illustration, we present a slice of four consecutive snapshots from a small simulation with $N^3 = 256^3$ grid points, $m/H_i = 10$, and $k_{\rm cut}/m = 5$ in Fig. \ref{fig:XiSim}. These snapshots illustrate that indeed after less than one $e$-fold, the dense network evolves into a simpler, less dense configuration, approaching the scaling regime with $\mathcal{O}(1)$ domain walls per Hubble volume.

\subsection{Comparison with the Velocity-dependent One-Scale (VOS) equations}\label{sec: VOS}

It is interesting to compare the results of our simulations regarding the approach to scaling with the predictions from the Velocity--dependent One--Scale model for domain walls 
\cite{
Avelino:2005kn,Avelino:2019wqd,Martins:2016ois, Avelino:2022zem,Leite:2011sc,Leite:2012vn,Martins:2016lzc}.

The VOS model can be deduced from the EoM of the scalar field describing the DW 
in the thin wall approximation, plus the condition of energy conservation. 
One obtains two equations involving the 
curvature of a single domain wall and its velocity. 
Taking then the average of these quantities over the entire network leads to the following differential equations:
\begin{align}
\begin{cases}
    \frac{d \ell}{dt} &= (1 + 3v^2)H\ell + c_w v + d\left[k_0 - k_w(v)\right]^r\\[6pt]
    \frac{d v}{dt} &= (1-v^2) \bigg(\frac{k_w(v)}{\ell}- 3Hv\bigg)
\end{cases} \qquad 
\text{(VOS equations)}\,,
\end{align} 
where 
\bea 
k_w(v) = k_0\frac{1-(qv^2)^\beta}{1+(qv^2)^\beta}\,.
\eea 
In these equations $\ell(t)$ represents the typical lenght scale of the domain wall network, and $v(t)$
the average velocity.
The average energy density of the network is $\rho_{\rm dw} = \sigma/\ell$ and the number of domain walls per Hubble volume is simply $\mathcal{A} = 1/(2H\ell)$.

The basic assumption of the VOS is that the entire dynamics of the network is captured only by the time evolution of $\ell(t)$ and $v(t)$. The terms on the right-hand side of the VOS differential equations account for Hubble friction, curvature effects on the velocity, and energy loss mechanisms such as domain wall interactions (chopping) and particle radiation.
We have used the following
numerical values taken from
\cite{Martins:2016lzc}: $c_w = 0, \, d = 0.26, \, r = 1.42, \, q = 3.35, \, \beta = 1.08, \, k_0 = 1.77$ and we set very small initial velocity $v_{\rm initial} = 10^{-5}$. The evolution of the $\mathcal{A}$ parameter is then obtained by using $\mathcal{A}(t) = t/\ell(t)$.
With this choice of parameters, the asymptotic value for the scaling in the VOS equations is $\mathcal{A} \simeq 0.9 $ and $v\simeq 0.38$, respectively.

We display in gray the parameter $\mathcal{A}$ obtained via the VOS in Fig. \ref{fig:approach_scaling} for each set of initial conditions, and observe that it provides a faithful description of the evolution towards scaling after a few $e$-folds, independently on the initialization of the fluctuations.

\subsection{An alternative algorithm for the area parameter}
\begin{figure}
    \centering
    \includegraphics[width=0.5\linewidth]{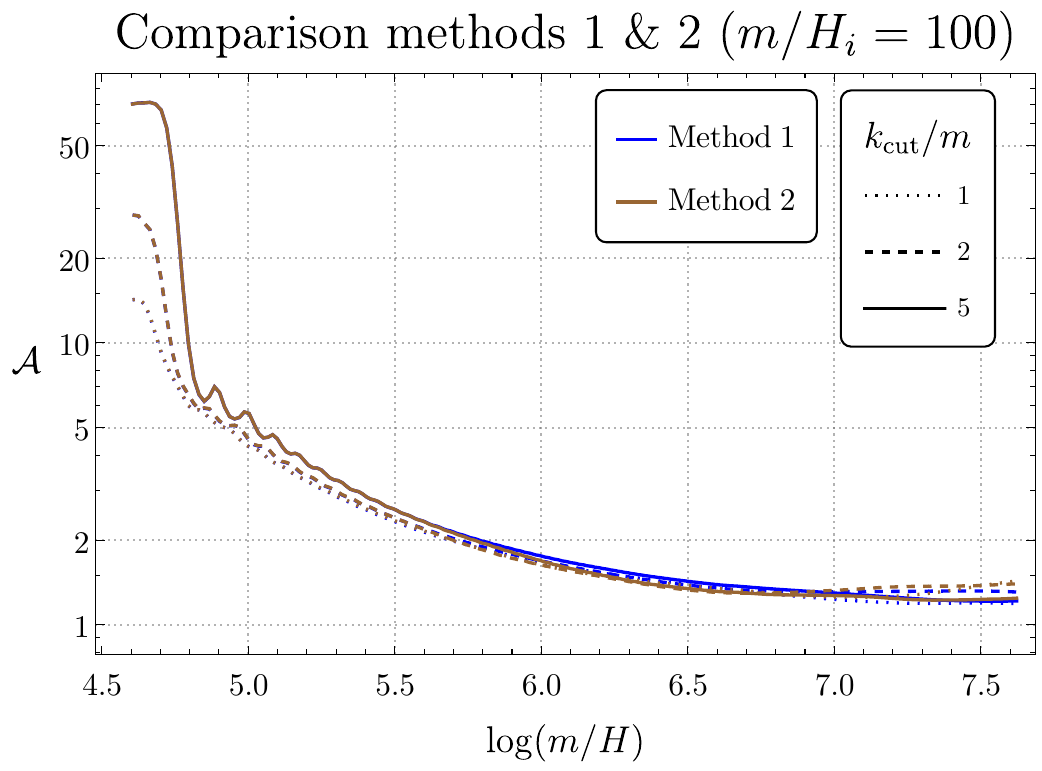}
    \caption{Comparison of the area parameter computed via methods 1 (blue) and 2 (brown) as explained in the text for the initial ratio $m/H_i = 100$ and three different cutoffs $k_{\rm cut}/m = 1$ (dotted), 2 (dashed) and 5 (solid).}
    \label{fig: Comparison Area}
\end{figure}

As mentioned previously, the area parameter $\mathcal{A}$ is expected to count the number of DWs per Hubble volume. Given a snapshot of the field theory simulation, we can hence define an alternative algorithm to estimate $\mathcal{A}$ based on the following physical intuition.
To get the number of walls within one Hubble radius, we fix a certain direction in the simulation box and count how many times the sign of the scalar field changes along this line, thus indicating the presence of a series of domain walls. 
By averaging over several random directions, this procedure gives the average number of DWs ($\equiv N_\text{dw}$) within a box of length $L$. Using this method, we can then find the area parameter as
\begin{equation}
\label{eq:line_of_sight}
    \mathcal{A} = N_\text{dw} \frac{\mathcal{H}^{-1}}{L}\,.
\end{equation}
We applied and compared the two methods
to calculate the area parameter $\mathcal{A}$,  
namely the one based on Eq.~\eqref{eq:A_total} 
({\bf Method 1}),
and the one just explained based on 
Eq.~\eqref{eq:line_of_sight} ({\bf Method 2}), where we considered $500$ random directions. 
Our analysis shows that both methods capture the same qualitative behaviour of $\mathcal{A}$, with differences limited to the percent level.
As a representative case, Fig. \ref{fig: Comparison Area} illustrates the comparison for $m/H_i = 100$.
The equivalence of the two methods
further underlines
the interpretation of the area parameter as the number of domain walls per Hubble volume.

\section{Gravitational waves from scaling domain walls}
\label{sec:GW_from_scaling}

In this section we study the production of gravitational waves from the motion of domain walls in the scaling regime in a radiation dominated (RD) Universe.
We first review basic aspects about the gravitational wave energy density, and introduce the parametrization that we use to describe the GW spectrum.
We will then show our results obtained with $N=2048$ and $N=4096$ grid sizes (hereafter denoted as $2K$ and $4K$) by employing the \CL ~package for gravitational waves, and compare with existing results in the literature. We will thus improve over former studies thanks to the higher resolution of our numerical simulations,
and set the stage for our novel investigation based on the Equal Time Correlator as well as different types of cosmologies to be discussed in Sec.\,\ref{sec: UTC} and \ref{sec: other cosmologies}.


\subsection{Emission of gravitational waves: theoretical background}
\label{eq:GW_emission_theory}

The production and the evolution of the GWs in the presence of a source is described by
\begin{equation}\label{eq: EoM h}
    \frac{\partial^2 h_{ij}}{\partial\tau^2} + 2\mathcal{H}\frac{\partial h_{ij}}{\partial\tau} - \nabla^2 h_{ij} = 16\pi GT_{ij}^{TT}\,,
\end{equation}
where $h_{ij}$ is the transverse-traceless (TT) part of the GW metric perturbation, $G$ is Newton's constant and $T_{ij}^{TT}$ is the TT part of the energy momentum tensor, which in this work is given by $T_{ij}^{TT} = (\partial_i\phi\partial_j\phi)^{TT}$.
The energy stored in GWs is then encoded in the GW energy density defined as
\bea 
\rho_{\rm gw} = \frac{1}{32\pi G a^2} \left\langle \frac{\partial h_{ij}}{\partial \tau}\frac{\partial h_{ij}}{\partial \tau} \right\rangle \, ,
\eea 
where $\langle \ldots\rangle$ denotes a spatial average over all relevant wavelengths of the perturbations $h_{ij}$ and summation over repeated indices is implied. The associated  energy fraction per logarithmic frequency interval is then
    \bea 
    \label{eq:energy_density_GW}
    \Omega_{\rm gw}(\tau, k) = \frac{1}{\rho_c(\tau)} \bigg(\frac{d \rho_{\rm gw}(\tau)}{d \ln k}\bigg) \, ,
    \eea 
where $\rho_c(\tau) = 3 M_{\rm pl}^2 H^2$ is the critical energy density of the Universe at conformal time $\tau$.

The GW energy density emitted at each time can be estimated from very basic principles, as shown in \cite{Hiramatsu:2013qaa}. To this end, we start from the quadrupole formula for the power injected in the GWs,
\bea 
\label{eq:power}
P_{\rm gw} = \frac{G}{5} \left\langle \dddot{Q}_{ij}\dddot{Q}_{ij} \right\rangle \, ,
\eea 
where a dot denotes a derivative w.r.t. \emph{cosmic time} $t$ and $Q_{ij}$  is the mass quadrupole moment tensor defined in the form 
\bea 
\label{eq:quad}
Q_{ij}(t) = \int \rho_S(\mathbf{x}, t) \left( x_i x_j - \frac{1}{3} \delta_{ij} \, r^2 \right) d^3x \,.
\eea 
Here, $\rho_S$ is the energy density of the source's mass distribution and $r$ the distance to the observer. We can now integrate the power $P_{\rm gw}$ over time to obtain the GW energy density in a volume $V$, 
\bea 
\label{eq:energ_d}
\rho_{\rm gw} \equiv \frac{1}{V} \int dt P_{\rm gw} \sim  P_{\rm gw}\frac{t}{V} \, .
\eea  
In the case of a DW network in scaling, there is only one relevant scale, $\ell \sim H^{-1}\sim t$, which governs the whole dynamics. Consequently, one can expect that $Q_{ij} \sim M_{\rm dw}\ell^2$, where $M_{\rm dw} \sim \mathcal{A}\sigma \ell^2$ is the mass of the wall, and $\dddot{Q}_{ij} \sim M_{\rm dw}/\ell$. 
Putting together Eqs.\eqref{eq:power} and \eqref{eq:energ_d} and considering $V/\ell^3$ emitters within the volume $V$, one obtains the following estimation for the energy density of the GWs:
\begin{equation}
    \rho_{\rm gw} \sim G\frac{M_{\rm dw}^2}{\ell^2}\times\frac{t}{V}\times\frac{V}{\ell^3} \sim G\mathcal{A}^2\sigma^2\,,
\end{equation}
which is hence constant in time. This implies that the associated (integrated) energy fraction grows as
\begin{equation}\label{eq: integrated rhogw}
    \int\Omega_{\rm gw}(\tau, k) {\rm d}(\ln k) =\frac{\rho_{\rm gw}}{\rho_c} \sim \tau^4\\,
\end{equation}
where we used $H = \mathcal{H}/a \sim \tau^{-2}$ in radiation domination.
We furthermore define the parameter $\epsilon_{\rm gw}$ that encapsulates the efficiency of GW production,
\begin{equation}
\label{eq:epsilonGW}
    \epsilon_{\rm gw} = \frac{\rho_{\rm gw}(\tau)}{G\mathcal{A}^2\sigma^2},
\end{equation}
which will remain constant in time as well after the DW network enters the scaling regime.

\subsection{Direct numerical computation of the GWs}\label{sec: DW Rad}

\CL ~allows us with a dedicated routine (we refer to \cite{GWTechNote} for the details)
to study the dynamics of the GW perturbations, thereby evolving Eq.\eqref{eq: EoM h} and extracting the GW energy density and spectrum. 

In what follows, a white noise spectrum with $k_{\rm cut} = m$ is used with the homogeneous value of the scalar field set to zero, and we choose the initial Hubble size to be given by the mass, $H_i = m$, with initial time $\tau_i = 1/m$ in a radiation dominated Universe. The benchmark considered here is the purple line in Fig. \ref{fig:approach_scaling}.
From a practical standpoint, simulations with smaller mass-to-Hubble ratios are also less time consuming, as indicated by Eq.~\eqref{eq: tend}, where the final simulation time scales as $m\tau_{\rm end} \sim \sqrt{N \times m/H_i}$.

We perform an average of $5$ simulations with $N^3=2048^3$ and one larger simulation with $N^3 = 4096^3$ with similar initial conditions.
The different $2K$ simulations allow us to take averages for the observable quantities, while the $4K$ one
has the advantage of stretching the time-range of validity of the simulation, thus exploring more deeply the scaling regime. For the $2K$ simulations, the box size $L$ is chosen such that the volume contains 8 Hubble patches at the end of the run, while still resolving the domain walls with at least two grid points. In terms of Eq.~\eqref{eq: box size}, this corresponds to setting $\alpha = \beta = 2$. For the $4K$ simulations, the box size $L$ is determined in a similar manner. However, we conservatively terminate the simulation slightly before the expected final time (specifically $m\tau = 40$ instead of $m\tau = 45$) to remain safely within the lattice resolution limits.

In the following, we will quantitatively determine the parameters associated to the scaling regime and the corresponding GW emission from our simulation data.

\begin{figure}
    \centering
     \includegraphics[scale=0.55]{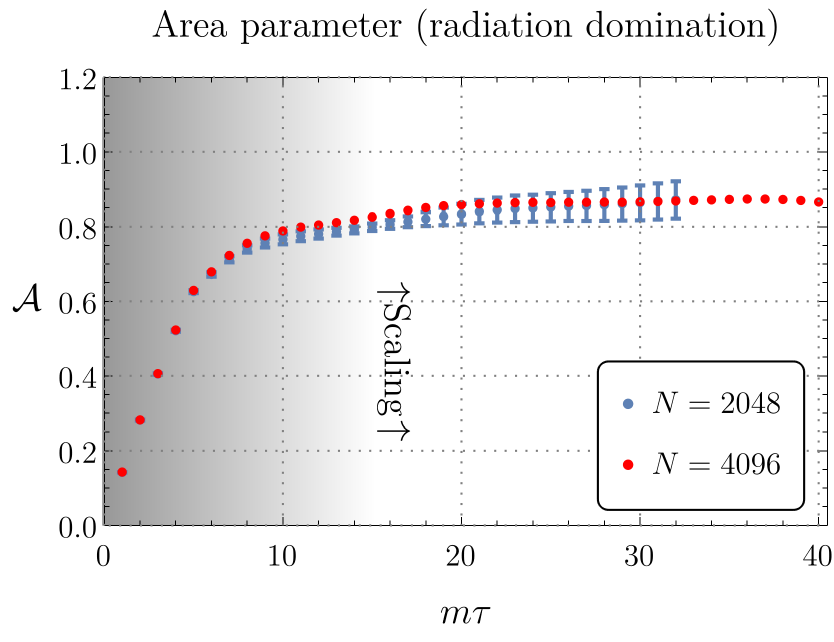}
    \includegraphics[scale=0.6]{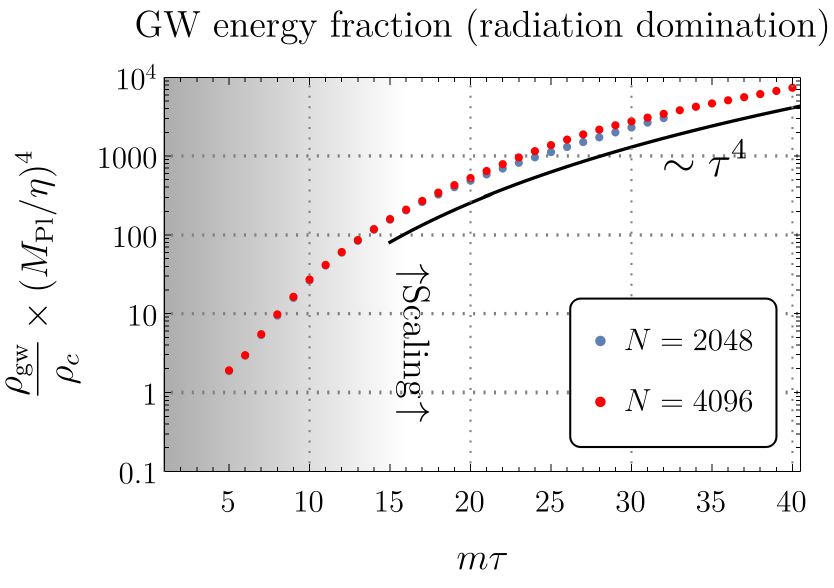}
    \caption{\textbf{Left:} Evolution of the area parameter $\mathcal{A}$ using method 1 as discussed in Section \ref{sec: scaling numerical results}. In blue, we show the results for simulations with $N^3 = 2048^3$ grid points, where the averages and error bars were taken from 5 simulations using different base seed values. The same data are also represented by the purple line in Fig. \ref{fig:approach_scaling}. Results from the $4K$ simulation are displayed in red. The estimated onset of scaling is around $m\tau = 15$.
    \textbf{Right:} Evolution of the GW energy fraction $\rho_{\rm gw}/\rho_c$ for the same simulations as in the left plot. The black line corresponds to the expected growth of the GW energy fraction in a RD Universe, \emph{i.e.} $\sim \tau^4$.}
    \label{fig:GW_energy}
\end{figure}

\paragraph{Estimate of scaling parameters.}
On the left panel of Fig. \ref{fig:GW_energy}, we display the evolution of the area parameter $\mathcal{A}$ using method 1 as discussed in Section \ref{sec: scaling numerical results}. 
For the $2K$ simulation we show the average over the $5$ simulations with corresponding error bars. The $4K$ simulation is displayed in red up to a larger time $m \tau=40$.

 We observe that we reach the scaling regime as the area parameter converges towards a constant value, for which the average over our $2K$ simulations gives
\begin{equation}
\label{eq:Aparam}
    \mathcal{A} = 0.87 \pm 0.05\,,
\end{equation}
consistently with values obtained in\,\cite{Hiramatsu:2013qaa, Dankovsky:2024zvs, Ferreira:2024eru, Cyr:2025nzf}. The value extracted from the $4K$ simulation at $m\tau = 40$ agrees with this result.
From Fig. \ref{fig:GW_energy}, we estimate that 
scaling is reached approximately from $m \tau \gtrsim 15$, denoted with a gray-to-white shading in the plot.

\begin{figure}
    \centering
    \includegraphics[width=0.48\linewidth]{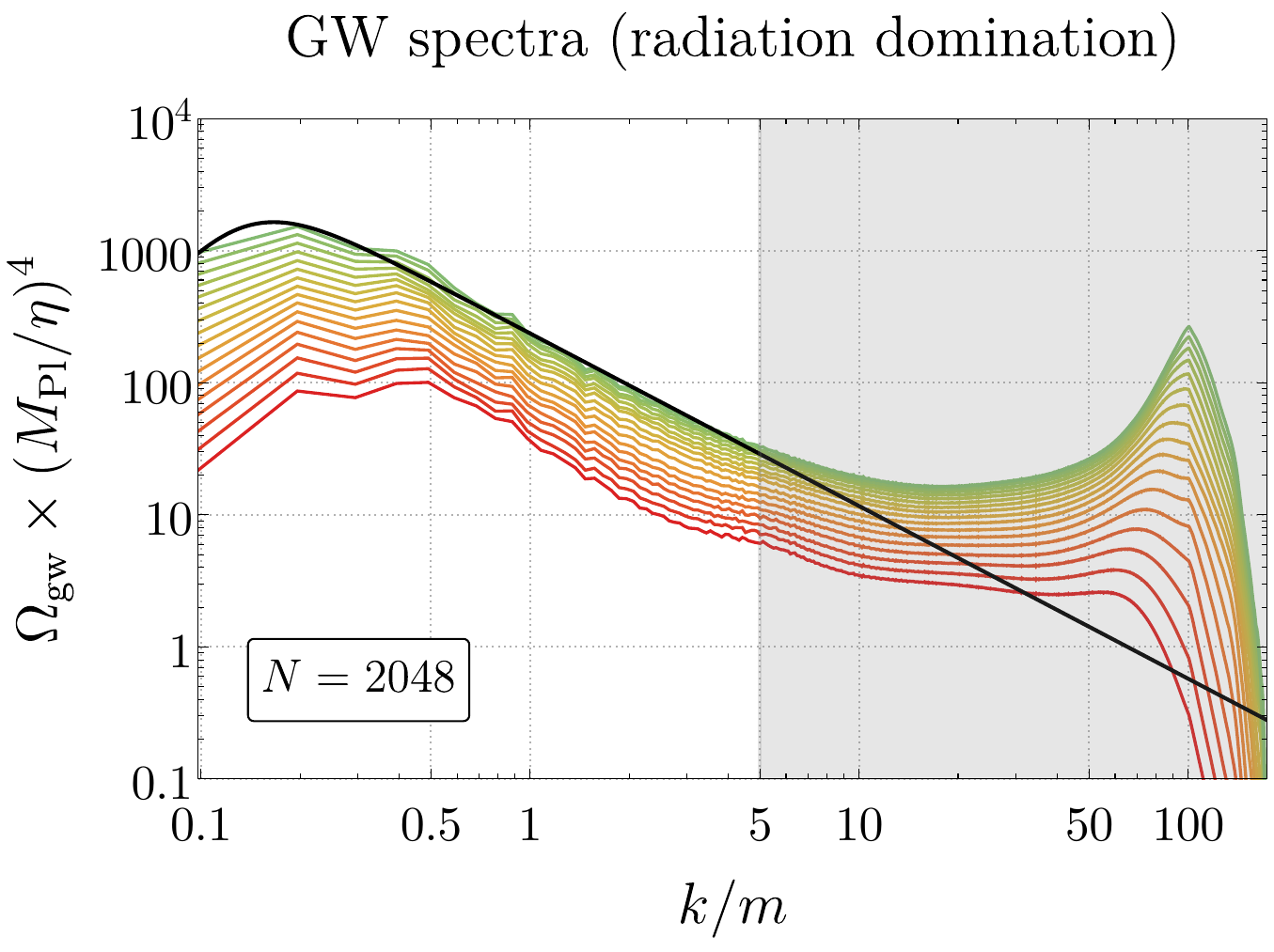}
     ~~~
     \includegraphics[width=0.48\linewidth]{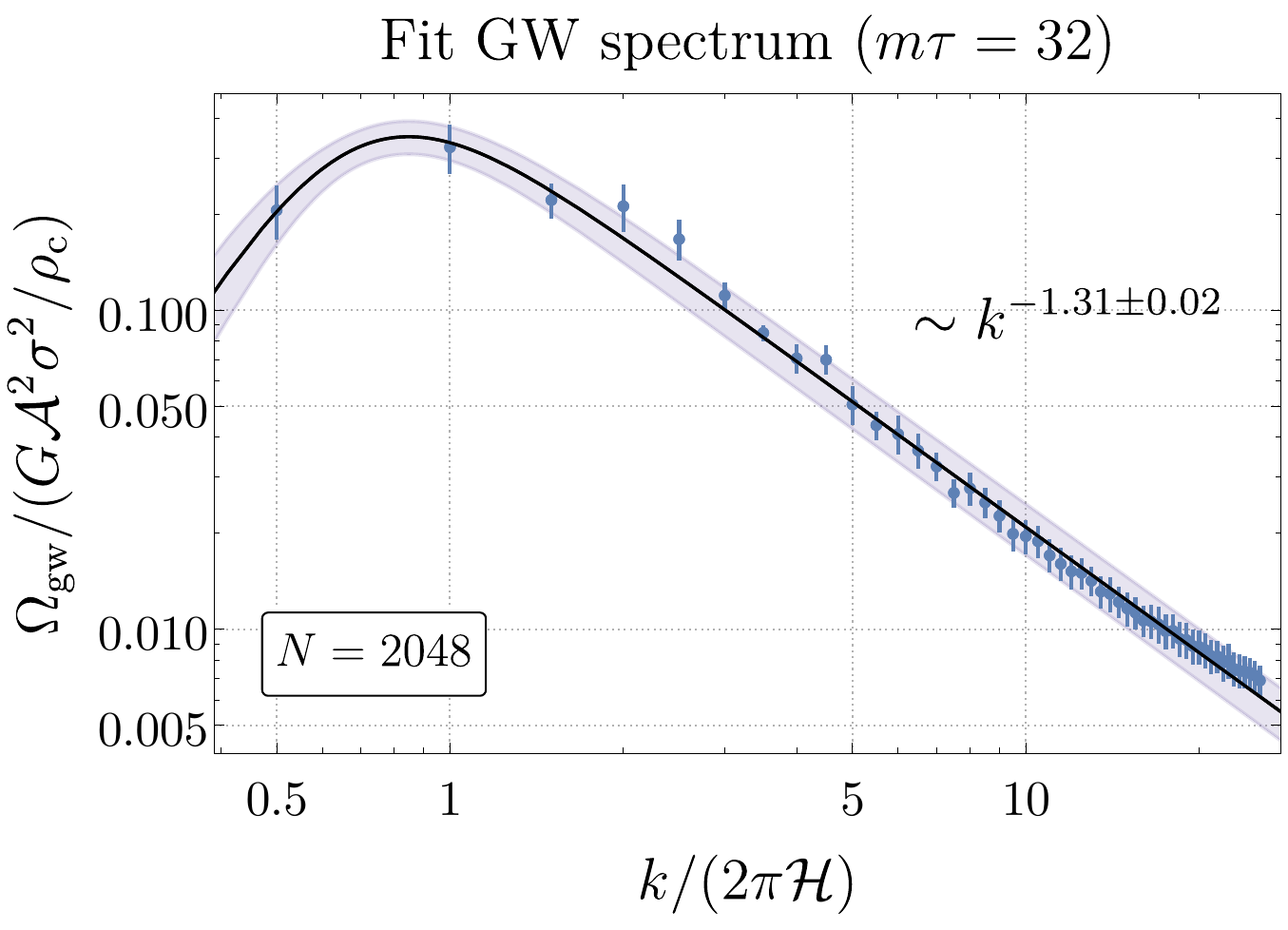}
    \includegraphics[width=0.48\linewidth]{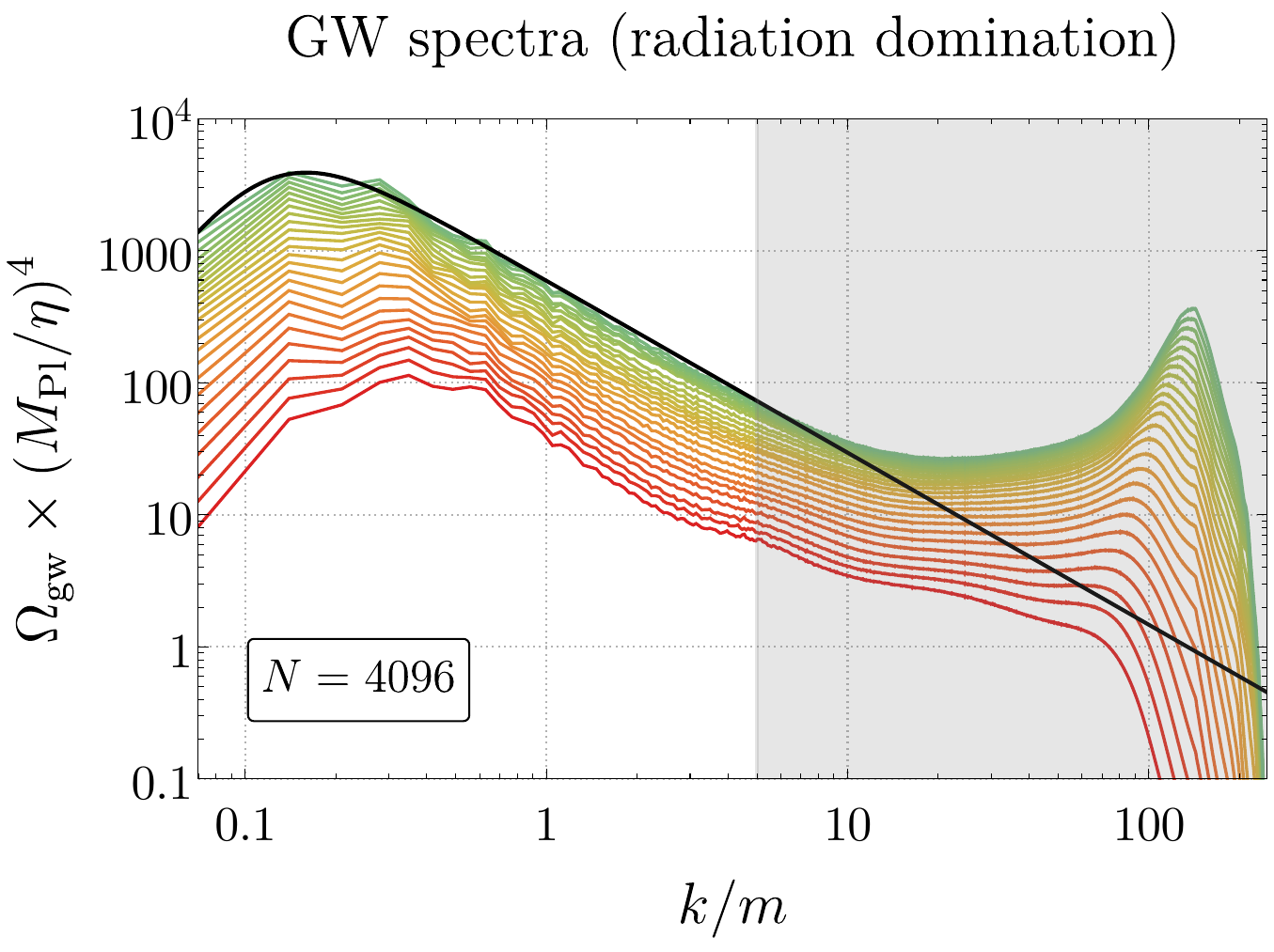}
    ~~~
    \includegraphics[width=0.48\linewidth]{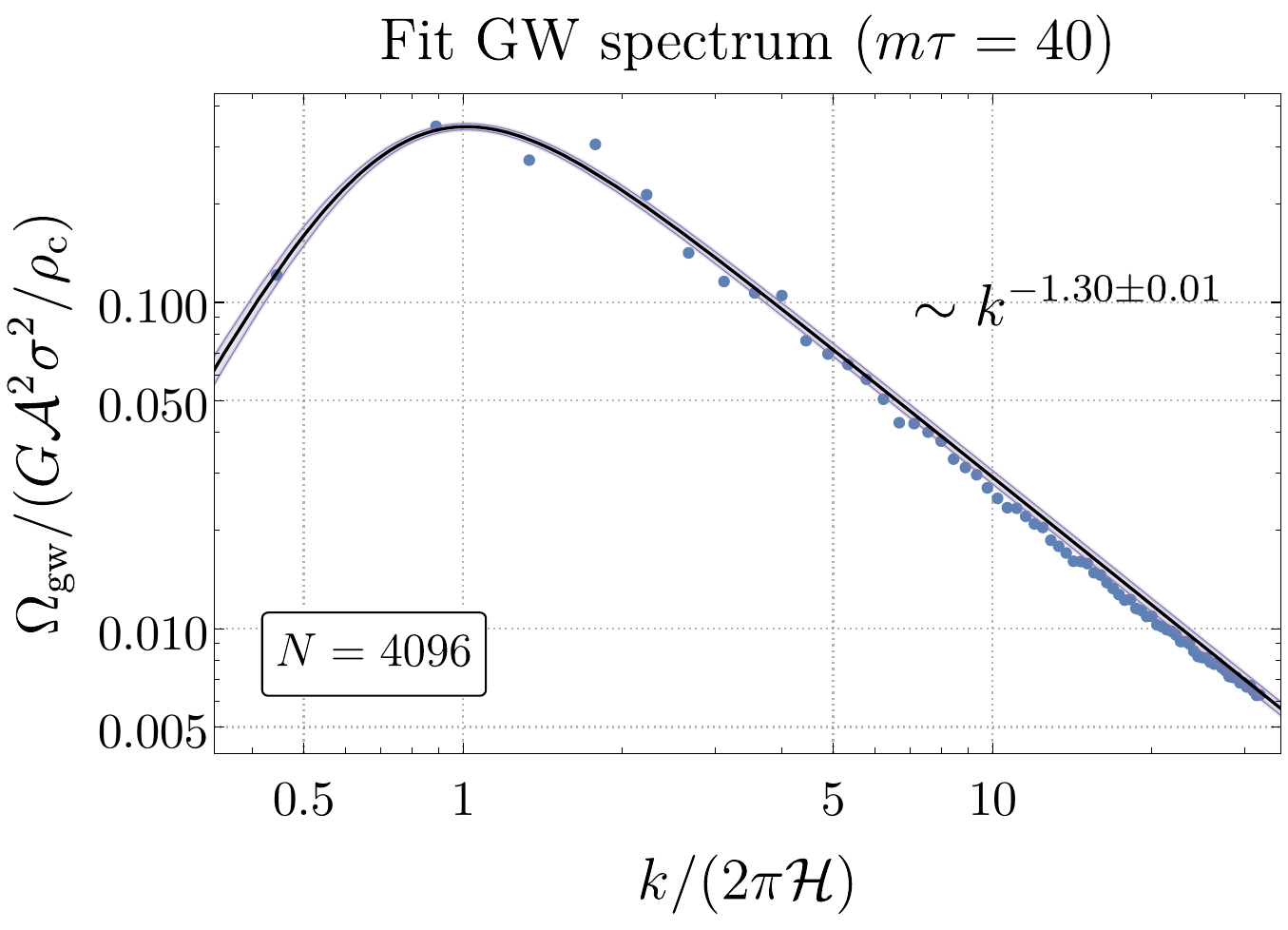}
    \caption{\textbf{Top Left:} Averaged GW spectra extracted from the same $2K$ simulations as in Fig. \ref{fig:GW_energy}, starting from $m\tau = 15$ (bottom red) up to $m\tau = 32$ (top green). The black curve corresponds to the fit on the right plot, while the gray region denotes modes that are omitted from the fit.
     \textbf{Top Right:} Fit of the spectral shape given in Eq.\eqref{eq: spectral shape} shown in black from the same $2K$ simulations as in Fig. \ref{fig:GW_energy} at the final simulation time (in this case $m\tau_{\rm end} = 32$). The $1\sigma$ error band is visualized by a purple band. \textbf{Lower Panels:} Same as above for one $4K$ simulation at $m\tau = 40$. The purple band represents the $1\sigma$ error band, where the error originates from the fit. 
        }
    \label{fig: GW2K}
\end{figure}

We then analyze the energy density in gravitational waves $\rho_{\rm gw}$ in the right panel of Fig. \ref{fig:GW_energy}, both for the $2K$ simulation set and for the $4K$ simulation.
Also in this case, the outcome is consistent with our expectations, namely with a fractional energy density growing as $\rho_{\rm gw}/\rho_c\sim \tau^4$ as soon as the scaling regime is established.
We can then determine the GW emission efficiency $\epsilon_{\rm gw}$ during scaling, defined in  Eq.\eqref{eq:epsilonGW},
by employing the $2K$ data-set,
which gives
\begin{equation}\label{eq: epsilon integrated}
    \epsilon_{\rm gw} = 0.54 \pm 0.08\,.
\end{equation}
The value we obtain is consistent with results from \cite{Hiramatsu:2013qaa, Dankovsky:2024zvs}, and a comparable result is also obtained from the $4K$ simulation at $m\tau = 40$. Note that the GW energy density is obtained by integrating the GW spectrum (which we show in Fig.~\ref{fig: GW2K}) over all modes. However, spurious effects in the UV arise due to a lack of resolution, as we will discuss in the following (specifically an enhancement of the spectrum at $k/m \sim 100$). We have checked that removing the unphysical UV region from the integration domain does not significantly affect the determination of $\epsilon_{\rm gw}$.

\paragraph{Gravitational wave spectrum.}
In Fig.\,\ref{fig: GW2K} (left column) we show several GW spectra taken from the simulations described above, at different times during the simulation run.
Concretely, we show the averaged GW spectra of the $2K$ simulations 
and the $4K$ simulation, considering only time snapshots where the system is in scaling.

Let us begin by discussing several key features of the GW spectrum, which has a characteristic broken-power law shape. Two distinct peaks emerge near the infrared (IR) and ultraviolet (UV) cutoffs of the simulation. The IR peak is associated with the Hubble scale, $k \sim 2\pi\mathcal{H}$, as we show below. This peak appears consistently across simulations, showing little dependence on the chosen grid size and resolution. In contrast, the UV peak, typically appearing at $k \gtrsim 50\,m$, can be attributed to discretization effects, as its position and amplitude vary significantly with the grid resolution. 
Notice that this issue distorts the spectrum also at momenta smaller than the UV peak, resulting in a 
plateau-like feature already observed in\,\cite{Notari:2025kqq, Dankovsky:2024zvs}.
As noticed in \,\cite{Notari:2025kqq},
this feature is in fact spurious and indeed due to low-resolution effects. 
In Appendix \ref{app:plateau}, we analyze these issues in detail by performing dedicated $2K$ simulations to be compared with the $4K$ one.
There we identify conservatively $k/m \simeq 5$ as the momentum where the finite-resolution effects becomes relevant.
In the following, we hence restrict our analysis to $k \lesssim 5\,m$, where the GW spectra remain stable and consistent across different grid sizes, and we have accordingly shaded the region outside of this range in the left panels of Fig.\,\ref{fig: GW2K}. 

We now present a fit of the GW spectrum restricting to momenta $ k/m \lesssim 5$. 
We parameterize the GW spectrum as in \cite{Notari:2025kqq}:
\begin{equation}\label{eq: GW fit}
\Omega_{\rm gw}(\tau, k) = \Omega^{\rm peak}_{\rm gw} (\tau) \times 
\mathcal{S}\left(\frac{k}{2\pi\mathcal{H}}\right)
 \, ,
\end{equation}
where the spectral function is parameterized by a broken power-law,
\begin{equation}\label{eq: spectral shape}
\mathcal{S}(x) \equiv \frac{(a+b)^c}{\left(b \left(\frac{x}{x_{\rm p}}\right)^{-a/c}+ a \left(\frac{x}{x_{\rm p}}\right)^{b/c}\right)^c}\,,
\end{equation} 
and we fix $a = 3$ from causality. The peak amplitude is given by 
\begin{equation}\label{eq: peak amplitude GW}
\Omega_{\rm gw}^\text{peak}(\tau) = \frac{\tilde\epsilon_{\rm gw} G \mathcal{A}^2 \sigma^2}{\rho_c(\tau)} = \frac{\tilde\epsilon_{\rm gw} G \mathcal{A}^2 \sigma^2}{3 H^2 M_{\rm Pl}^2} \,.
\end{equation} 
In this formulation, the comoving momentum $k$ is rescaled with respect to the Hubble momentum, $2\pi \mathcal{H}$. This choice reflects the fact that for a scaling network of domain walls, the Hubble scale is the only relevant physical scale governing the dynamics, as discussed in Section~\ref{sec:approach_scaling}. Accordingly, the peak of the spectrum is expected to lie near $k_{\rm peak} \sim 2\pi \mathcal{H}$. In addition, we allow for a different efficiency parameter $\tilde{\epsilon}_{\rm gw}$ for the peak amplitude compared to the one obtained from the full integrated GW energy density in Eq.\eqref{eq: epsilon integrated}. 

We have then carried out a best-fit analysis of the GW spectra from the simulations. For the averaged spectra of the $2K$ runs, the fit was performed at the final simulation time, $m\tau_{\rm end} = 32$, with the standard deviation across realizations included as a weight in the procedure. The resulting best-fit parameters are listed in Table \ref{tab: fit 2K}, and the corresponding fit is shown as the black curve in the top-right panel of Fig. \ref{fig: GW2K}. For the $4K$ simulation, only a single realization was available, so deviations between runs could not be taken into account. Given that our analysis relies on a single simulation and that, by the end of the run, the simulation box contains only a limited number of superhorizon regions to sample from, we first restricted the $4K$ fit to modes with momenta above twice the Hubble scale, $k > 4\pi \mathcal{H}$. In this regime, the spectrum was approximated by a simple power-law, and the resulting exponent was then used as input for the parameter $b$ in Eq.\eqref{eq: spectral shape}. The outcome is shown in the bottom-right panel of Fig. \ref{fig: GW2K} for $m\tau = 40$, with the corresponding best-fit parameters again summarized in Table \ref{tab: fit 2K}.
\begin{table}
    \centering
    \renewcommand{\arraystretch}{1}
    \begin{tabular}{c|c|c|c|c}
   & \multicolumn{4}{c} {Radiation domination}
    \\ 
    \hline \hline
    N &  $\tilde{\epsilon}_{\rm gw}$ & $b$ & $c$ & $x_{\rm p}$ \\
    \hline \hline
       2048 &   $0.35 \pm 0.04$ & $1.31 \pm 0.02 $ &
          $1.12 \pm 0.33$ &
          $0.85 \pm 0.04$ \\ 
         \hline
          4096 &   $0.35 \pm 0.01$ & $1.30 \pm 0.01 $ &
          $1.41 \pm 0.10$ &
          $1.01 \pm 0.02$\\ 
         \hline
    \end{tabular}
    \caption{Best-fit values for the parameters reconstructing the GW spectral shape as given in Eq.\eqref{eq: spectral shape} for $N = 2048$ simulations and one $N = 4096$ simulation.}
    \label{tab: fit 2K}
\end{table}

We observe that the peak frequency $f_{\rm peak} = k_{\rm peak}/(2\pi a)$ is indeed close to the Hubble scale, which we also report in the entry $x_{\rm p}$ of Table \ref{tab: fit 2K},
\bea
  N = 2048: \qquad \qquad   f_{\rm peak} = (0.85 \pm 0.04)\, H\,,
  \\
   N = 4096: \qquad \qquad  f_{\rm peak} = (1.01 \pm 0.02)\, H \,.
\eea 
These values are slightly shifted relative to the results of Refs.\cite{Dankovsky:2024zvs, Notari:2025kqq}.
In both simulations, the high-frequency slope of the spectrum, reported in Table~\ref{tab: fit 2K} as the entry $b$, is consistent across resolutions:
\bea
  N = 2048: \qquad \qquad   \Omega_{\rm gw} \sim k^{-1.31 \pm 0.02} \qquad (2\pi\mathcal{H} < k)\,,
  \\
   N = 4096: \qquad \qquad   \Omega_{\rm gw} \sim k^{-1.30 \pm 0.01} \qquad (2\pi\mathcal{H} < k)\,,\label{eq: UV slope 4K}
\eea 
and similarly for the efficiency parameter $\tilde{\epsilon}_{\rm gw}$. On the other hand, the parameter $c$ which controls the broadness of the peak is less robust to the modification of the grid size.  

Let us stress here that the error bars that we quote for the fit parameters are based only on the statistics of each set of simulations. By comparing simulations with different number of grid points $N$, we however notice that the dominant source of uncertainty is actually related to systematics and likely due to the numerical resolution of the lattice. The same consideration also applies to all subsequent results where error bars are reported.

\paragraph{Comparison with previous studies.}

\begin{table}
    \centering
    \renewcommand{\arraystretch}{1}
    \begin{tabular}{c|c|c|c|c}
    Reference  & $N$ & Initial conditions & IR slope & UV slope \\
    \hline \hline
     \cite{Hiramatsu:2013qaa}  &  1024 & Vacuum &  3 &  $-1$ \\
          \hline
      \cite{Kitajima:2023kzu} & 1024 & White noise & 3 & $-1$ \\
    \hline
    \cite{Kitajima:2023kzu} & 1024 & Scale invariant & 3 & $-1$ \\
    \hline
    \cite{Li:2023yzq}$^\dagger$ & 1024 & Thermal & 3 & $-2.14$ ($\pm$ unknown) \\
    \hline
     \cite{Ferreira:2023jbu}$^*$  &  1024 & White noise &  $2.49 \pm 0.09$ & $-2.07 \pm 0.04$  \\ 
          \hline
     \cite{Ferreira:2023jbu}$^*$  &  1024 & Thermal & $2.6 \pm 0.1$ &  $-1.772 \pm 0.008$ \\
          \hline
     \cite{Dankovsky:2024zvs}$^\dagger$  &  2048 & Vacuum &  $2.66 \pm 0.06$ & $-1.53 \pm 0.04$   \\
          \hline
     \cite{Dankovsky:2024zvs}$^\dagger$  &  2048 & Thermal &  $2.69 \pm 0.11$ & $-1.28 \pm 0.01$  \\
           \hline
    \cite{Cyr:2025nzf} & 2048 & Vacuum & 3 & $-1.56 \pm 0.201$ \\
    \hline
    \cite{Babichev:2025stm}$^\dagger$ & 2048 & Vacuum & $2.36 \pm 0.37$ & $-1.45 \pm 0.07$ \\
    \hline
     \cite{Notari:2025kqq}$^\dagger$  &  3060 & White noise &  3 & $-1.19 \pm 0.02$   \\
     \hline
     Our work$^\dagger$ & 2048 & White noise & 3 & $-1.31\pm 0.02$ \\
     \hline
     Our work$^\dagger$ & 4096 & White noise & 3 & $-1.30\pm 0.01$
    \end{tabular}
    \caption{Table summarizing the results reported in previous studies, listed in the first column. All studies assumed a radiation dominated background, except those marked with an asterisk $^*$, which assumed a matter dominated background. Studies that used \CL ~are marked with a dagger $^\dagger$. The second column shows the number of grid points per spatial dimension used in the simulations. The third column describes the type of initial conditions applied (see the corresponding references for details on their setup). The final two columns present the slopes of the GW spectrum in the IR and UV regions, respectively. Where available, fitted slopes are accompanied by the uncertainties reported in the cited studies. 
    }
    \label{tab:summary_old_results}
\end{table}

The first study addressing gravitational waves generated by a collapsing domain wall network \cite{Gleiser:1998na} presented an analytical investigation of the peak GW amplitude using the Einstein quadrupole formula. The analysis was limited to estimating the GW energy density, with no simulations for the spectrum included. More than a decade later, the first numerical simulations of the GW spectrum from a DW network were performed in \cite{Hiramatsu:2010yz, Kawasaki:2011vv}. In \cite{Hiramatsu:2010yz}, both stable and collapsing DW networks in a radiation dominated background were examined, while \cite{Kawasaki:2011vv} extended the study to include both radiation and matter dominated backgrounds. Furthermore, the latter work not only computed the GW spectrum directly from numerical simulations, but also obtained it indirectly using the Unequal Time Correlator, an approach that we will discuss and apply later in Section~\ref{sec: UTC}. Both studies, however, contained an error in the numerical implementation that led to an artificially flat spectrum, which was subsequently corrected in \cite{Hiramatsu:2013qaa}, resulting in a GW spectrum scaling as $\sim k^{-1}$ for subhorizon modes. This $k^{-1}$ UV behaviour was recently revisited by new simulations using different initial fluctuations \cite{Kitajima:2023kzu, Li:2023yzq, Ferreira:2023jbu, Dankovsky:2024zvs, Notari:2025kqq, Cyr:2025nzf} (and in the context of melting domain walls, where the vacuum expectation value decreases with temperature, in \cite{Dankovsky:2024ipq}), as well as by analytical arguments presented in \cite{Gruber:2024pqh}. A concise summary of the results from these earlier studies is presented in Table~\ref{tab:summary_old_results}. 

In addition, several recent works have revisited the production of GWs during the collapsing phase of the network \cite{Kitajima:2023cek, Ferreira:2024eru, Babichev:2025stm, Notari:2025kqq, Cyr:2025nzf}, for both temperature dependent and constant bias. These studies generally find that the GW emission is enhanced due to a delayed DW wall annihilation and the final stage of GW production. However, it remains unclear how, or to what extent, the annihilation phase modifies the GW spectrum. On the one hand, Ref.~\cite{Kitajima:2023cek} reports no significant deviation from the conventional $\sim k^{-1}$ behaviour. On the other hand, Refs.~\cite{Cyr:2025nzf, Babichev:2025stm} observe a less steep UV slope compared to the unbiased case, which they attribute to the formation of a larger number of small, closed walls that soften the high frequency GW component, yielding a spectrum scaling approximately as\,\footnote{For the unbiased case, the UV slopes in Refs.~\cite{Cyr:2025nzf, Babichev:2025stm} are steeper than $\sim k^{-1}$, see Table~\ref{tab:summary_old_results}.} $\sim k^{-1}$. In contrast, Ref.~\cite{Notari:2025kqq} finds that a doubly broken power law model provides the best fit to the spectrum during annihilation, featuring a slope that is initially shallower than in the unbiased case but becomes significantly steeper beyond a certain wavenumber (about 3 times the peak frequency). The discrepancies among these results highlight the need for a more detailed and systematic investigation of the collapsing phase, which we however leave for future work.

\section{A window over gravitational wave production: the Unequal Time Correlator}
\label{sec: UTC}

In this section, we introduce the domain wall Unequal Time Correlator (UTC) and relate it to the emitted GW spectrum. 
We will then extract the Equal Time Correlator (ETC) from numerical simulations in Section \ref{sec: ETC DW Rad},
where the scaling properties of the DW network will become apparent through the special dependence of the ETC on its time and momentum variables. As we shall see, this scaling behaviour of the ETC will be violated only at scales that can resolve the DW width.
In section \ref{sec: GW from ETC} we will extend the ETC to the UTC by employing common assumptions on the coherence of the source along the lines of \cite{Caprini:2009fx}, and recompute the GW spectrum within this framework.
We will also introduce a new ansatz for connecting the ETC to the UTC such that the temporal coherence of the source decreases at larger momenta, as this behavior has been actually observed from numerical simulations of other defects. As we shall see, this assumption matches reasonably well the spectral shape of the GWs from scaling DWs, thus shedding light on the coherence properties of the network.

\paragraph{Unequal Time Correlator.}
Let us consider the EoM of the GW perturbation $h_{ij}$ given in Eq.\eqref{eq: EoM h}, which we rewrite for convenience:
\begin{equation}\label{eq: eom h}
    \frac{\partial^2 h_{ij}}{\partial\tau^2} + 2\mathcal{H}\frac{\partial h_{ij}}{\partial\tau} - \nabla^2 h_{ij} = 16\pi GT_{ij}^{TT}\,.
\end{equation}
Here, $T_{ij}^{TT}$ denotes the transverse-traceless part of the energy-momentum tensor, which can be expressed in terms of the dimensionless anisotropic stress tensor $\Pi^{TT}_{ij}$ as
\begin{equation}\label{eq: T equals Pi}
T^{TT}_{ij} = a^2 \rho_S \Pi^{TT}_{ij}\,,
\end{equation}
where $\rho_S$ is the energy density of the source \cite{Caprini:2009fx}. This parameterization is particularly convenient as $\rho_S$ is expected to contain all the relevant dependence on the microscopic symmetry-breaking scale characterized by the scalar field VEV $\eta$.

Rewriting the EoM in terms of the rescaled perturbation $\tilde h_{ij} = ah_{ij}$ in Fourier space,  we obtain the following equation for the GW emission:
\begin{equation}
\label{eq:EoM}
    \tilde h_{ij}''(\mathbf{k},\tau) + \left(k^2 - \frac{a''}{a}\right)\tilde h_{ij}(\mathbf{k},\tau) = 16\pi G a^3 \rho_S\Pi^{TT}_{ij}(\mathbf{k},\tau)\,.
\end{equation}
In a radiation dominated Universe, the scale factor satisfies $a'' = 0$, in which case Eq.\eqref{eq:EoM} admits the solution 
\begin{equation}
\label{eq:solution_EoM}
    \tilde h_{ij}(\mathbf{k},\tau) = 16\pi G\int_{\tau_i}^\tau d\tau' \mathcal{G}(k,\tau, \tau') a^3(\tau') \rho_S(\tau')\Pi^{TT}_{ij}(\mathbf{k}, \tau')\,,
\end{equation}
where, for the subhorizon modes, the retarded Green's function $\mathcal{G}$ takes the simple form
\begin{equation}\label{eq: Green function}
    \mathcal{G}(k, \tau, \tau') = \frac{\sin(k(\tau-\tau'))}{k}\,.
\end{equation}

The GW energy density, previously defined in Eq.\eqref{eq:energy_density_GW}, can be written in terms of the rescaled perturbation $\tilde{h}_{ij}$ as
\begin{equation}
    \rho_{\rm gw}(\tau) = \frac{1}{32\pi G a^4(\tau)}\left\langle \left(\tilde{h}_{ij}' - \mathcal{H}\tilde{h}_{ij}\right)^2(\mathbf{x},\tau)\right\rangle\,,
\end{equation}
or, equivalently, in Fourier space as
\begin{equation}
    \rho_{\rm gw}(\tau) \approx \frac{1}{32\pi Ga^4(\tau)}\int\frac{d^3\mathbf{k}}{(2\pi)^3}\frac{d^3\mathbf{q}}{(2\pi)^3}e^{-i\mathbf{x\cdot(\mathbf{k}-\mathbf{q}})}\langle \tilde h_{ij}'(\mathbf{k},\tau)\tilde h_{ij}'^\star(\mathbf{q},\tau)\rangle\,,
\end{equation}
where we only consider subhorizon modes, for which $\tilde h_{ij}' \sim k\tilde h_{ij} \gg \mathcal{H}\tilde h_{ij}$. Inserting the GW perturbation from Eq.\eqref{eq:solution_EoM}, we obtain the GW spectrum 
\begin{equation}\label{eq: rho_gw Caprini}
    \frac{\mathrm{d}\rho_\text{gw}}{\mathrm{d}\ln k}(\tau,k) = \frac{2 G}{\pi}\frac{k^3}{a^4(\tau)}\int_{\tau_i}^\tau\mathrm{d}\tau_1\int_{\tau_i}^\tau
\mathrm{d}\tau_2\ a^3(\tau_1)a^3(\tau_2)\rho_S(\tau_1)\rho_S(\tau_2)\cos\left[k(\tau_1 - \tau_2)\right]\Pi^2(k,\tau_1,\tau_2)\,, 
\end{equation}
where a temporal averaging over time-scales of order $1/k$ or smaller is implicitly assumed to extract the coarse-grained signal \cite{Caprini:2009fx, Figueroa:2012kw}. We furthermore defined the Unequal Time Correlator (UTC)  $\Pi^2(k,\tau_1,\tau_2)$ as
\bea \label{eq: power spectrum Pi}
\langle \Pi^{TT}_{ij}(\mathbf{k}, \tau_1) \Pi^{TT\star}_{ij}(\mathbf{q}, \tau_2) \rangle \equiv (2\pi )^3 \delta^{(3)} (\mathbf{k}-\mathbf{q})\Pi^2(k,\tau_1,\tau_2) \, ,
\eea 
where statistical homogeneity and isotropy of the source are assumed. As $\Pi^{TT}_{ij}(\mathbf{x},\tau)$ is dimensionless (in mass units), the UTC $\Pi^2(k,\tau_1,\tau_2)$ has dimension $-3$ in mass units.

In the following, we will study the properties of the UTC for a network of domain walls and their implications for the GW spectrum. 

\subsection{UTC and GWs from a scaling network of domain walls}\label{sec: UTC DW}

As discussed in Section \ref{sec:approach_scaling}, the DW network rapidly enters the scaling regime soon after formation. In this regime, the dependence of the UTC on the momentum $k$ is expected to arise solely through the dimensionless combination $x = k\tau$, at least for scales that cannot resolve the DW width, $k^{-1} \gg \delta_{\rm dw}/(2 \pi a(\tau))$.
By dimensional analysis, and since $\Pi^2(k,\tau_1,\tau_2)$ has dimension $-3$, one can single out a factor of $\tau^3$ without loss of generality, which is then symmetrized by taking $(\tau_1 \tau_2)^{3/2}$. The UTC takes then the general form:
\begin{equation}\label{eq: Pi to CT}
\Pi^2 (k,\tau_1,\tau_2) = (\tau_1\tau_2)^{3/2}C^T(x_1, x_2)\,,
\end{equation}
where $C^T(x_1, x_2)$ is a dimensionless function that encodes the properties of the specific network of defects under consideration (in our case, domain walls), and can in principle depend on the background cosmology through the EoS parameter $\omega \equiv P/\rho$. 

Notice that in Eq.\,\eqref{eq: Pi to CT} there is no dependence on the energy density of the source or its redshift, as these have already been taken into account by the prefactor $a^2\rho_S$ in Eq.\eqref{eq: T equals Pi}. 
Our parameterization then differs from the one adopted in Ref.\,\cite{Figueroa:2012kw}, where the microscopic scale associated to the scalar field VEV, $\eta$, was effectively factored out in the RHS of \eqref{eq: T equals Pi} as $\eta^2$. This implicitly assumes that the energy density of the scaling source remains a constant fraction of the critical energy density during the expansion of the universe, which is however not the case for a DW network\footnote{In fact, as for DWs $\rho_S \propto \sigma \sim \eta^3$ with $\sigma$ the DW tension, one should actually postulate $\Pi^2 \propto \eta^6$ rather than $\Pi^2 \propto \eta^4$ within the parameterization of Ref.\,\cite{Figueroa:2012kw}.}.

Let us finally stress that dimensional analysis alone cannot exclude an explicit dependence of $\Pi^2$ on the scale factor, $a(\tau)$. However, by taking advantage of the fact that Weyl rescaling can be made a classical symmetry of the action by allowing spurious transformation of the Lagrangian parameters, we argue in Appendix \ref{app:rescaling} that no such dependence should arise due to the symmetry properties of the system, and that our parameterization in Eq.\,\eqref{eq: Pi to CT} captures the most general structure of $\Pi^2$ in the (scaling) regime of interest.

To evaluate the GW emission from scaling domain walls, we start from the expression  for the GW energy density as given in Eq.\eqref{eq: rho_gw Caprini}, where we substitute
\begin{equation}
\label{eq:scalingenergy}
\Pi_{\rm dw}^2(k,\tau_1,\tau_2) = (\tau_1\tau_2)^{3/2}C_{\rm dw}^T(x_1,x_2)\,, \qquad 
a(\tau) = \sqrt{\Omega^0_\text{rad}}H_0\tau\,, \qquad \rho_{\rm dw} = 2\mathcal{A}\sigma H(\tau)\,,
\end{equation}
where the subscript ``dw'' refers to domain walls specifically. Here, $\Omega^0_{\rm rad}$ and $H_0$ are the radiation density parameter and Hubble rate today respectively. We remind that the conformal Hubble is given by
$\mathcal{H} = aH $, so that $\rho_{\rm dw} = 2\mathcal{A}\sigma \mathcal{H}/a$, with $\mathcal{H} = a'/a = \tau^{-1}$ for radiation domination. 
The GW energy density is then given by 
\begin{equation}
    \frac{\mathrm{d}\rho_\text{gw}^{\rm dw}}{\mathrm{d}\ln k}(k,\tau) = \frac{8 G\mathcal{A}^2\sigma^2}{\pi}(\Omega^0_\text{rad})^2H_0^4\frac{k^3}{a^4(\tau)}\int_{\tau_i}^\tau\mathrm{d}\tau_1\int_{\tau_i}^\tau
\mathrm{d}\tau_2\, (\tau_1  \tau_2)^{5/2} \cos\left[k(\tau_1 - \tau_2)\right] C^T_{\rm dw}(x_1, x_2)\,.
\end{equation}
Expressing this in terms of the GW energy density parameter, we obtain
\begin{align}
\label{eq:main_result_for_ETC}
    \Omega^{\rm dw}_\text{gw}(\tau, k) & \equiv \frac{1}{\rho_c} \frac{\mathrm{d}\rho_\text{gw}^{\rm dw}}{\mathrm{d}\ln k} = \frac{64}{3} G^2 \mathcal{A}^2 \sigma^2 \Omega_{\rm rad}(\tau) \Omega^0_{\rm rad} H_0^2 \tau^4 F^T_{\rm dw}(x)\,, 
    \\
    \label{eq:FTdw}
    F^T_{\rm dw}(x) &= \frac{1}{x^4}\int_{x_i}^x\mathrm{d}x_1\int_{x_i}^x\mathrm{d}x_2\, (x_1 x_2)^{5/2}\cos(x_1 - x_2) C^T_{\rm dw}(x_1,x_2)\,,
\end{align}
where we have
\begin{equation}
    \Omega_{\rm rad}(\tau) = \frac{\Omega^0_{\rm rad}}{a^4(\tau)}\left(\frac{H_0}{H(\tau)}\right)^2\,,
\end{equation}
which remains constant during a RD Universe epoch as $a\sim \tau$ and $H = \mathcal{H}/a\sim \tau^{-2}$.  The function $F^T_{\rm dw}(x=k \tau)$ fully encodes the spectral shape of the GW background, while the overall amplitude during radiation domination scales as $\sim \tau^4$, in agreement with the scaling expected from Eq.\eqref{eq: integrated rhogw}:
\begin{equation}
\label{eq:rhogwdw}
    \frac{\rho_{\rm gw}^{\rm dw}}{\rho_c} = \frac{64}{3} G^2 \mathcal{A}^2 \sigma^2 \Omega_{\rm rad}(\tau) \Omega^0_{\rm rad} H_0^2 \tau^4 \int \text{d}\,\text{ln}\,x \, F^T_{\rm dw}(x) \propto \tau^4,
\end{equation}
where we have used that $\text{d}\,\text{ln}\,k = \text{d}\,\text{ln}\,x$.
This prediction for the scaling of the GW energy density has been already verified numerically in the right panel of Fig.\,\ref{fig:GW_energy}.

Let us now discuss the implications of Eq.\eqref{eq:main_result_for_ETC}. As we can see, the GW emission from a scaling network of DWs is controlled by $F^T_{\rm dw}(x)$, which in turn depends on the UTC encoded in the function $C_{\rm dw}^T(x_1,x_2)$. As already pointed out in other studies \cite{Figueroa:2012kw, Fenu:2009qf, Bevis:2010gj, CamargoNevesdaCunha:2022mvg}, we expect $C_{\rm dw}^T(x_1,x_2)$ to be sharply peaked along the diagonal $x_1 = x_2$ and to decay rapidly away from it, reflecting the short time coherence of the source \cite{Albrecht:1995bg, Durrer:2001cg, CamargoNevesdaCunha:2022mvg}. Furthermore, we expect $C_{\rm dw}^T(x,x)$ to decay as a power law along the diagonal at large $x$ \cite{Durrer:2001cg}, 
\begin{equation}
\label{eq:CT}
    C^{T}_{\rm dw}(x,x) \propto x^{-q}, \quad x \gg 1.
\end{equation}

Let us at this point make a first comparison with the expression for $F^T(x)$ given in Ref.\,\cite{Figueroa:2012kw} for a scaling source in the early universe, even though specifically derived for a network of (in general, non--topological) defects arising from the $O(N)/O(N-1)$ coset for $N>2$, with the special case $N=2$ corresponding to a network of global strings. We can see how the integral in Eq.\eqref{eq:FTdw} for domain walls contains a different power of $x_1$ and $x_2$ (namely $5/2$ rather than $1/2$), and also a different $x^{-4}$ prefactor, compared to \cite{Figueroa:2012kw}. This can be traced back to the different scaling of the energy density of the DW network, see e.g. Eq.\,\eqref{eq:scalingenergy}, compared to the defects of Ref.\,\cite{Figueroa:2012kw} for which the energy density is $\propto H^2$. 

In the following, we will extract $C_{\rm dw}^T(x,x)$ from our numerical simulations in radiation domination, thus determining its shape and in particular the value of $q$ for a network of domain walls. Our analysis will provide a further characterization of the scaling regime beyond the one based on the area parameter, and will set the stage for understanding the general structure of the UTC.

\subsection{Numerical study of the ETC of the domain wall network}
\label{sec: ETC DW Rad}

We investigate here for the first time the behaviour of the ETC, $C_{\rm dw}^T(x,x)$, for a domain wall network via numerical simulations in radiation domination\,\footnote{A similar study of the ETC was presented already in Ref.\,\cite{Kawasaki:2011vv}, but the numerical results therein were affected by a coding error, as pointed out later on in Ref.\,\cite{Hiramatsu:2013qaa}.}, including the approach to scaling where $C^T_{\rm dw}$ does become a function of the combination $x = k \tau$ only, thus aligning with the general structure of $\Pi^2(k,\tau,\tau)$ postulated in Eq.\,\eqref{eq: Pi to CT}.
To this end, we modified the \CL ~public code to compute the power spectrum of the TT part of the energy-momentum tensor at equal times, \emph{i.e.}
\begin{equation}
    \left\langle T_{ij}^{\rm TT}(\mathbf{k},\tau)\, T_{ij}^{\rm TT\star}(\mathbf{q},\tau) \right\rangle = (2\pi)^3 \delta^{(3)}(\mathbf{k}-\mathbf{q})\, T^2_{\rm dw}(k, \tau, \tau)\,.
\end{equation}
From the definition of the energy-momentum tensor in Eq.~\eqref{eq: T equals Pi}, and using Eqs.~\eqref{eq: power spectrum Pi} and~\eqref{eq: Pi to CT}, we extract $C^T_{\rm dw}(x,x)$ via
\begin{equation}\label{eq: T2}
    T^2_{\rm dw}(k,\tau,\tau) = a^4(\tau)\, \rho_{\rm dw}^2(\tau)\, \tau^3\, C^T_{\rm dw}(x,x)\,.
\end{equation}
Further details of our implementation are provided in Appendix~\ref{app: Implementation in CosmoLattice}.

\begin{figure}
    \centering
    \includegraphics[scale=0.31]{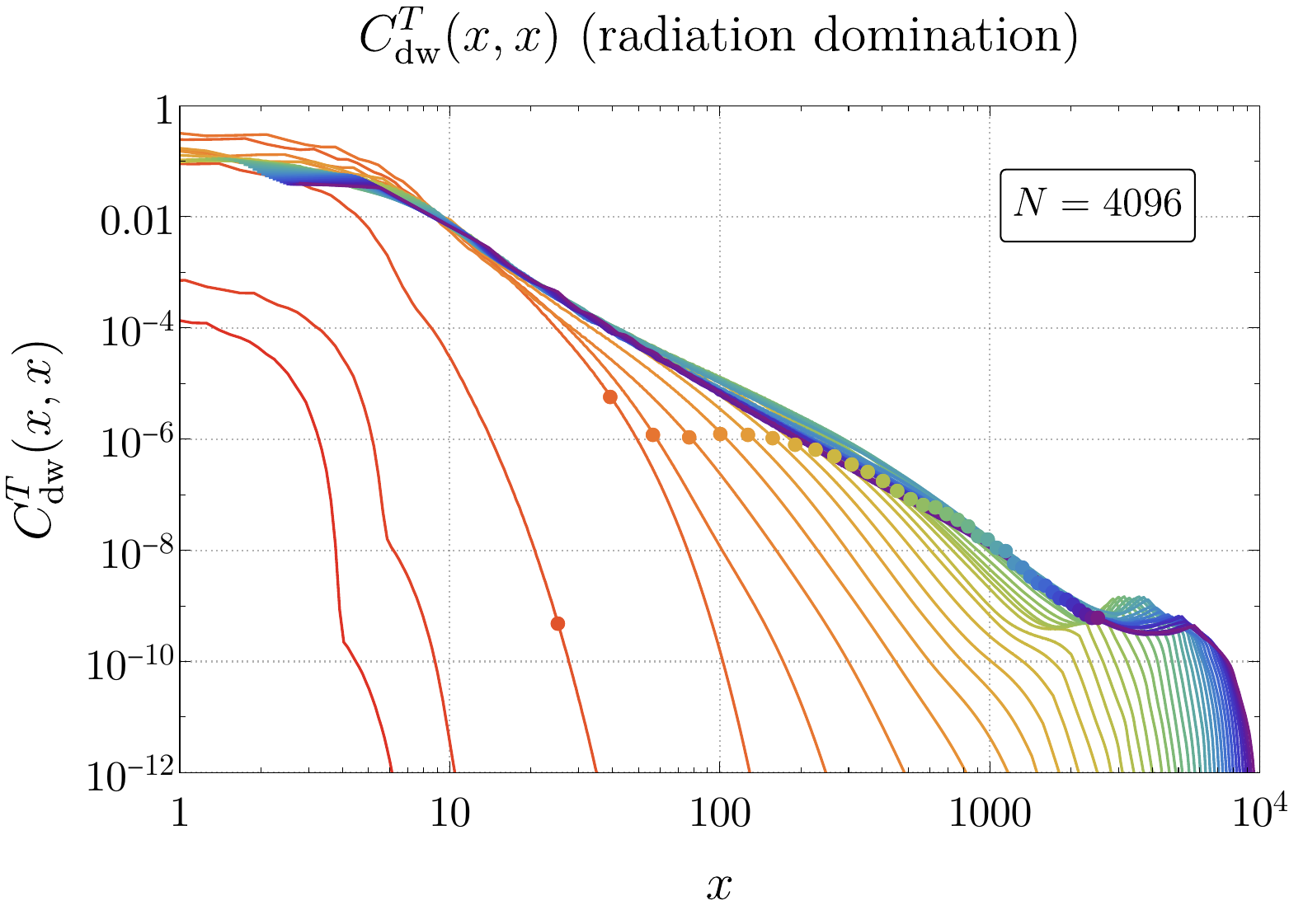}
    ~
    \includegraphics[scale=0.31]{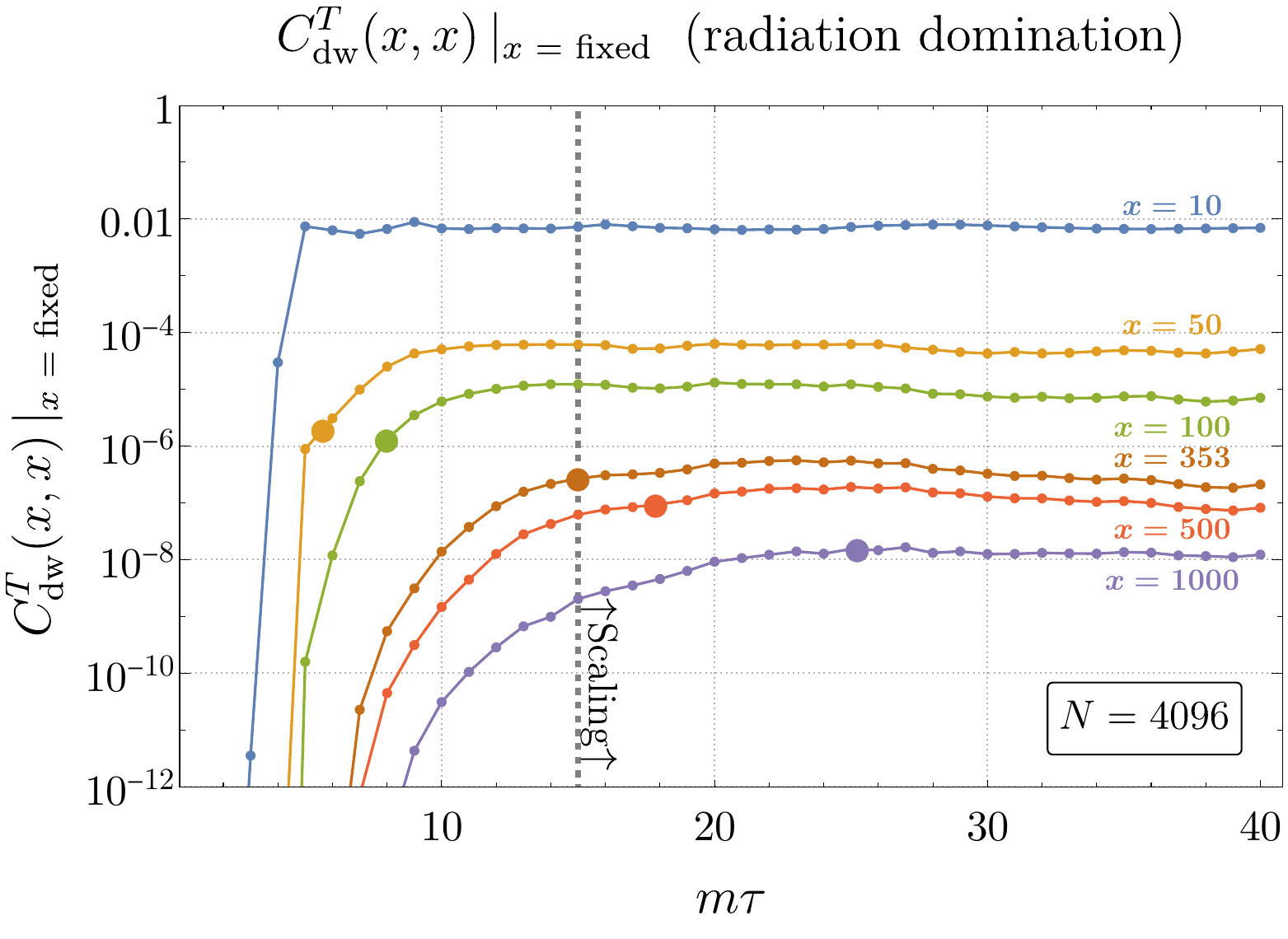}
    \caption{\textbf{Left:} The ETC $C^T_{\rm dw}(x,x)$ as a function of $x$ for times between $m \tau = 2$ and $m \tau = 40$ (from red to purple) for the $4K$ simulation. The dots represent the momentum associated to twice the domain wall width. \textbf{Right:} Evolution of $C^T_{\rm dw}(x,x)$ for fixed values of $x$. The enlarged dots correspond to the momentum related to twice the domain wall width, and the gray dashed line indicates the onset of the scaling regime based on the area parameter. For the special value of $x \approx 350$ (brown line), the onset of the scaling regime coincides with the crossing of the comoving wall width (indicated by the dot), so that the function $C^T_{\rm dw}$ is expected to remain constant right after this point.}
    \label{fig:ETC_scalng}
\end{figure}

We display our results in Fig.~\ref{fig:ETC_scalng},
where we show only the outcome of the $4K$ simulations for concreteness (the $2K$ simulations give the same qualitative result).
On the left panel we show 
$C^T_{\rm dw}(x,x)$ extracted at different times, while on the right panel
we show the time dependence of $C^T_{\rm dw}(x,x)$ for fixed values of $x$.
In both panels, we indicate by colored circles the momentum associated to (twice) the domain wall width, namely $k = a(\tau)2\pi/2\delta_{\rm dw}$, as a reference.

First of all, we notice that the ETCs evaluated at different times evolve such that they eventually converge to a unique function $C_{\rm dw}^T$ that depends only on the combination $x = k \tau$.
This feature is clearly visible in both panels of 
Fig.~\ref{fig:ETC_scalng}, and demonstrates that the ETC does indeed take the form expected from general arguments for a network in the scaling regime.

From our analysis, we can also inspect the effect of the finite domain wall width on the ETC, as this will induce a violation of scaling at large momenta, $k \gtrsim a(\tau)\pi/\delta_{\rm dw}$.
This effect can be seen from the behaviour of the red and (more clearly) purple line in the right panel of  Fig.~\ref{fig:ETC_scalng}, where the ETCs at fixed $k \tau =500$ and $k \tau =1000$ do show an explicit dependence on $m\tau$ even for $m \tau \gtrsim 15 $ (namely after the network has entered the scaling regime, as determined for instance from the convergence of the area parameter in Fig.\,\ref{fig:GW_energy}) but only for momenta that are indeed larger than the inverse domain wall width, namely points to the left of the colored circles.
This feature is visible also in the left panel, where the various ETCs never converge for momenta larger the inverse wall width (here values of $x$ to the right of the colored circles). 

Let us also comment on the conjecture made in Ref.\,\cite{Dankovsky:2024zvs} that a new intermediate scale between the Hubble horizon and the inverse wall width, $k_\ast \simeq 2 \pi \sqrt{H/\delta_w}a$, would show up in the GW spectrum from a scaling domain wall network as a flat region (plateau). By looking at the ETC results in Fig.\,\ref{fig:ETC_scalng}, we find no evidence for scaling violation at scales around $k_\ast$. Conversely, the scaling behaviour appears to be well established all the way down to the inverse domain wall width. We then interpret the non--appearance of the scale $k_\ast$ in the ETC as evidence against the conjecture above. This also aligns with the results in Ref.\,\cite{Notari:2025kqq}, where a possible plateau in the GW spectrum was found to disappear when increasing the precision in evaluating spatial gradients on the lattice.

We conclude that, for times following the onset of the scaling regime, the ETC does converge to the functional form as predicted in Eq.\eqref{eq: Pi to CT}, allowing us to determine the characteristic function 
$C^T_{\rm dw}(x,x)$ for momenta smaller than the inverse wall width.

\begin{figure}
    \centering
    \includegraphics[width=0.49\linewidth]{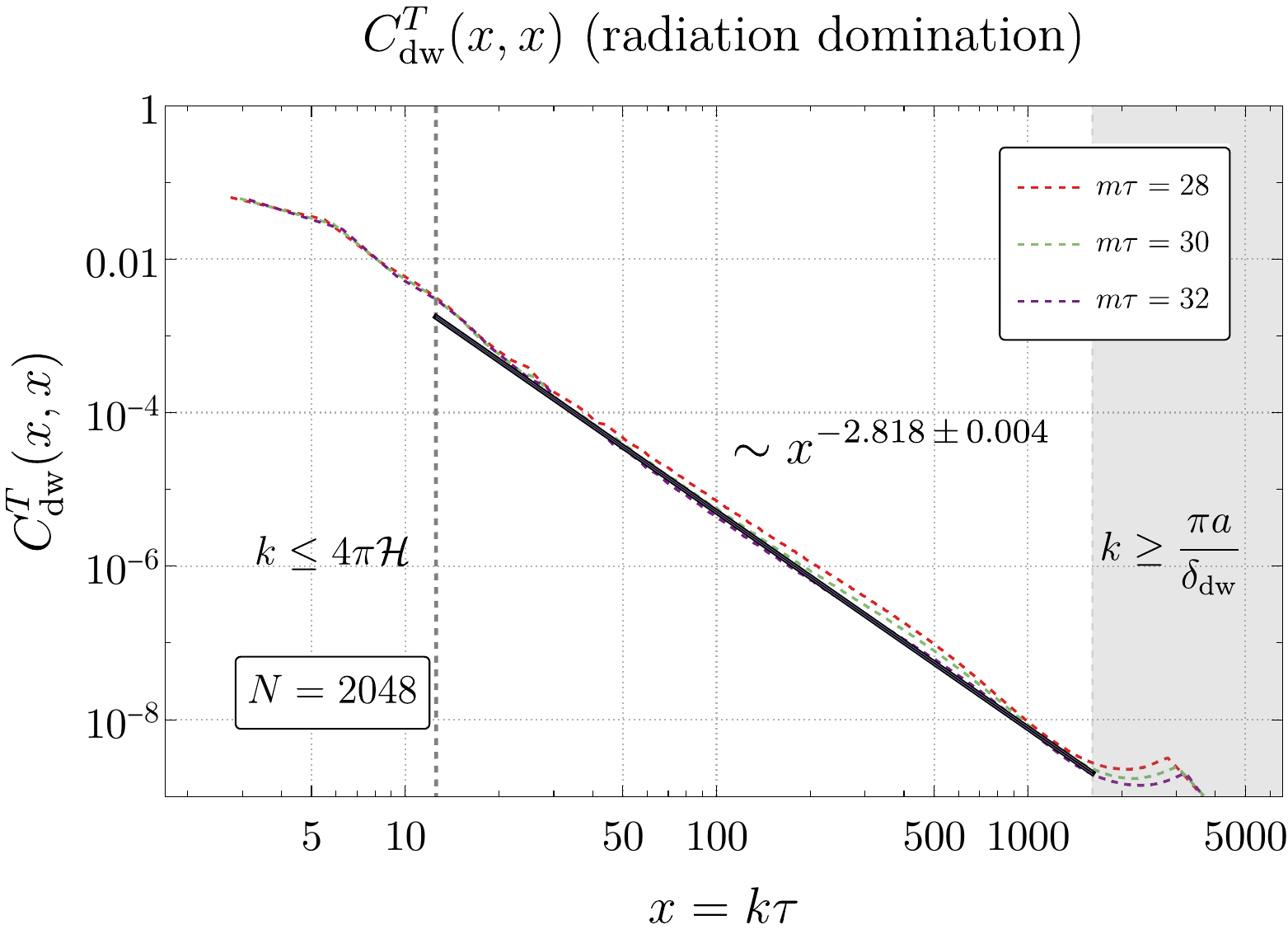}
     ~
     \includegraphics[width=0.49\linewidth]{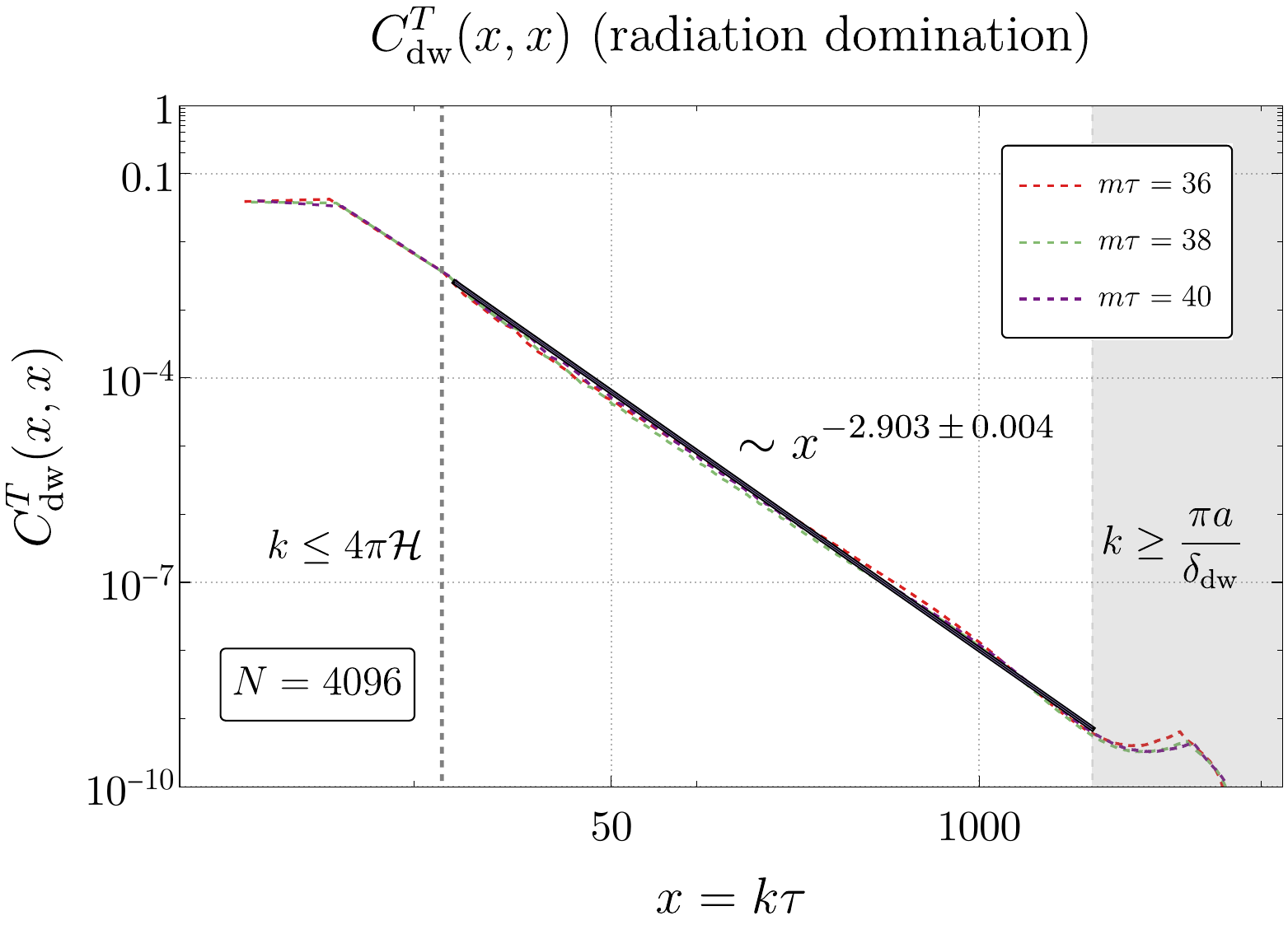}
    \caption{\textbf{Left:} Fit of the function $C^T_{\rm dw}(x,x)$ from the averaged $2K$ simulations at the final simulation time ($m\tau_{\rm end} = 32$), shown in black. The data follow a power-law behaviour consistent with $C^T_{\rm dw}(x,x) \sim x^{-2.818}$. Dashed curves correspond to the averaged $C^T_{\rm dw}(x,x)$ at late times, $m\tau = 28$, $30$ and $32$, when the domain wall network is fully in the scaling regime. The fit includes only modes larger than twice the Hubble scale and smaller than half the inverse domain wall width; the latter range already violates scaling, as indicated by the gray-shaded region. 
    \textbf{Right:} Same fit as the left plot, but for the $4K$ simulation, shown at $m\tau_{\rm end} = 40$ in black, together with $C^T_{\rm dw}(x,x)$ curves at later times ($m\tau = 36$–$40$). The data exhibit a power-law behaviour consistent with $C^T_{\rm dw}(x,x) \sim x^{-2.903}$. 
    }
    \label{fig: CT_RAD}
\end{figure}

Let us now move to discuss the properties of $C^T_{\rm dw}(x,x)$. For this purpose, we average the results of the five $2K$ simulations and fit them with a simple power law at the final simulation time ($m\tau_{\rm end} = 32$). The fit is restricted to modes larger than the expected peak momentum $k_{\rm peak} \sim 2\pi\mathcal{H}$, starting from momenta above twice the Hubble scale ($k > 4\pi\mathcal{H}$) and extending up to the momentum corresponding to half the inverse domain wall width ($k < \pi a/\delta_{\rm dw}$). 
In addition, we fitted the higher-resolution $4K$ simulation at $m\tau = 40$ in a similar manner. Our data points are shown in Fig.\,\ref{fig: CT_RAD} for both the $2K$ and $4K$ simulations. The results show a consistent power-law behaviour close to the simple shape $q \simeq 3$,
\bea
N = 2048: \qquad & C^T_{\rm dw}(x,x) \propto x^{-2.818 \pm 0.004}\,, \\
N = 4096: \qquad & C^T_{\rm dw}(x,x) \propto x^{-2.903 \pm 0.004}\,,
\eea
indicating in particular a falloff steeper than $\sim x^{-2}$, which, as we shall see in the next section, guarantees a regular behaviour for $F_{\rm dw}^T(x)$ as well as its integral in Eq.\,\eqref{eq:rhogwdw} for $x \rightarrow \infty$. We recall that the quoted error bars reflect only statistical uncertainties, as discussed previously below Eq.\eqref{eq: UV slope 4K}.

\subsection{From the ETC back to the gravitational wave spectrum}\label{sec: GW from ETC}

The determination of the GW spectrum requires the knowledge of the UTC of the energy momentum tensor encoded in the function $C^T_{\rm dw}(x_1,x_2)$ away from the special case of the ETC where $x_1=x_2$ studied in the previous section. As a direct extraction of the UTC from the simulation data is computationally expensive, in the following we shall explore the implications of commonly used assumptions regarding the coherence of the GW source in the spirit of \cite{Caprini:2009fx}, namely totally incoherent and totally coherent, as well as a new scenario interpolating between the two, and compare the results with the actual GW output from \CL.

\medskip
\noindent \textbf{Totally incoherent.}
For a DW network in the scaling regime, the coherence of the source cannot be maintained over a long time,
as each mode is coupled in a strongly non-linear manner to all the others. 
Intuitively, this coupling can be viewed as imparting ``random'' 
kicks to a given mode, which in turn drives the UTC to vanish for $\tau_1 \neq \tau_2$. This picture partially agrees with what has been actually observed in numerical simulations of cosmic strings\,\cite{Albrecht:1995bg, Durrer:2001cg, Bevis:2010gj}.

The assumption of random uncorrelated events can be implemented at the level of the UTC as:
\begin{equation}
    \langle \Pi^{TT}_{ij}(\mathbf{k}, \tau_1) \Pi^{TT\star}_{ij}(\mathbf{q}, \tau_2) \rangle = (2\pi)^3\delta^{(3)} (\mathbf{k}-\mathbf{q})\frac{\delta(\tau_1 - \tau_2)}{k}\Pi^2_{\rm dw}(k,\tau_1,\tau_1)\,.
\end{equation}
The factor $1/k$ ensures the correct dimensionality and consistency with an intermediate approximation called the \emph{top hat approximation} \cite{Caprini:2009fx}, which we however leave out in our discussion. This implies
\begin{equation}
   \Pi^2_{\rm dw}(k,\tau_1,\tau_2) = \frac{1}{k}\delta(\tau_1-\tau_2)\Pi^2_{\rm dw}(k,\tau_1,\tau_1)\,,
\end{equation}
so that
\begin{equation}
\label{eq:incoh}
    C^T_{\rm dw}(x_1, x_2 ) = C^T_{\rm dw}(x_1, x_1) \delta(x_1-x_2) = C^T_{\rm dw}(x_1, x_1) \frac{1}{k} \delta(\tau_1 - \tau_2)\,.
\end{equation}

Using this ansatz, we can compute the asymptotic behaviour of $F^T_{\rm dw}(x)$ at large $x$ by referring to Eq.\,\eqref{eq:FTdw} and assuming a power law for $C^T_{\rm dw}(x,x)$ as in Eq.\,\eqref{eq:CT}. We find:
\begin{equation}
\label{eq:FTdwinc}
    F^T_{\rm dw}(x)\big|_{x \to \infty, \,\text{incohe.}} \propto  
    \begin{cases}
        x^{2 - q} & q < 6\\
        \dfrac{\log x}{x^4} & q = 6 \\
        x^{-4} & q > 6 
    \end{cases}\,.
\end{equation}
By recalling that $x = k \tau$, and that $F_{\rm dw}^T(x)$ encodes all the $k$ dependence of the GW spectrum, Eq.\,\eqref{eq:FTdwinc} indicates that this is in general far from flat (except for the special value of $q=2$). 
In particular, by taking $q \simeq 3$ as found in the previous section, we see that the incoherent approximation would predict a fall--off of the GW power spectrum roughly given by $k^{-1}$ (we will provide a direct comparison with the GW spectrum obtained with \CL ~at the end of this section).

We notice that $F^T_{\rm dw}(x)$ is dominated by its upper limit of integration, $x$, as opposed to the behaviour presented in Ref.\,\cite{Figueroa:2012kw}, where $F^T(x)$ becomes increasingly insensitive to $x$, leading to a saturated, scale-invariant GW spectrum.
In Appendix\,\ref{app: UTC CS} we discuss how this behaviour is ultimately linked to the fact that \cite{Figueroa:2012kw} considers only defects (or in general, sources) whose energy density remains a constant fraction of the critical density in the scaling regime.
Therefore, while domain walls do exhibit the key scaling properties for the area parameter and the functional dependence of the ETC as discussed in the previous sections, they evade the conclusion in \cite{Figueroa:2012kw} regarding the generation of a scale--invariant spectrum during radiation domination. Our study clarifies precisely the interplay of these different aspects.

We then conclude that the statement that the GW spectrum from \emph{any scaling source} is scale--invariant during radiation domination is actually not applicable to domain walls or other scaling networks whose energy density redshifts non--trivially compared to the background cosmology, as this plays a crucial role not only on the amplitude but also on the GW spectral shape\,\footnote{This discussion has been carried out within the totally incoherent approximation for the UTC. However, we shall see that the occurrence of a peaked GW spectrum is independent of this assumption.}.

\medskip
\noindent \textbf{Totally coherent.} On the opposite side of the incoherent approximation, one may consider the idealized scenario where the source maintains its coherence at all times.
In this approximation, the DW network at different times is considered to be completely correlated:
    \begin{equation}
        \langle \Pi^{TT}_{ij}(\mathbf{k}, \tau_1) \Pi^{TT\star}_{ij}(\mathbf{q}, \tau_2) \rangle = (2\pi )^3 \delta^{(3)} (\mathbf{k}-\mathbf{q})\sqrt{\Pi^2_{\rm dw}(k,\tau_1,\tau_1)}\sqrt{\Pi^2_{\rm dw}(k,\tau_2,\tau_2)}\,,
    \end{equation}
so that we can identify the UTC as
\begin{equation}
    \Pi^2_{\rm dw}(k,\tau_1,\tau_2) = \sqrt{\Pi^2_{\rm dw}(k,\tau_1,\tau_1)}\sqrt{\Pi^2_{\rm dw}(k,\tau_2,\tau_2)}\,,
\end{equation}
and finally:
    \begin{equation}
    \label{eq:coh}
        C^T_{\rm dw}(x_1,x_2) = \sqrt{C^T_{\rm dw}(x_1,x_1)} \sqrt{C^T_{\rm dw}(x_2,x_2)}.
    \end{equation}
This choice naturally leads to a fast decay at large $x$ and to oscillations at small $x$, which are reminiscent of the \emph{Sakharov oscillations} in the context of coherent perturbations from inflation \cite{Dodelson:2003ip,Baumann:2009ds}. 

One can then obtain an expression for $F^T_{\rm dw}$ similar to Eq.\eqref{eq:FTdwinc} for $x \to \infty$:
\bea 
\label{eq:cohe_anal}
F^{T}_{\rm dw}(x)\big|_{x \to \infty, \,\text{cohe.}} \propto x^{1-q}.
\eea 
Taking again $q\simeq3$ from the previous section, one obtains $\Omega_{\rm gw}(\tau,k) \propto k^{-2}$ at subhorizon scales if the DW network turned out to be totally coherent.
  
\medskip
\noindent
\textbf{Interpolating function.} In reality, we expect the DW network to be a GW source which is neither totally coherent nor totally incoherent, but rather something in between. Drawing intuition from the numerical simulations available for the computations of the UTCs of a network of strings \cite{Durrer:2001cg, Bevis:2010gj}, one observes that the typical behaviour of the UTC $\Pi(k, \tau, \tau')$ is to decay exponentially fast away from the line  $\tau= \tau'$ with a slope which is however strongly dependent on the comoving wave vector $k$ (see for example Figures 4 and 5 of \cite{Bevis:2010gj}). 
An ansatz that would mimic this behaviour is obtained by multiplying the integrand of the coherent case with the exponential $e^{-|x_1 - x_2|/|x_1 + x_2|^r}$, where $r$ is some positive real number,
   \begin{equation}
   \label{Eq:inter}
        C^T_{\rm dw}(x_1,x_2) = \sqrt{C^T_{\rm dw}(x_1,x_1)} \sqrt{C^T_{\rm dw}(x_2,x_2)} \times e^{-|x_1 - x_2|/|x_1 + x_2|^r}.
    \end{equation}
    
\begin{figure}
    \centering
    \includegraphics[scale=0.55]{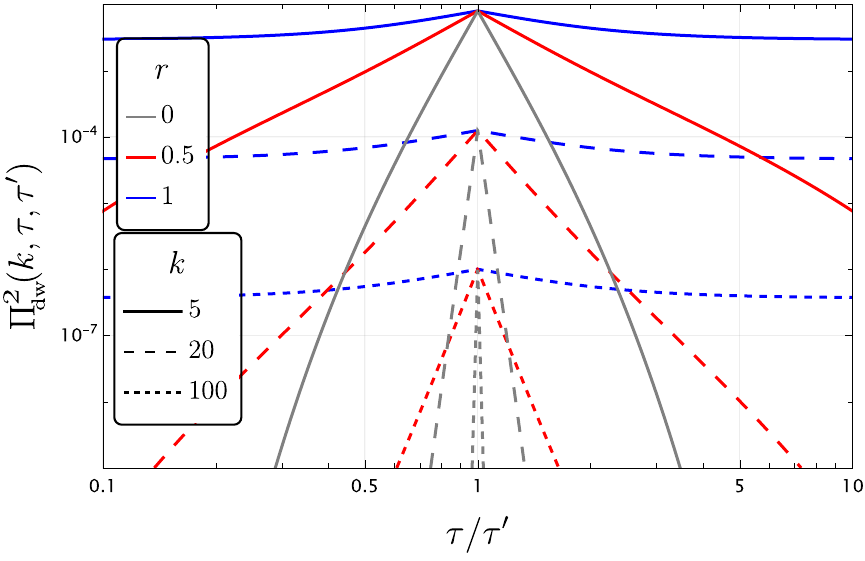}
    \includegraphics[scale=0.55]{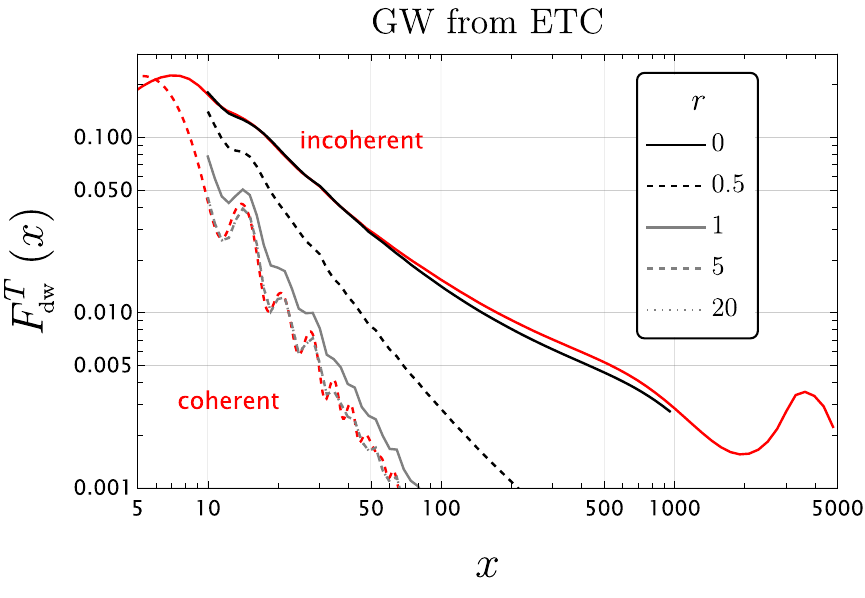}
    \caption{\textbf{Left}: The UTC across the the diagonal $\tau/\tau'=1$ for the different values of $r = 0, 0.5, 1$ and different values of $k$ assuming the form in Eq.\eqref{Eq:inter}.  Interestingly, we observe that the falloff of the UTC away from the diagonal does not depend on $k$ for the $r=1$ (close to coherent case), while it depends strongly on $k$ for the highly incoherent $r=0$ case. This mimics the $k$-dependent falloff obtained for string networks in the simulations of \cite{Durrer:2001cg, Bevis:2010gj}, justifying our choice of the function. \textbf{Right}: The function $F^T_{\rm dw}(x)$ using 1) the incoherent source (upper solid red line), 2) the coherent source (lower dashed red line) and 3) the interpolating function for different values of $r$ (the several black and gray lines).}
    \label{fig:ETC_GW_FTa}
\end{figure}

The UTC resulting from the ansatz above for $C_{\rm dw}^T$ is illustrated on the left panel of Fig.\,\ref{fig:ETC_GW_FTa}. 
We observe that this function interpolates approximately (but actually quite precisely) between the coherent and the incoherent case. 
Numerically, we notice that varying the factor $r \in[0, \infty]$, the GW spectrum slope interpolates between the case of a coherent source for $r = \infty$ (in practice, $r \sim $ a few is enough to recover the coherent behaviour) and the incoherent case for $r = 0$. The results for $r = 0, 0.5, 1, 2, 5, 20$ are shown in the right panel of Fig.\,\ref{fig:ETC_GW_FTa}. 

\begin{figure}
    \centering
    \includegraphics[scale=0.45]{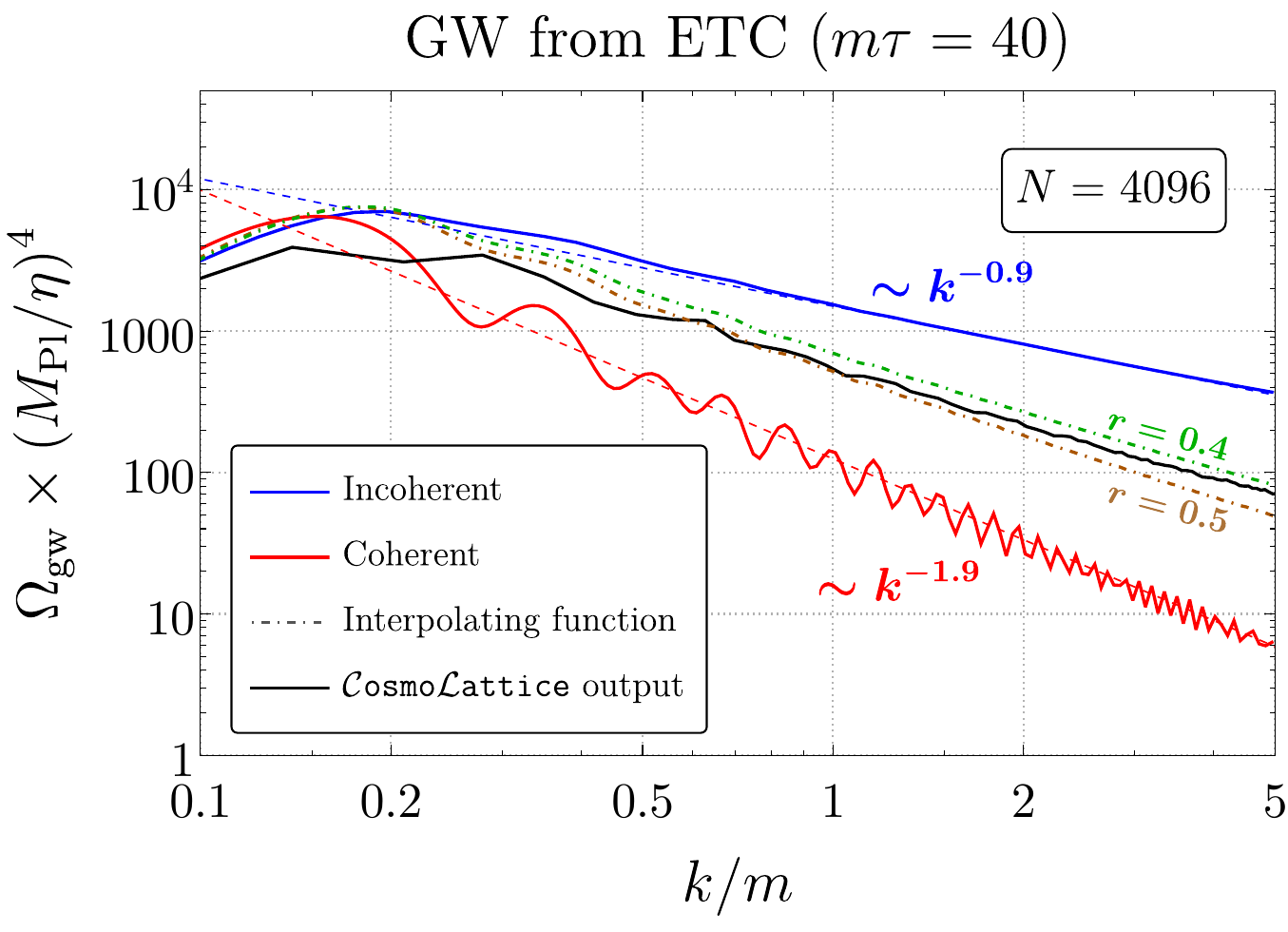}
   \caption{
   GW spectrum for the incoherent (solid blue) and coherent (solid red) cases using the $C^T_{\rm dw}(x,x)$ data for the $4K$ simulation at $m\tau = 40$. The blue dashed line is the function $k^{-0.9}$ while the dashed red one is $k^{-1.9}$, which correspond to the expected large $x$ behaviours as discussed around Eqs.\eqref{eq:FTdwinc} and \eqref{eq:cohe_anal}. The green and brown dot-dashed curves represent the interpolating function Eq.\,\eqref{Eq:inter} with $r = 0.4$ and $0.5$ respectively. The black line is the spectrum provided directly by \CL. }
 
    \label{fig:ETC_GW}
\end{figure}

The implications of our interpolating ansatz for the GW spectrum are shown in Fig.\,\ref{fig:ETC_GW}. The black line indicates the $4K$ GW spectrum obtained from \CL ~at $m\tau = 40$. As we can see, this lies between the totally incoherent and totally coherent approximations given by the blue and red curves, respectively, and can actually be described (for modes with $k > k_{\rm peak}$, for which Eq.\,\eqref{eq: rho_gw Caprini} applies) by the interpolating ansatz in Eq.\,\eqref{Eq:inter} with $r \sim 0.4-0.5$, indicating a partial coherence of the source. We consider this result as a first step for understanding the actual degree of coherence of a DW network in the scaling regime and the corresponding dynamics at small scales.

Let us finally notice that, as the definition of $\Pi_{ij}^{TT}$ in Eq.\,\eqref{eq: T equals Pi}, and therefore Eq.\,\eqref{eq: Pi to CT}, already factors out most of the dependence on the cosmology, we expect the function $C^T_{\rm dw}$ to be largely independent of the background expansion (while of course depending strongly on the type of defect). In the next section we will test this expectation explicitly for different cosmologies.

\section{Gravitational waves from domain walls during exotic cosmologies }
\label{sec: other cosmologies}

We now turn to the study of GW emission from a DW network during the scaling regime in different cosmological backgrounds. Specifically, we depart from the standard radiation dominated Universe and consider two alternative scenarios: an epoch of kination, characterized by an EoS $\omega = 1$ (corresponding to $p = 0.5$), and an exotic cosmological model with $\omega = 2/3$ ($p = 2/3$)\footnote{We do not consider a matter-dominated era, as in our simulations the domain wall system did not have sufficient time to reach the scaling regime before the simulation ended. 
}, for which we will recompute all the GW and scaling parameters.

Before discussing our numerical results, let us first consider the GW spectrum as obtained from the UTC to understand the possible outcomes. First of all, within the cosmological epoch under consideration, the scale factor and DW energy density evolve as
\begin{equation}
a(\tau) =a_i\left(\frac{\mathcal{H}_i\tau}{p}\right)^p \quad \text{and} \quad \rho_{\rm dw}(\tau) = \frac{\mathcal{A}\sigma H(\tau)}{\tilde{p}}\,,
\end{equation}
where $a_i$ and $\mathcal{H}_i$ are the scale factor and conformal Hubble at some initial conformal time $\tau_i$ and
\begin{equation}
    \mathcal{A} = \frac{\tilde{p}A}{\mathcal{H}V} \quad \text{with} \quad \tilde{p} = \frac{2}{3(1+\omega)} = \frac{p}{1+p}\,,
\end{equation}
where $\mathcal{A}$ is similarly defined as in Eq.\eqref{eq:method 1}. The factor $\tilde{p}$ ensures the correct scaling behaviour in terms of cosmic time $t$ (see the discussion below Eq.~\eqref{eq: rho dw wiggle}). Starting from Eq.~\eqref{eq: rho_gw Caprini}, the GW spectrum then takes the form:
\begin{align}
\label{eq:FT}
    \Omega_{\rm gw}^{\rm dw}(\tau, k) &= \frac{16}{3}\frac{G^2\mathcal{A}^2\sigma^2}{\tilde{p}^2}a_i^2\left(\frac{\mathcal{H}_i}{p}\right)^{2p}\tau^{2+2p}F^T_{\rm dw}(x)\,,\\
    \label{eq:FT2}
    F^T_{\rm dw}(x) &= \frac{1}{x^{4p}}\int_{x_i}^x \mathrm{d}x_1 \int_{x_i}^x\mathrm{d}x_2 \left(x_1 x_2\right)^{2p+1/2}\cos\left(x_1-x_2\right) C^T_{\rm dw}(x_1,x_2)\,.
\end{align}
Note that Eq.\eqref{eq: rho_gw Caprini} is only satisfied for subhorizon modes. Indeed, one assumes $k^2 \gg a''/a \sim \tau^{-2}$ in the EoM of the GW perturbation in Eq.\eqref{eq:EoM}.

Starting from the equations above, one can show that in the totally coherent, Eq.\,\eqref{eq:coh}, and totally incoherent, Eq.\,\eqref{eq:incoh}, approximation the large $x$ behaviour of the function $F^T_{\rm dw}(x)$ is actually $p-$independent. We have checked numerically that this remains true also for the interpolating ansatz in Eq.\,\eqref{Eq:inter}.

For the totally incoherent case, and assuming that $C^T_{\rm dw}(x_1,x_2)$ is independent of the cosmological background at large $x_{1,2}$, one finds:
\begin{equation}
    F^T_{\rm dw}(x)\big|_{x \to \infty} \propto  x^{2-q} \qquad (2 < q < 4p+2)\,.
\end{equation}
In the previous section, we found that $q \approx 3$ in radiation domination, which would actually fall in the range specified above for all epochs with $\omega \in [0, 1]$, as then $p \in [0.5, 2]$. If $q\approx3$ was to hold independently of $p$, \emph{i.e.} on the background evolution, as we will confirm below, then the GW spectrum from scaling DWs will exhibit the same power-law behaviour for subhorizon modes for a large set of cosmological backgrounds. This may be contrasted for instance with the case of cosmic strings or scaling sources whose energy density remains a constant fraction of the critical density, where a different cosmological evolution implies a different GW spectral shape, see e.g. \cite{Caprini:2018mtu,Cui:2018rwi,Gouttenoire:2019kij,Blasi:2020wpy,Figueroa:2020lvo,Gouttenoire:2021jhk}. In particular, the GW spectrum associated to the scaling source considered in Ref.\,\cite{Figueroa:2020lvo} was found to be tilted for subhorizon modes ($\propto k^{-2}$) in matter domination, while being scale invariant ($\propto k^0$) in radiation domination. The reason for the different behavior expected from a DW network can again be traced back to the function $F^T_{\rm dw}(x)$ and the fact that it is dominated by the upper limits of integration.

In the following sections, we will examine this matter in detail by focussing on the specific cases of kination and the exotic cosmological model introduced above. As in radiation domination, we initialize the field with a white-noise spectrum, set the initial Hubble parameter to $H_i = m$, and choose the conformal time origin as $m\tau_i = 0$. We have performed five simulations for each cosmology on a $N^3=2048^3$ grid, with the comoving box size, $L$, chosen to enclose eight Hubble volumes at the end of the simulation. The DW width is similarly resolved by at least two grid points. 

Let us anticipate that our findings, summarized at the end of this section in Fig.\,\ref{fig:CTallcosmologies}, are compatible with $C^T_{\rm dw}(x,x)$ being in fact independent of the background cosmology for $x \gg 1$, and support the conclusion of a universal GW spectrum in the UV from scaling domain walls.

\subsection{superhorizon modes}

Let us here discuss the GW spectrum for modes that are superhorizon at a given conformal time $\tau$, namely those with comoving momenta $k$ such that $x= k \tau \ll 1$. For these modes, one needs to consider the full Green function taking into account the $a''(\tau)$ in the EoM \eqref{eq:EoM}. First of all, it should be noted that the Green's function given in Eq.\eqref{eq: Green function} is only valid for subhorizon modes. Following the steps outlined in \cite{Figueroa:2020lvo}, it is observed that for an arbitrary cosmology with $a \propto \tau^p$ and for superhorizon modes, one simply needs to replace $\cos(x_1-x_2)$ within the function $F^T_{\rm dw}(x)$ by
\begin{equation}
    \cos(x_1-x_2) \longrightarrow \left[1-p + p\left(\frac{x_1}{x}\right)\right]\times\left[1-p + p\left(\frac{x_2}{x}\right)\right]\,.
\end{equation} 
In addition, it is assumed that the UTC $C^T_{\rm dw}(x_1,x_2)$ is constant for superhorizon modes and $x_1 \simeq x_2$, and becomes negligible for $x_1 \neq x_2$ as this is required for random, uncorrelated sources of stress–energy on scales much larger than the Hubble horizon. Expanding at small $x$, we then find 
\begin{equation}
    F^T_{\rm dw}(x) \propto \frac{1}{x^{4p}}\left[\int^{x\ll 1}\mathrm{d}x_1 x_1^{2p+1/2}\left(1-p-p\frac{x_1}{x}\right)\right]^2
    \propto \frac{x^{4p+3}}{x^{4p}}
    \propto x^3\,,
\end{equation}
thereby showing the expected behavior $\Omega \propto k^{3}$ for modes that are still outside of the horizon at the conformal time $\tau$.

\subsection{Kination cosmology}

\begin{figure}
    \centering
     \includegraphics[scale=0.55]{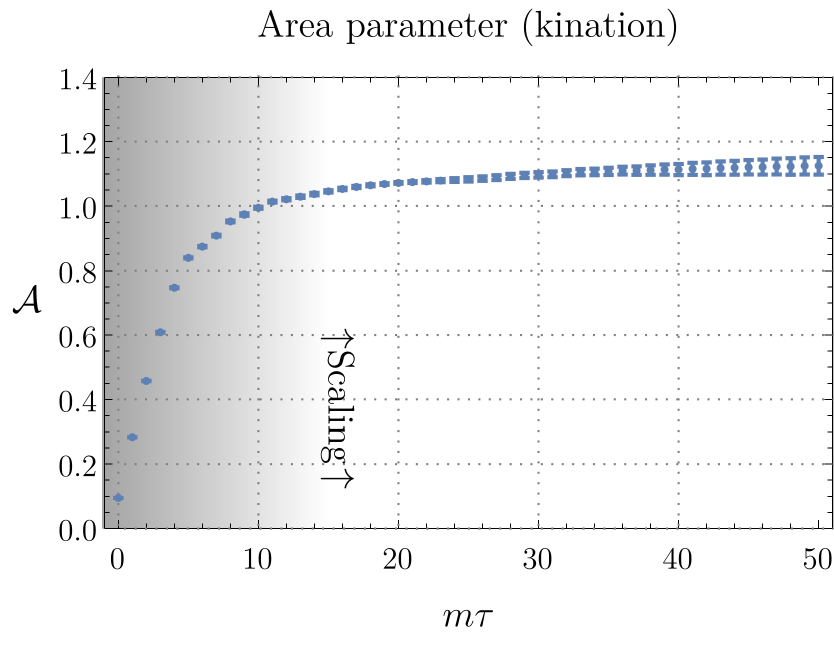}
    \includegraphics[scale=0.59]{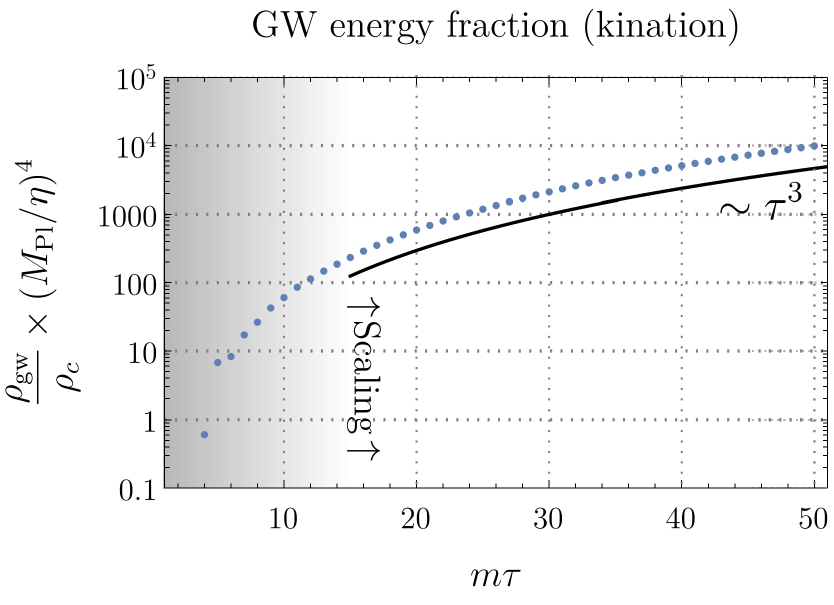}
    \caption{Evolution of the area parameter $\mathcal{A}$ (left) and the GW energy fraction $\rho_{\rm gw}/\rho_c$ (right) for kination, similarly to the radiation dominated case in Fig. \ref{fig:GW_energy}. In this specific cosmological setting, the GW energy fraction grows as $\sim \tau^3$, as expected from Eq.\,\eqref{eq:FT} with $p = 1/2$.}
    \label{fig: kin_gw_energy}
\end{figure}

\begin{figure}
    \centering
    \includegraphics[width=0.45\linewidth]{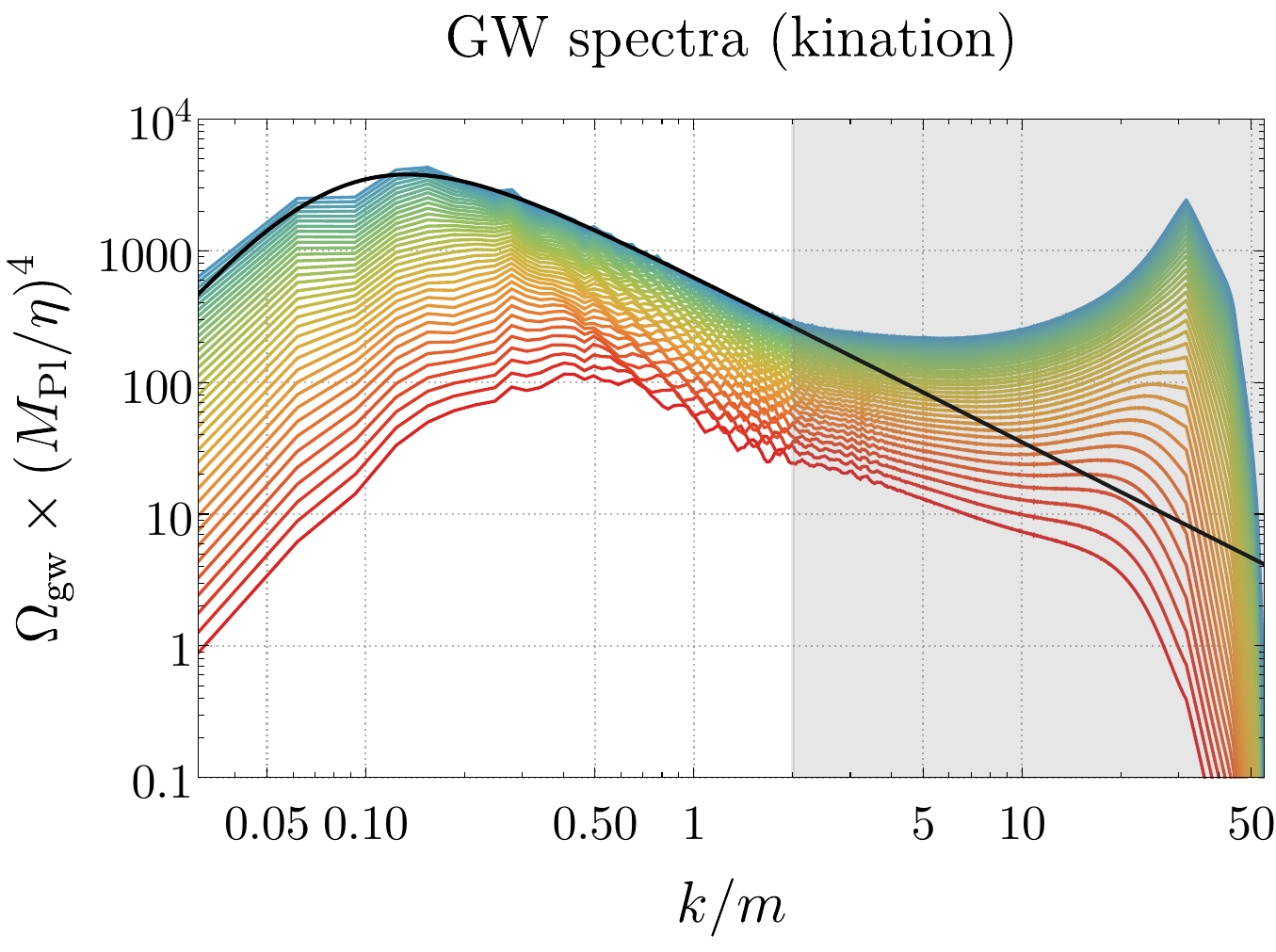}
    \includegraphics[width=0.45\linewidth]{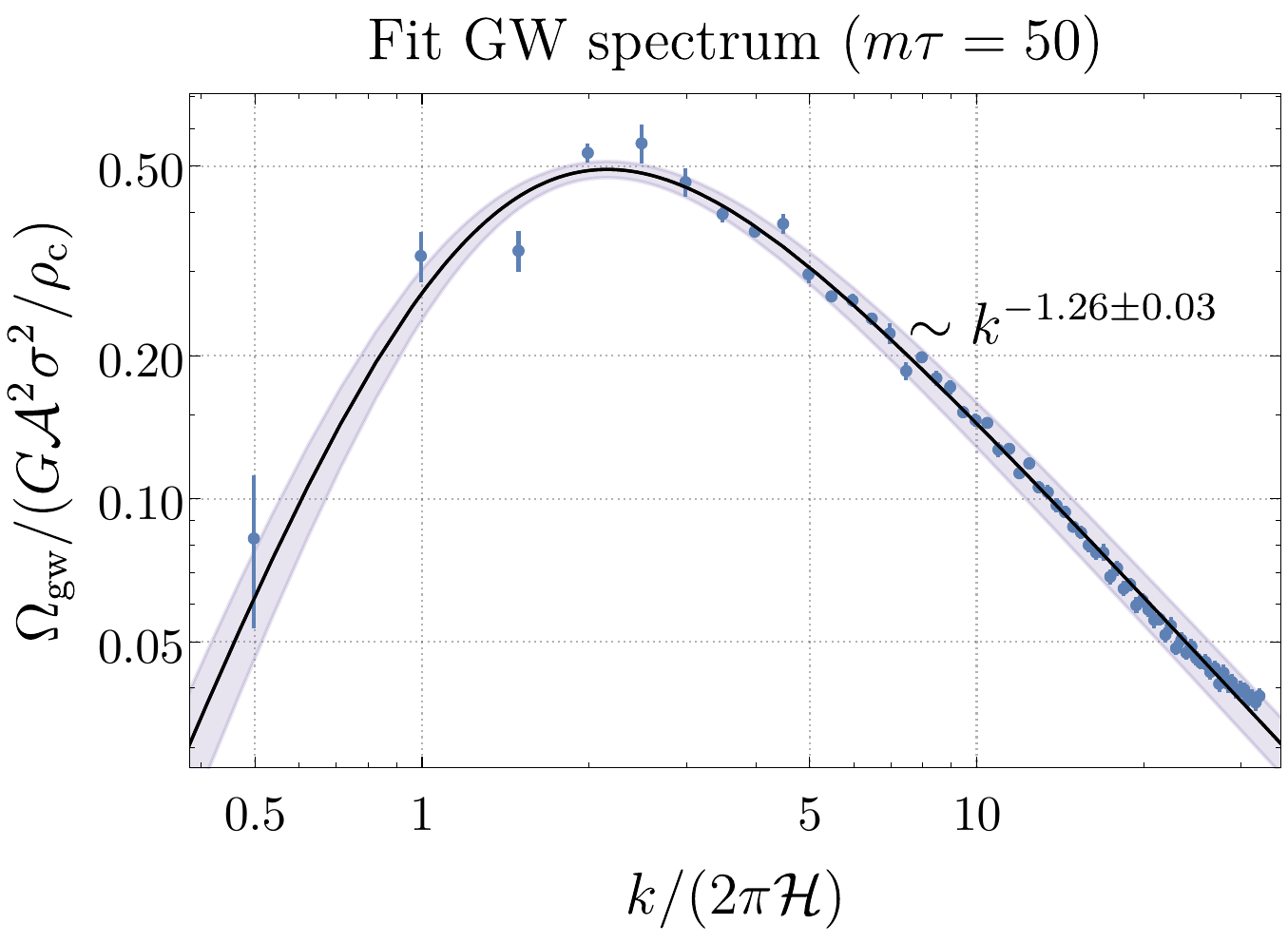}
    \caption{\textbf{Left:} Averaged GW spectra starting from $m\tau = 15$ (bottom red) up to $m\tau = 50$ (top green). The black curve corresponds to the fit on the right plot, while the gray region denotes modes that are omitted from the fit. \textbf{Right:} Similar fit as in Fig.~\ref{fig: GW2K} for the spectral shape given in Eq.\eqref{eq: spectral shape}, shown in black at the final simulation time (in this case $m\tau_{\rm end} = 50$) during kination. The $1\sigma$ error band is visualized by a purple band.}
    \label{fig: GW2Kkin}
\end{figure}

We start our discussion on kination\,\footnote{GW production during kination has been studied in Ref.\,\cite{Gouttenoire:2021jhk} for several early Universe sources such as cosmic strings, first order phase transitions, and inflation
(see also e.g. \cite{
Gouttenoire:2021wzu,
Co:2021lkc,Li:2016mmc,Li:2021htg,
Kuroyanagi:2011fy,Kuroyanagi:2013ns,
Figueroa:2019paj,Bernal:2019lpc,
Bernal:2020ywq,Duval:2024jsg}). However, the case of a domain wall network has not been previously considered. 
Constraints on a stiff epoch (including the case of kination) deriving from the enhancement of the GW spectrum from inflation
have been studied using LVK data in \cite{Duval:2024jsg,LIGOScientific:2025kry}.
Notice also that our study does not require a long duration of the kination epoch, which is bounded by observations\,\cite{Eroncel:2025bcb}, but rather only a few $e$-folds for the DW network to consistently enter the scaling regime.
}, for which $p = 1/2$, by determining the area parameter $\mathcal{A}$ and the GW production efficiency $\epsilon_{\rm gw}$. Our results are shown in Fig.~\ref{fig: kin_gw_energy}, where the scaling regime is reached around $m \tau=15$ with converging values of
\begin{equation}
    \mathcal{A} = 1.12 \pm 0.03 \quad \text{and} \quad \epsilon_{\rm gw} = 0.89 \pm 0.06.
\end{equation}
Concerning the fit of the GW spectrum, we perform a similar analysis as for the $2K$ simulations in radiation domination presented in section \,\ref{sec: DW Rad}. We restrict our analysis to $k \lesssim 2m$, and parameterize the GW spectrum as in Eq.~\eqref{eq: spectral shape}. We show the fit in Fig.~\ref{fig: GW2Kkin}, and the values of the fit parameters in Table\,\ref{tab: fit 2K kin}. We find that the spectral behaviour for the subhorizon modes, parameterized by the variable $b$, is close to what we have found in radiation domination.

\begin{table}
    \centering
    \renewcommand{\arraystretch}{1}
    \begin{tabular}{c|c|c|c}
       \multicolumn{4}{c} {Kination domination}
    \\ 
    \hline \hline
     $\tilde{\epsilon}_{\rm gw}$ & $b$ & $c$ & $x_{\rm p}$ \\
    \hline \hline
          $0.49 \pm 0.02$ & $1.26 \pm 0.03 $ &
          $2.10 \pm 0.28$ &
          $2.16 \pm 0.11$ \\ 
         \hline
    \end{tabular}
    \caption{Best-fit values for the parameters reconstructing the GW spectral shape as given in Eq.\,\eqref{eq: spectral shape} during kination.}
    \label{tab: fit 2K kin}
\end{table}

\begin{figure}
    \centering
    \includegraphics[width=0.49\linewidth]{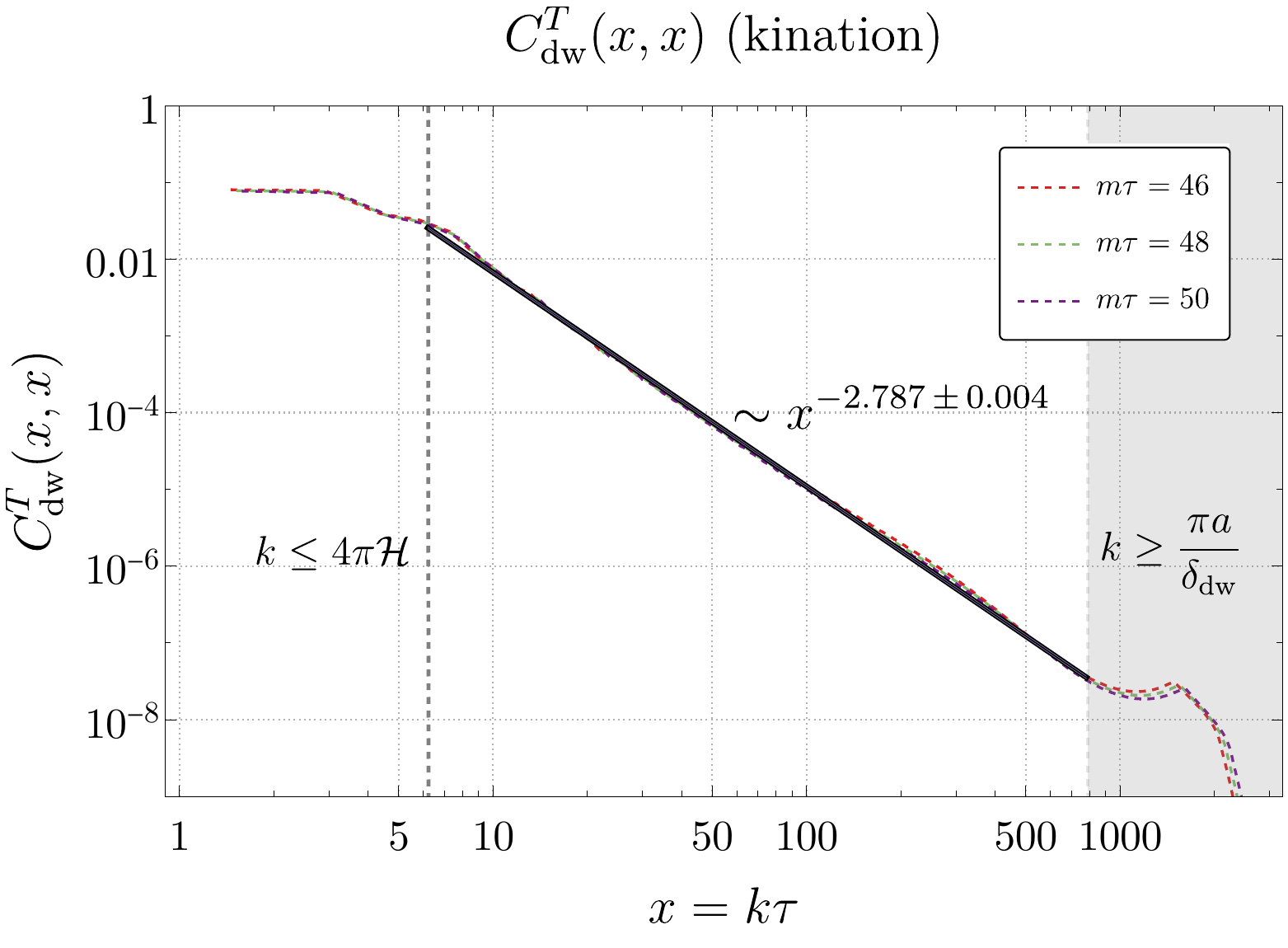}
    \includegraphics[width=0.49\linewidth]{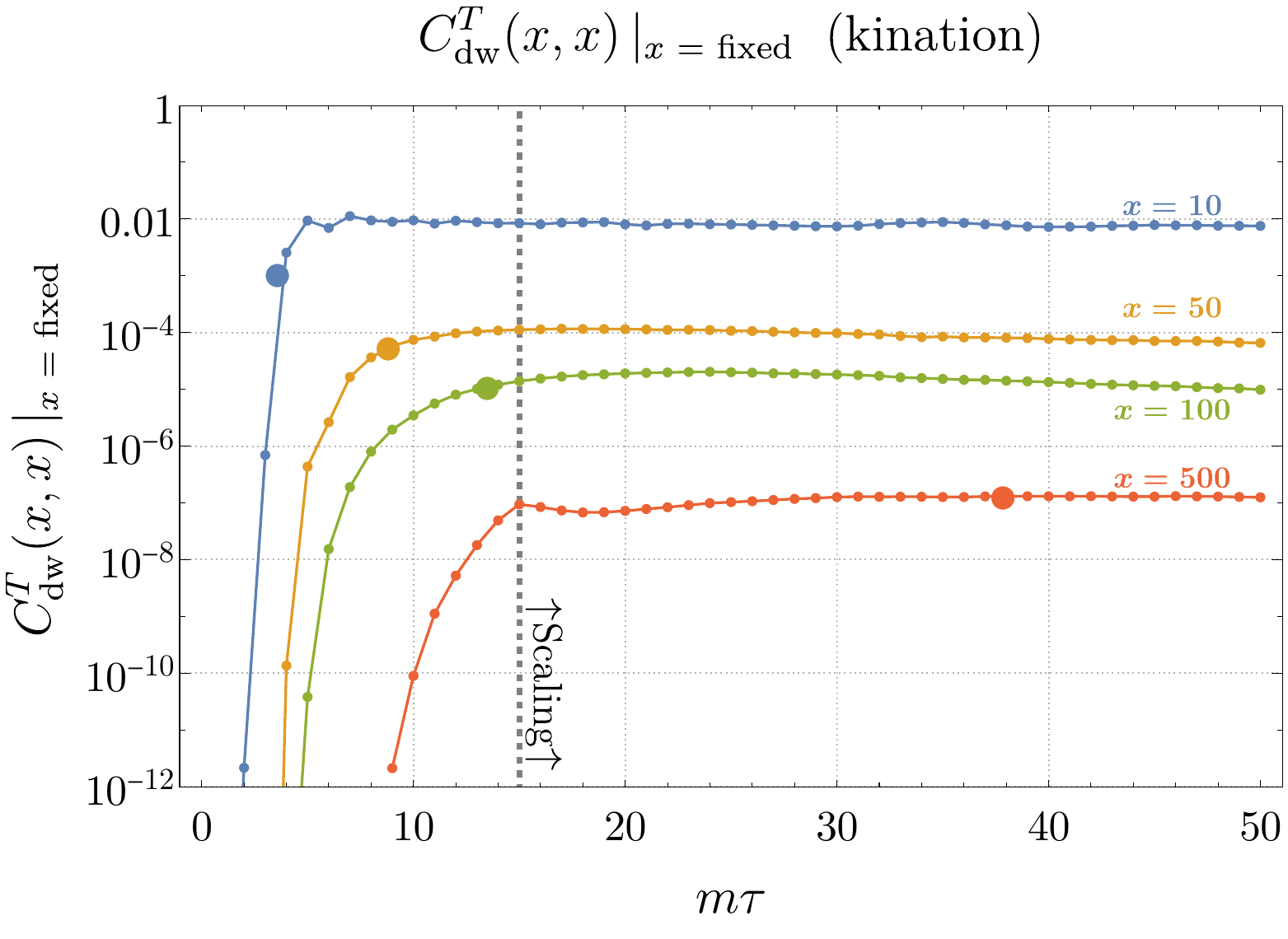}
    \caption{\textbf{Left:} Fit of the function $C^T_{\rm dw}(x,x)$ shown at the final simulation time ($m\tau_{\rm end} = 50$) in black during kination. The data exhibit a power-law behaviour consistent with $C^T_{\rm dw}(x,x) \sim x^{-2.787}$. The dashed curves show the averaged $C^T_{\rm dw}(x,x)$ from late times, $m\tau = 46$, $48$ and $50$, during which the domain wall system has fully entered the scaling regime. For the fit, only modes larger than twice the Hubble scale and smaller than twice the inverse width of the wall are considered. \textbf{Right:} Evolution of the ETC $C^T_{\rm dw}(x,x)$ for fixed values of $x$ during kination. The enlarged dots correspond to the momentum related to twice the domain wall width, and the gray line indicates the onset of the scaling regime.}
    \label{fig: CT_KIN}
\end{figure}

Moving on to the ETC during kination, we again determine numerically the scaling of the function $C^T_{\rm dw}(x,x)$ at the final simulation time, $m\tau_{\rm end} = 50$, as shown in the left panel of Fig.~\ref{fig: CT_KIN}, and find:
\begin{equation}
    C^T_{\rm dw}(x,x) \propto x^{-2.787 \pm 0.004}\,.
\end{equation}
We emphasize that this result is compatible with our expectation that the ETC is largely background (or $p$) independent, \emph{i.e.} the same as in the radiation domination era. In addition, by comparing the GW spectrum from kination to the RD case, we see that both UV slopes are remarkably similar, thus hinting to a universal shape.

\subsection{Exotic cosmological model}

\begin{figure}
    \centering
     \includegraphics[scale=0.55]{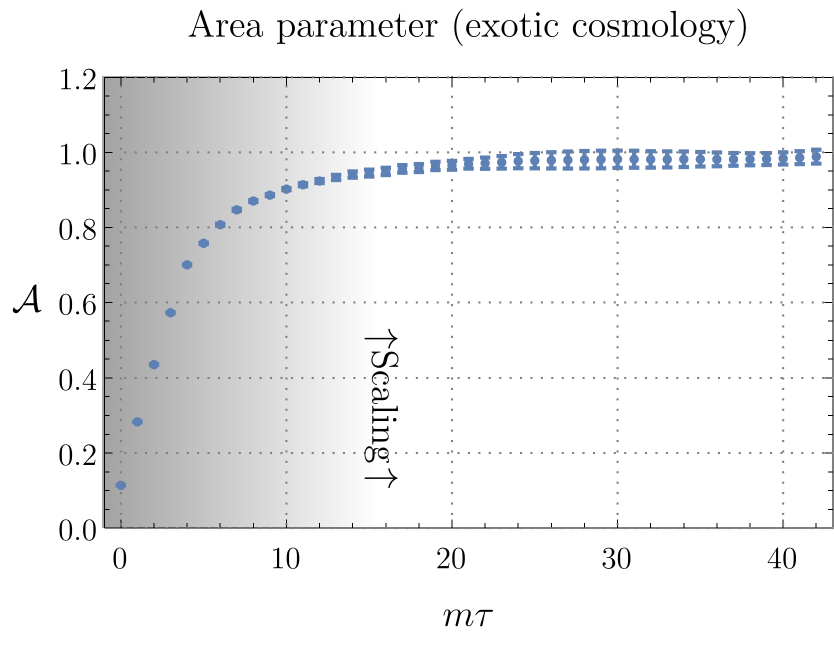}
    \includegraphics[scale=0.59]{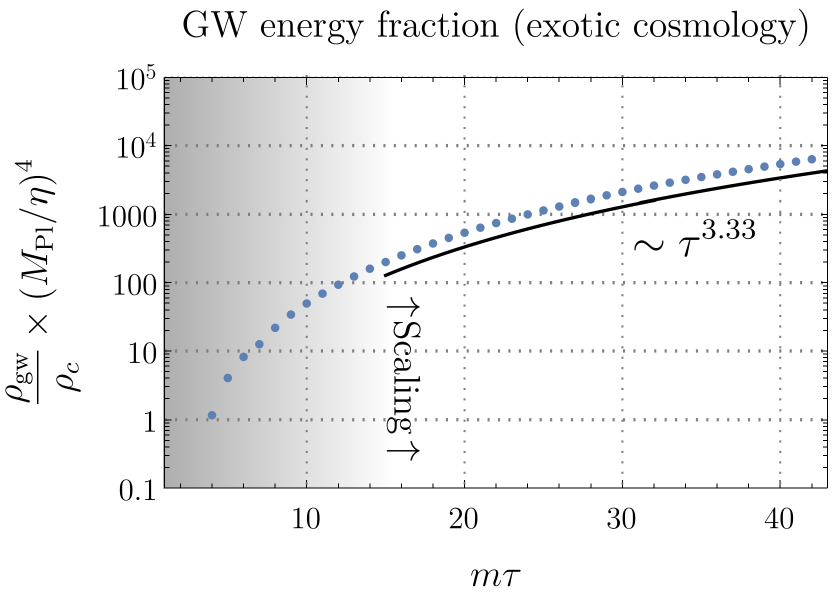}
    \caption{Evolution of the area parameter $\mathcal{A}$ (left) and the GW energy fraction $\rho_{\rm gw}/\rho_c$ (right) for the exotic cosmology with $p=2/3$, similarly to the radiation dominated case in Fig.\,\ref{fig:GW_energy}. In this specific cosmological setting, the GW energy fraction grows as $\sim \tau^{3.33}$ as expected from Eq.\eqref{eq:FT} by taking $p = 2/3$.}
    \label{fig: exotic cosmological_gw_energy}
\end{figure}

\begin{figure}
    \centering
    \includegraphics[width=0.45\linewidth]{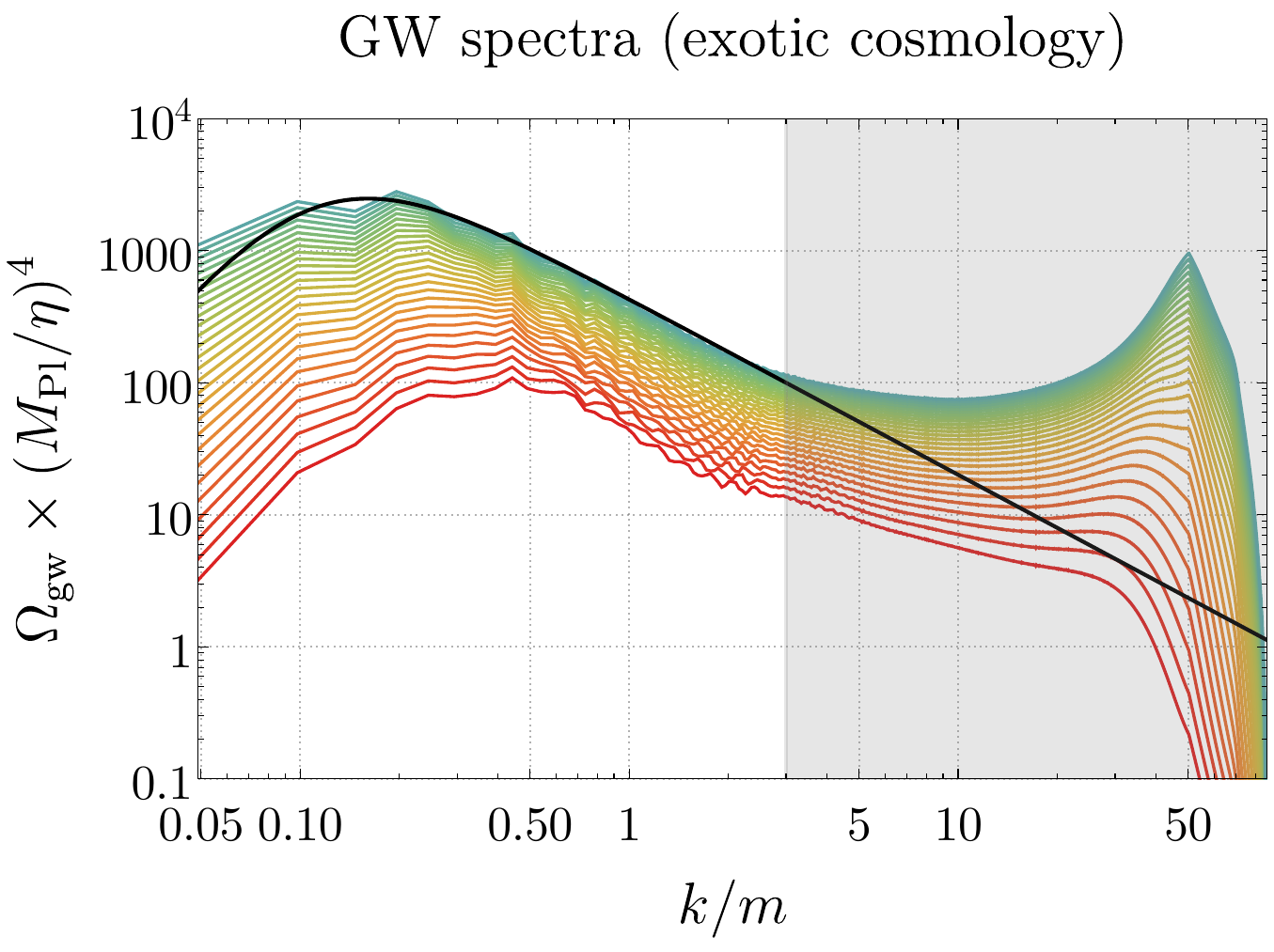}
    \includegraphics[width=0.45\linewidth]{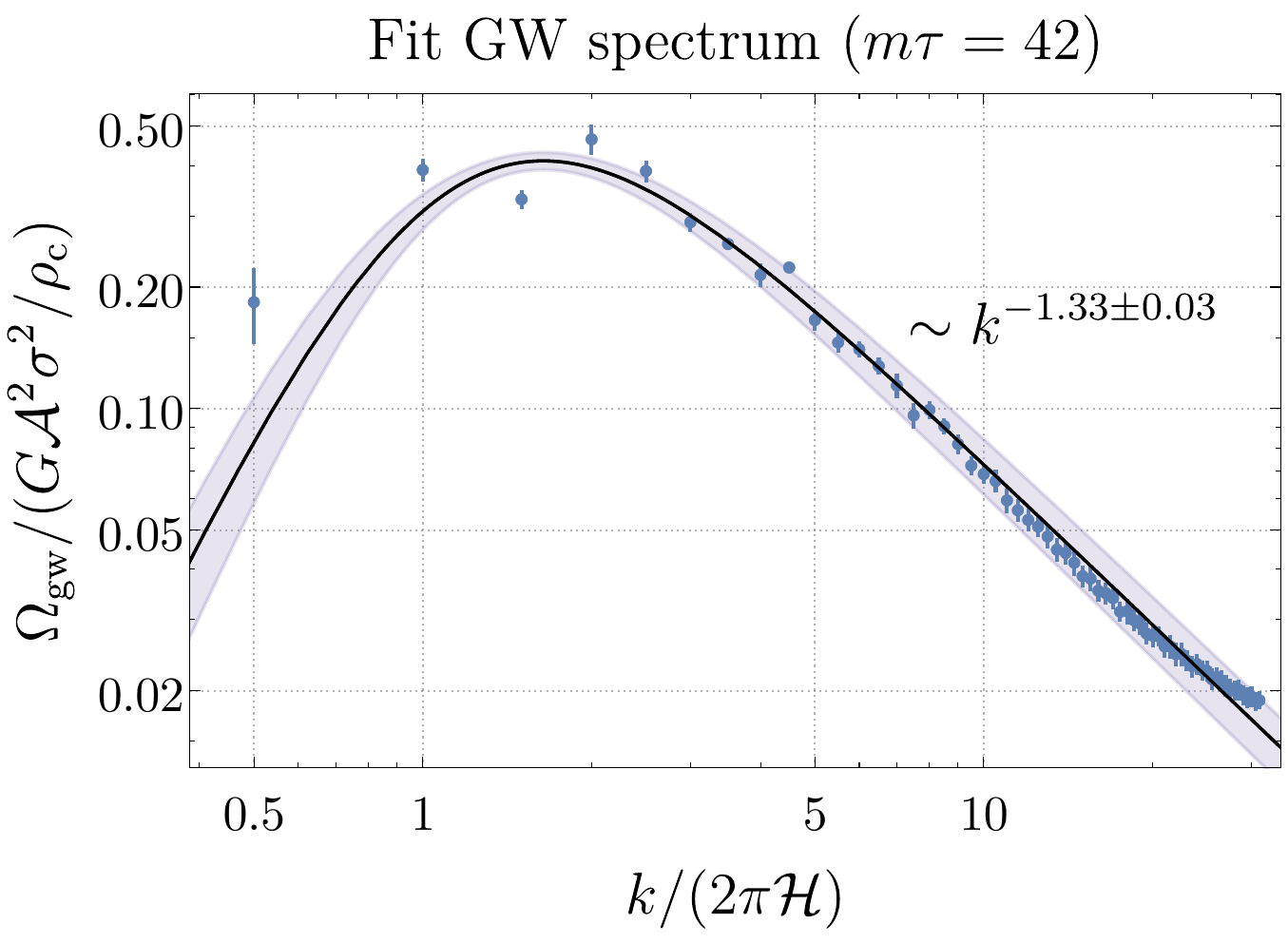}
    \caption{\textbf{Left:} Averaged GW spectra starting from $m\tau = 15$ (bottom red) up to $m\tau = 42$ (top green) in the exotic cosmology with $p=2/3$. The black curve corresponds to the fit on the right plot, while the gray region denotes modes that are omitted from the fit. \textbf{Right:} Similar fit as in Fig. \ref{fig: GW2K} for the spectral shape given in Eq.\eqref{eq: spectral shape} shown in black at the final simulation time (in this case $m\tau_{\rm end} = 42$) for the exotic cosmological model. The $1\sigma$ error band is visualized by a purple band.}
    \label{fig: GW2Kexotic cosmological}
\end{figure}

\begin{figure}
    \centering
    \includegraphics[width=0.49\linewidth]{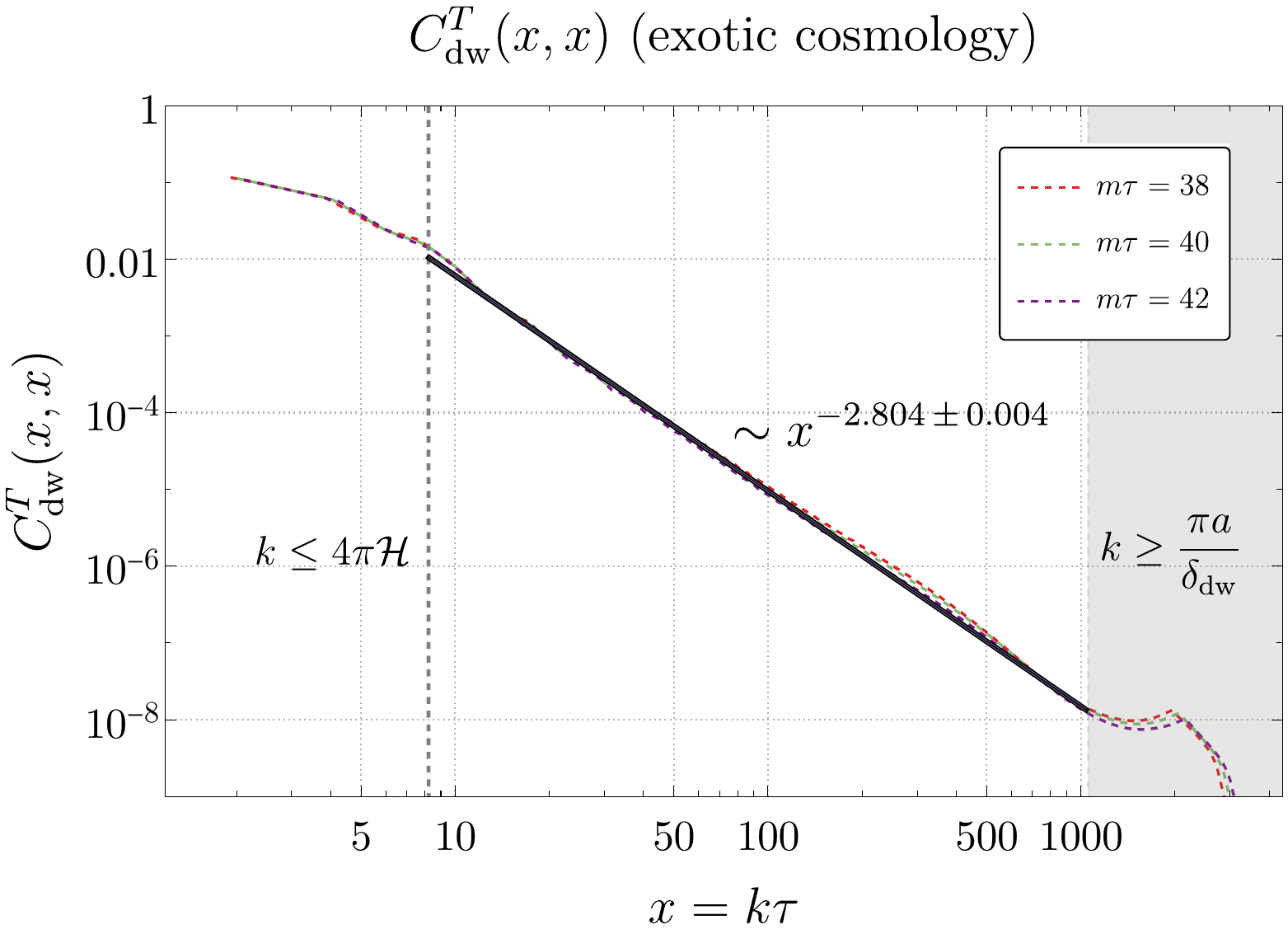}
    \includegraphics[width=0.49\linewidth]{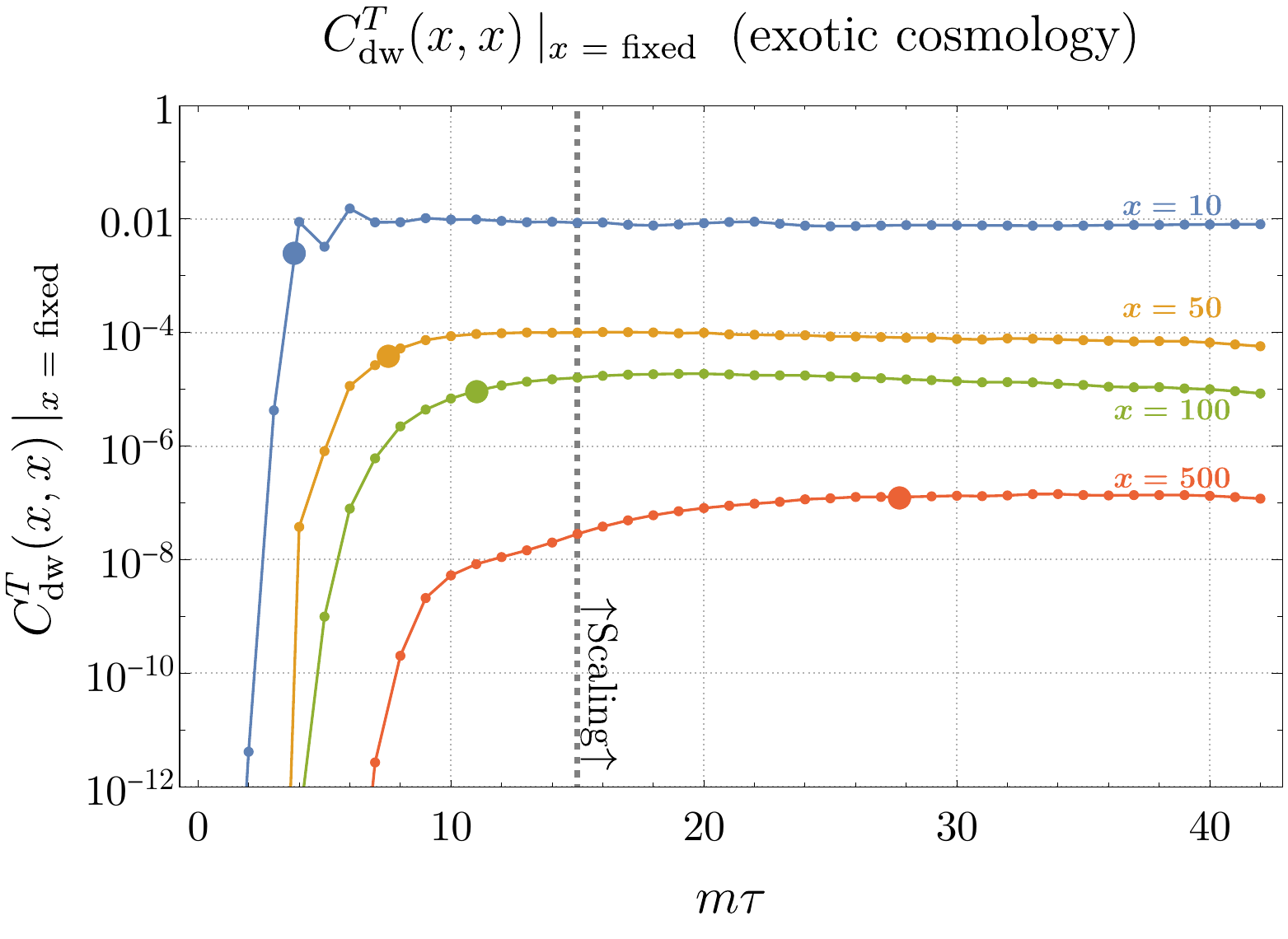}
    \caption{\textbf{Left:} Fit of the function $C^T_{\rm dw}(x,x)$ shown at the final simulation time ($m\tau_{\rm end} = 42$) in black for the exotic cosmological model. The data exhibit a power-law behaviour consistent with $C^T_{\rm dw}(x,x) \sim x^{-2.804}$. The dashed curves show the averaged $C^T_{\rm dw}(x,x)$ from late times, $m\tau = 38$, $40$ and $42$, during which the domain wall system has fully entered the scaling regime. For the fit, only modes larger than twice the Hubble scale and smaller than twice the inverse width of the wall are considered. \textbf{Right:} Evolution of the ETC $C^T_{\rm dw}(x,x)$ for fixed values of $x$ for the exotic cosmological model. The enlarged dots correspond to the momentum related to twice the domain wall width, and the gray line indicates the onset of the scaling regime.}
    \label{fig: CT_exotic cosmological}
\end{figure}

In the case of a cosmological epoch with $\omega = 2/3$ and thus $p=2/3$, we obtain the following numerical values for the area parameter and the GW efficiency:
\begin{equation}
    \mathcal{A} = 0.99 \pm 0.02 \quad \text{and} \quad \epsilon_{\rm gw} = 0.78 \pm 0.04\,,
\end{equation}
with data plotted in Fig.~\ref{fig: exotic cosmological_gw_energy}.
For the GW spectrum, we fit only momenta $k \lesssim 3m$, and parameterize the spectrum as in Eq.\,\eqref{eq: spectral shape}. The GW spectral fit is shown in Fig.~\ref{fig: GW2Kexotic cosmological} and the best-fit parameters in Table \ref{tab: fit 2K exotic cosmological}, where a similar slope as in the RD case is found.
Finally, we show the scaling of the function $C^T_{\rm dw}(x,x)$ in Fig. \ref{fig: CT_exotic cosmological}, with the numerical fit given by
\begin{equation}
    C^T_{\rm dw}(x,x) \propto x^{-2.804 \pm 0.004}\,.
\end{equation}
As we can see, both the slope of the GW spectrum and the ETC, $C^T_{\rm dw}(x,x)$, at $x \gg 1$ are compatible with those found in the radiation and kination cases.

\subsection{Summary}

\begin{figure}
    \centering
    \includegraphics[width=0.65\linewidth]{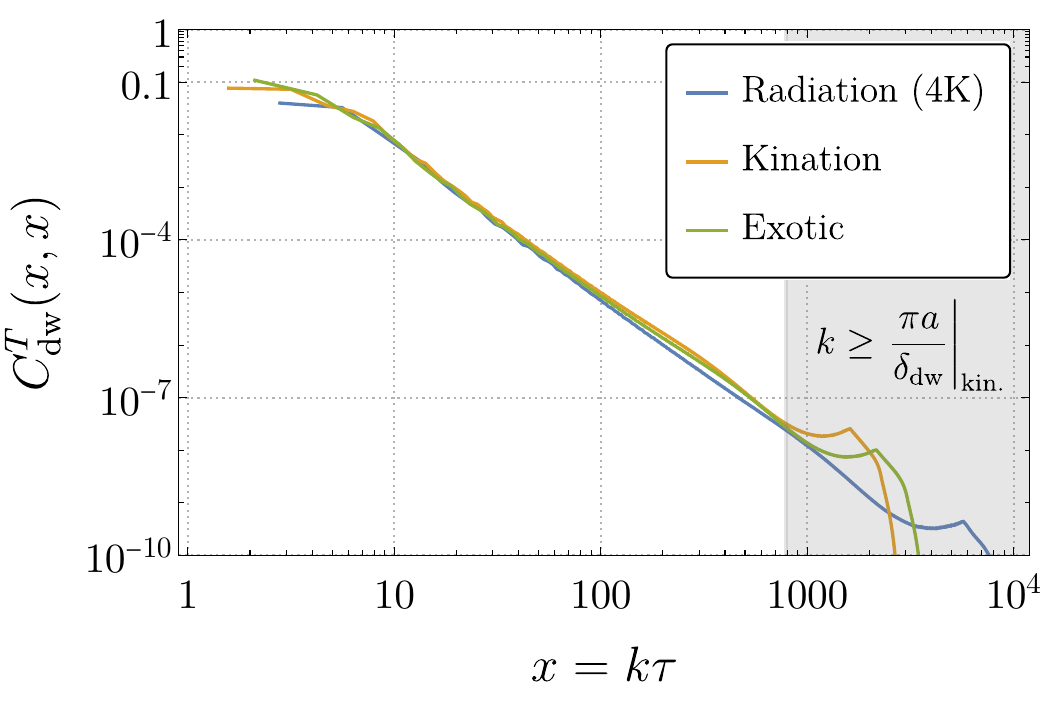}
    \caption{Combined plot of the $C^T_{\rm dw}(x,x)$ curves for the three different cosmologies considered in this work. Curves are shown at the final simulation time for kination (orange) and the exotic cosmology (green), \emph{i.e.} $m\tau = 50$ and $m\tau = 42$ respectively. For the radiation dominated scenario, we show the result from the $4K$ simulation at $m\tau = 40$ (blue). These results illustrate the background independence of the correlator $C^T_{\rm dw}$. The gray shaded band corresponds to modes larger than half the inverse DW width for the case of kination.
    }
    \label{fig:CTallcosmologies}
\end{figure}

We here summarize the results of this section, and combine them with the ones obtained in radiation domination. The comparison of the ETC in the various background cosmologies is shown in Fig.\,\ref{fig:CTallcosmologies}. The function $C^T_{\rm dw}(x,x)$ appears to be universal, $C_{\rm dw}^T(x,x) \simeq C/x^q$, with $C=\mathcal{O}(1)$ and $q \approx2.8$ for $x \gg1$. A residual dependence of $C^T_{\rm dw}(x,x)$ on the cosmology still remains at (super)horizon scales, even though the overall qualitative behaviour remains unchanged.

This universal ETC hints toward a universal shape of the GW spectrum from a scaling domain wall network for subhorizon modes. In particular, we have checked that the momentum dependence through the self-similar variable $x=k\tau$ resulting from the integral in Eq.\eqref{eq:FT2} is actually independent of $p$ for all the different Ans\"atze for the UTC (incoherent, coherent, and intermediate) considered in Sec.\,\ref{sec: GW from ETC}. This provides a possible explanation for the close similarity of the UV slope for the GW spectrum observed for all the various background cosmologies explored in this paper. This should be contrasted to the scaling source considered in Ref.\,\cite{Figueroa:2020lvo} where matter and radiation domination were found to lead to a different spectral shape in the UV.

Let us finally mention that, while we expect a GW spectrum $\propto k^3$ for modes that are superhorizon at the end of our numerical simulations, the actual IR shape as observed today may differ depending on the subsequent cosmology. In fact, as the DW network needs to eventually annihilate to avoid the DW problem, the evolution of the modes that are superhorizon at the time of annihilation can impart a different IR tilt to the spectrum\,\cite{Barenboim:2016mjm,Cai:2019cdl,Ellis:2020nnr,Hook:2020phx,Gouttenoire:2021jhk}, potentially allowing to pinpoint the cosmological epoch when annihilation occurred.

\begin{table}
    \centering
    \renewcommand{\arraystretch}{1}
    \begin{tabular}{c|c|c|c}
    \multicolumn{4}{c} {Exotic cosmology}
    \\ 
    \hline \hline
     $\tilde{\epsilon}_{\rm gw}$ & $b$ & $c$ & $x_{\rm p}$ \\
    \hline \hline
          $0.41 \pm 0.02$ & $1.33 \pm 0.03 $ &
          $1.88 \pm 0.37$ &
          $1.64 \pm 0.09$ \\ 
         \hline
    \end{tabular}
    \caption{Best-fit values for the parameters reconstructing the GW spectral shape as given in Eq.\eqref{eq: spectral shape} for the exotic cosmological model.}
    \label{tab: fit 2K exotic cosmological}
\end{table}

\section{Cosmic archaeology with domain walls}
\label{sec:pheno}

Let us discuss here some examples highlighting the features in the GW spectrum from scaling domain walls that can arise in connection with a non-standard expansion history. 

In particular, we will consider the case of a DW network that annihilates during a cosmological epoch characterized by a generic EoS. We shall assume that the bias term annihilating the network is temperature independent, so that the Hubble horizon at annihilation is given by equating the vacuum pressure from the bias, $\epsilon$, to the tension force, $\sigma H$. One then obtains $H_{\rm ann}=\sigma/\epsilon$ independently of the underlying cosmology. We will then assume that the non-standard epoch lasts for a fixed number of $e$-folds, $N_e$, after annihilation takes place, and that it will end in standard radiation domination. 

The consequences for the emitted GW spectrum are as follows: \emph{i)} a suppression (or enhancement) of the GW peak amplitude coming from the last moment of scaling\,\footnote{We will neglect here the intrinsic contribution from the phase of network collapse.}; \emph{ii)} a different redshift of the peak amplitude due to the expansion during the non-standard epoch; \emph{iii)} a tilt of the IR spectrum compared to $f^3$ for scales that enter the horizon during the non-standard epoch (notice that the tilted IR spectrum will actually start rather close to the peak frequency, as this is set by the Hubble horizon).

\paragraph{GW spectrum today from radiation domination.} 
We begin by considering the present-day GW spectrum produced by a DW network during radiation domination. To this end, we make use of the two following identities,
\begin{equation}
    g_{\star s} T^3 a^3 = {\rm const.}\,, 
    \qquad \text{and} \qquad
    H = \frac{\pi}{3}\left(\frac{g_\star}{10}\right)^{1/2}
        \frac{T^2}{M_{\rm Pl}}\,,
\end{equation}
where the first expresses conservation of entropy per comoving volume, and the second gives the Hubble rate during radiation domination. Here, $T$ is the temperature of the radiation bath, and $g_\star$ and $g_{\star s}$ denote the effective numbers of relativistic degrees of freedom for the energy and entropy densities, respectively.

As the GW energy density redshifts like radiation, the present-day peak amplitude is related to the one at emission as:
\begin{align}
    \Omega_{\rm gw, \,rad}^{\rm peak}(t_0) &= \Omega_{\rm rad}^0 
       \left(\frac{g_\star}{g_{\star 0}}\right)
       \left(\frac{g_{\star s0}}{g_{\star s}}\right)^{4/3}
       \Omega_{\rm gw, \,rad}^{\rm peak}\nonumber \\
       &\simeq 4.58 \times 10^{-5} \left(\frac{g_\star}{10}\right)
       \left(\frac{g_{\star s}}{10}\right)^{-4/3}
       \Omega_{\rm gw, \,rad}^{\rm peak}\,,
\end{align}
where a subscript ``0’’ denotes quantities evaluated today. In the second line, we used $g_{\star 0} = 3.36$ and $g_{\star s0} = 3.91$, and $\Omega_{\rm rad}^0 \simeq 5.38 \times 10^{-5}$ for the present radiation fraction \cite{ParticleDataGroup:2024cfk}. 
The peak frequency on the other hand is redshifted as:
\begin{align}
    f_{\rm rad,0}^{\rm peak} &= \left(\frac{a}{a_0}\right)f_{\rm rad}^{\rm peak}\nonumber \\
    &\simeq 171 \text{ Hz } \left(\frac{g_\star}{10}\right)^{1/4}
       \left(\frac{g_{\star s}}{10}\right)^{-1/3}x_{\rm p}^{\rm rad}\sqrt{\frac{H_{\rm ann}}{\rm GeV}}\,,
\end{align}
where $a_0 = 1$ is the present scale factor. Here, we used the numerical result $f_{\rm rad}^{\rm peak} = x_{\rm p}^{\rm rad}\, H_{\rm ann}$ and the present-day CMB temperature $T_0 \simeq 2.7255\,$K \cite{ParticleDataGroup:2024cfk}.
 
\paragraph{GW spectrum today from non-standard cosmologies.}
Redshifting the GW amplitude from the time of annihilation specified by $H_{\rm ann}$, and including $N_e$ $e$-folds of non-radiation expansion, leads to the following expression relating the new GW peak amplitude to the one obtained in radiation domination:
\begin{equation}
    \Omega_{{\rm gw},p}^{\rm peak}(t_0) = e^{2\left(\frac{1}{p}-1\right)N_e} \cdot 
    \frac{\tilde \epsilon_{{\rm gw},p}\,\mathcal{A}^2_p}{\tilde \epsilon_{{\rm gw,\, rad}}\, \mathcal{A}_{\rm rad}^2} \cdot \Omega_{\rm gw, \,rad}^{\rm peak}(t_0),
\end{equation}
where we have factored in the possible differences in the GW and scaling parameters for the non-radiation cosmologies compared to the ones in RD, indicated above by $\tilde\epsilon_{\rm gw,\, rad}$ and $\mathcal{A}_{\rm rad}$.

In addition, the peak frequency today will be blue- or redshifted compared to the case of the same DW network annihilating during radiation domination, $f^{\rm peak}_{\rm rad}$, by the following amount:
\begin{equation}
    f^{\rm peak}_{p,0} = e^{\frac{1}{2}\left( \frac{1}{p}-1\right)N_e} \cdot \frac{x_{\rm p}^p}{x_{\rm p}^{\rm rad}} \cdot f^{\rm peak}_{{\rm rad},0},
\end{equation}
where we have taken into account a small intrinsic change in the position of the peak compared to $H_{\rm ann}$ in non-radiation cosmologies by $x_{\rm p}^p/x_{\rm p}^{\rm rad}$, according to our simulations.

\begin{figure}
    \centering
    \includegraphics[scale = 0.51]{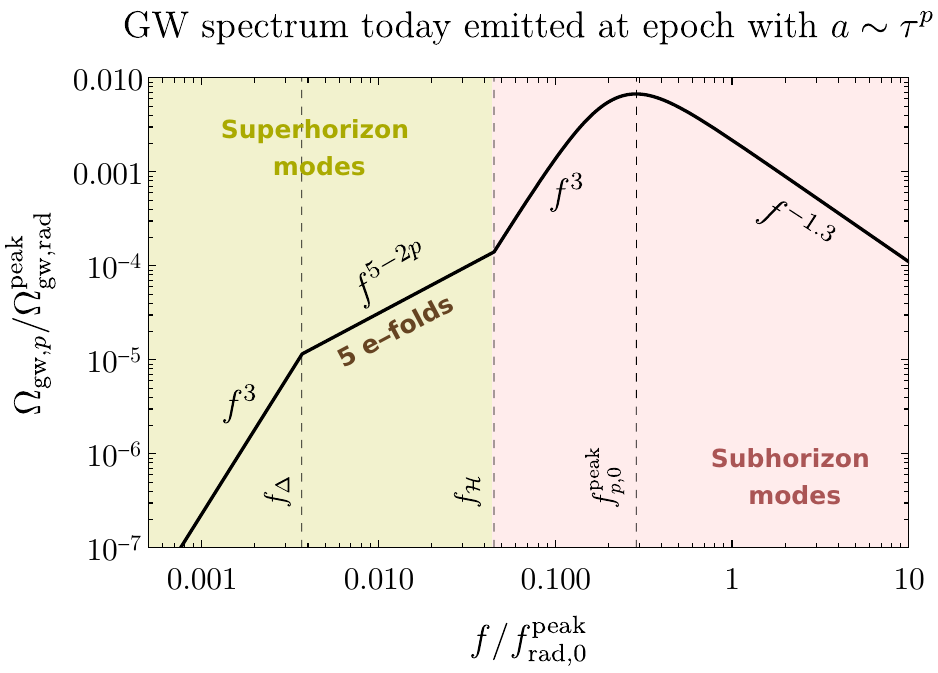}
    \includegraphics[scale = 0.55]{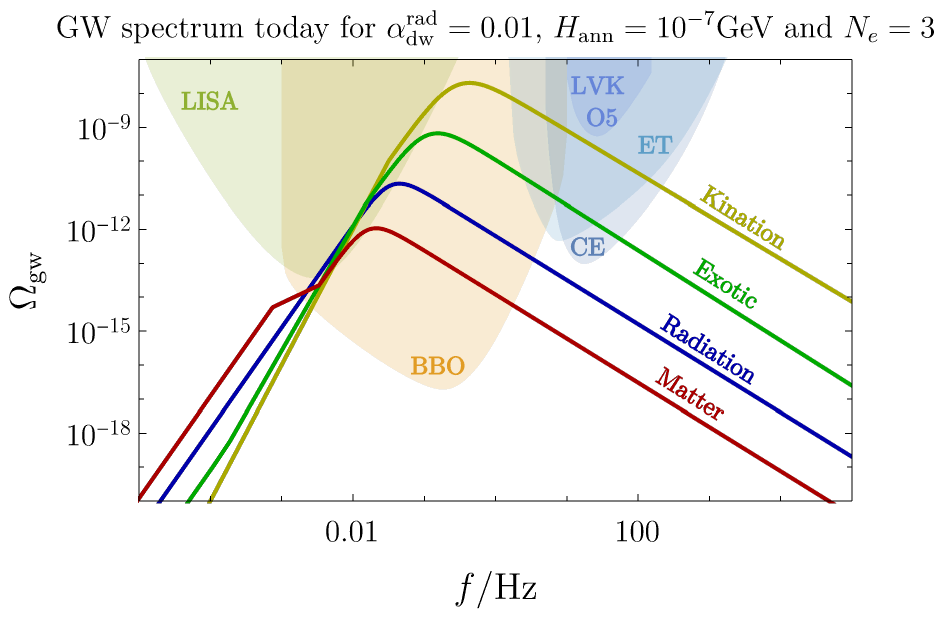}
    \caption{\textbf{Left:} Sketch of the GW spectrum today from a DW network emitted during an epoch with $a \sim \tau^p$ lasting for 5 $e$-folds. For illustration, we assume a matter dominated epoch with $p = 2$ and take $\mathcal{A}_{\rm mat}$, $\tilde{\epsilon}_{\rm gw,\, mat}$, and $x_{\rm p}^{\rm mat}$ equal to their RD values. \textbf{Right:} GW spectra observed today and produced during epochs dominated by radiation (blue, $p=1$), matter (red, $p=2$), kination (yellow, $p=1/2$), and an exotic component (green, $p=2/3$), for fixed $H_{\rm ann} = 10^{-7}$ GeV (for which $g_\star = g_{\star s} = 106.75$) and $\alpha_{\rm dw}^{\rm rad} = 0.01$. Numerical values for $\mathcal{A}_{\rm rad}$, $\tilde{\epsilon}_{\rm gw,\, rad}$, and $x_{\rm p}^{\rm rad}$ are taken from the $4K$ simulation, and for simplicity we assume similar values for the matter dominated spectrum (as in the left plot). Each non-standard epoch is assumed to last for $N_e = 3$ $e$-folds. Shaded regions indicate the projected sensitivities of LISA, BBO, ET, CE, and the fifth observing run of LVK. Sensitivity curves are taken from~\cite{Schmitz:2020syl}, except for the LVK curve from~\cite{LIGOScientific:2025bgj}.
}
    \label{fig: cosmologies prospects}
\end{figure}

Finally, the superhorizon modes at the last moment of GW emission from the DW network will exhibit, upon entering the horizon in the non-standard era, an IR power-law given by \cite{Barenboim:2016mjm,Cai:2019cdl,Ellis:2020nnr,Hook:2020phx,Gouttenoire:2021jhk}
\begin{equation}
    \Omega_{{\rm gw},p}(f)\propto f^{5-2p}, \quad \frac{f_{p,0}^{\rm peak}}{2\pi x_{\rm p}^p} =  f_\mathcal{H} > f > f_\Delta = e^{-\frac{N_e}{p}} \frac{f_{p,0}^{\rm peak}}{2\pi x_{\rm p}^p},
\end{equation}
and $\Omega_{{\rm gw},p}(f)\propto f^3$ for $f_\mathcal{H} < f < f_{p,0}^{\rm peak}$, and the same for $f<f_\Delta$. Here, $f_\mathcal{H}$ is the frequency today corresponding to the mode that enters the horizon at annihilation, \emph{i.e.}\ $k = \mathcal{H}_{\rm ann}$, whereas $f_\Delta$ denotes the present-day frequency of the mode entering the horizon at the end of the non-standard era\,\footnote{Notice that the horizon will in general depend on the cosmology, given that we here consider the same fixed number of $e$-folds for all the non-radiation scenarios.}.
The UV spectrum for $f > f_{p,0}^{\rm peak}$ is instead universal according to our results, and, considering only the contribution from the scaling regime, well approximated by $\Omega_{{\rm gw},p}(f) \propto f^{-1.3}$, independently of the cosmology. On the left plot of Fig.~\ref{fig: cosmologies prospects}, we show a sketch of the effect of the non-standard epoch.

It is instructive to explore the detectability of GWs from a DW network across different cosmological histories. In the right panel of Fig.~\ref{fig: cosmologies prospects}, we show as a reference the spectrum produced during RD (blue), assuming $H_{\rm ann} = 10^{-7}$~GeV and fixing the DW energy fraction to $\alpha_{\rm dw}^{\rm rad} = 0.01$, defined as
\begin{equation}\label{eq: DW abundance}
    \alpha_{\rm dw}^{\rm rad} \equiv \frac{\rho_{\rm dw}^{\rm rad}}{\rho_c}
    = \frac{2\mathcal{A}_{\rm rad} \sigma}{3 H_{\rm ann} M_{\rm Pl}^2}\,.
\end{equation}

Using this RD spectrum as a baseline, we construct the corresponding spectra for the non-standard cosmological eras. For illustration, each non-standard epoch is taken to last for $N_e = 3$ $e$-folds, while keeping the same value of $H_{\rm ann}$. Notice that changing $H_{\rm ann}$ while keeping the same $\alpha_{\rm dw}^{\rm rad}$ will induce an overall shift of all the GW spectra in Fig.\,\ref{fig: cosmologies prospects}, so that different experiments will be able to probe the UV and IR part of the spectrum. Our results thus demonstrate the possibility of detecting GWs from domain walls, while potentially also resolving the IR tilt from a non-standard cosmological era, also for relatively small values of $\alpha_{\rm dw}^{\rm rad}$, especially in in combination with few $e$-folds of a stiff epoch such as kination.

\section{Conclusion}
\label{sec:conclusion}

In this work, we have studied the scaling regime of domain wall networks and their associated GW spectra using large-scale lattice field theory simulations. Particular emphasis was given to the Equal Time Correlator (ETC) of the stress-energy tensor, which provides a powerful diagnostic of the network’s scaling properties and its GW emission. We here summarize our main results and outline possible future directions (all numerical results and fitted parameters can be found in Table~\ref{tab:summary}).

Neglecting particle friction, the domain wall network rapidly evolves toward a scaling regime in which its 
energy density scales proportionally to the Hubble parameter, \emph{i.e.} $\rho_{\rm dw} \propto \sigma/t$ with $t$ the cosmic time, and where the number of domain walls per Hubble volume (quantified by the area parameter $\mathcal{A}$) is of order unity.
By performing simulations with grid size $N^3=1250^3$ and $N^3=2048^3$, we have numerically studied the approach to scaling for various initial conditions, parameterized by different ratios $m/H_i$ and UV cutoffs $k_{\rm cut}/m$ 
(see Fig.~\ref{fig:approach_scaling}). 
In doing so, we have also explored scenarios with highly overdense networks at formation, 
$\mathcal{A}_{\rm initial} \gg 1$. 
Our analysis confirms that the DW network attains a scaling regime with $\mathcal{A} \sim \mathcal{O}(1)$ after only a few $e$-folds, independently of the initial conditions.  While our determination of the scaling parameters is consistent with previous results in the literature, extending simulations to larger volumes and later times would allow to reduce the current $\mathcal{O}(10\%)$ uncertainty, see \emph{e.g.} Eq.\,\eqref{eq:Aparam} and \eqref{eq: epsilon integrated}.

The GW spectrum emitted by the domain walls was obtained by
employing the dedicated package in \CL.
Using five simulations of grid size $N^3 = 2048^3$ and one of grid size $N^3=4096^3$, we have confirmed that the spectrum can be described by a broken power-law with IR slope $k^a$ with $a = 3$ by causality, and UV slope $k^{-b}$. 
In our analysis, we carefully assess the range of validity of the GW spectrum in the UV by comparing numerical simulations with different resolutions. We then obtain $b = 1.31 \pm 0.02$  for the $N^3 = 2048^3$ simulations and $b = 1.30 \pm 0.01$ for the $N^3=4096^3$ simulation. 

To gain further insights into the properties of the domain wall network, we have numerically extracted for the first time the Equal Time Correlator (ETC) of the scalar energy momentum tensor, $C^T_{\rm dw}(x,x)$ in Eq.\,\eqref{eq: Pi to CT}. This provides key information on the domain wall dynamics for what concerns the emission of GWs, and thus allows for a more confident extrapolation of the results from numerical simulations.
The onset of scaling is reflected in the ETC by its dependence solely on the dimensionless combination $x \equiv k\tau$, for momenta smaller than the inverse domain wall width. We find that $C^T_{\rm dw}(x,x) \propto x^{-q}$ with $q \approx 2.8$ at large $x$. Relating the ETC to the Unequal Time Correlator (UTC) requires an assumption about the coherence of the source. We have explored three possibilities: fully coherent, fully incoherent, and an interpolating case with momentum-dependent coherence. We have found that neither of the two limiting cases adequately reproduces the GW spectrum. The partially coherent model, with 
$r \approx 0.4-0.5$ in Eq.\,\eqref{Eq:inter}, provides instead a good fit, suggesting that domain walls are partially coherent GW sources with coherence decreasing at higher momenta. 

We have also simulated DW networks in different cosmological backgrounds, including radiation domination, kination ($\omega = 1$) and an exotic cosmology with $\omega = 2/3$ as Equation of State. The extracted slopes of $C^T_{\rm dw}(x,x) \propto x^{-q}$ remain consistent across all cosmologies at $x \gg 1$, see Fig.\,\ref{fig:CTallcosmologies}, demonstrating that both the scaling properties of the network ETC and the GW spectrum for subhorizon modes (see again Table~\ref{tab:summary}) are largely universal. On the other hand, the peak amplitude today and the IR tail of the GW spectrum do depend on the underlying cosmology, see Fig.\,\ref{fig: cosmologies prospects} where we show for instance how $N_e$ $e$-folds of kination greatly enhance the GW amplitude by a factor of $e^{2 N_e}$ compared to the same DW network in radiation domination.

A natural extension of our work would be to extract the Unequal Time Correlator directly from the numerical simulations. This would allow one to test the degree of coherence of the domain wall network as a GW source against the assumptions considered in this paper (such as the partially coherent one), and to possibly provide an explanation for the rather strong emission in the UV with index $k^{-1.3}$ during the scaling regime.

Another very interesting direction would involve a similar study of the phase of network collapse in setups that include an explicit bias term in the scalar potential. In this case, a numerical determination of the ETC and UTC would provide additional information on the departure from the scaling regime when the collapse begins, highlighting the possible appearance of additional scales and features in the GW spectrum such as the doubly--broken power law observed in Ref.\,\cite{Notari:2025kqq}.

\begin{table}
    \centering
    \renewcommand{\arraystretch}{1}
    \begin{tabular}{c|c|c|c|c|c|c|c|c|c}
     Cosmo & N & $\omega$ & $\tilde{\epsilon}_{\rm gw}$ & $b$ & $c$ & $x_{\rm p}$ & $q$  & Fig. \\
    \hline \hline
       Rad & $4096$ & $1/3$ &   $0.35 \pm 0.01$ & $1.30 \pm 0.01 $ &
          $1.41 \pm 0.10$ &
          $1.01 \pm 0.02$ &  $2.903 \pm 0.004$& \ref{fig:GW_energy},\ref{fig: GW2K},\ref{fig: CT_RAD}  \\ 
         \hline
         Rad & $2048$ & $1/3$ & $0.35 \pm 0.04$ & $1.31 \pm 0.02 $ &
          $1.12 \pm 0.33$ &
          $0.85 \pm 0.04$  &
          $2.818 \pm 0.004$ & \ref{fig:GW_energy},\ref{fig: GW2K},\ref{fig: CT_RAD} \\ 
         \hline
         Kin & $2048$ & $1$ &   $0.49 \pm 0.02$ & $1.26 \pm 0.03 $ &
          $2.10 \pm 0.28$ &
          $2.16 \pm 0.11$  &
          $2.787 \pm 0.004$ & \ref{fig: kin_gw_energy},\ref{fig: GW2Kkin},\ref{fig: CT_KIN} \\ 
         \hline
         Exotic & $2048$ & $2/3$ &   $0.41 \pm 0.02$ & $1.33 \pm 0.03 $ &
          $1.88 \pm 0.37$ &
          $1.64 \pm 0.09$  &
          $2.804 \pm 0.004$ &\ref{fig: exotic cosmological_gw_energy},\ref{fig: GW2Kexotic cosmological},\ref{fig: CT_exotic cosmological}\\ 
         \hline
    \end{tabular}
    \caption{Table summarizing our main results. All the entries are defined in the main text. The fitting function we have used for the GW spectral shape is $\mathcal{S}=(a+b)^c/(b \left(x/x_{\rm p}\right)^{-a/c}+ a \left(x/x_{\rm p}\right)^{b/c})^c$, where we have set $a =3$. We also used $C^T_{\rm dw} \propto x^{-q}$.
    The last entry, ``Fig.'', refers to the figures containing the corresponding results in the main text. The entry $\omega$ refers to the Equation of State of the background cosmology.}
    \label{tab:summary}
\end{table}

\section*{Acknowledgments}

We thank Riccardo Argurio, Daniel Figueroa, Yann Gouttenoire, Adriana Menkara and Géraldine Servant for discussions. The resources and services used in this work were provided by the VSC (Flemish Supercomputer Centre),
funded by the Research Foundation - Flanders (FWO) and the Flemish Government. AR, MV and AM are supported in part by the Strategic Research Program High-Energy Physics
of the Research Council of the Vrije Universiteit Brussel and by the iBOF ``Unlocking the
Dark Universe with Gravitational Wave Observations: from Quantum Optics to Quantum
Gravity'' of the Vlaamse Interuniversitaire Raad. AR is supported
by FWO-Vlaanderen through grant number 1152923N. M.V. is also funded by the European Union (ERC, HoloGW, Grant Agreement No. 101141909). Views and opinions expressed are, however, those of the authors only and do not necessarily reflect those of the European Union or the European Research Council. Neither the European Union nor the granting authority can be held responsible for them. M.V. also acknowledges financial support from Grant CEX2024-001451-M funded by MICIU/AEI/10.13039/501100011033, from Grant No. PID2022-136224NB-C22 from the Spanish Ministry of Science, Innovation and Universities, and from Grant No. 2021-SGR-872 funded by the Catalan Government. SB is partially supported by the Deutsche Forschungsgemeinschaft under Germany’s Excellence Strategy - EXC 2121 Quantum Universe - 390833306.

\newpage

\appendix

\section{Resolution effects in the GW spectrum}
\label{app:plateau}

In this Appendix, we investigate how finite resolution effects distort the GW spectrum at momenta significantly smaller than the naive expectation set by the lattice spacing. To quantify this effect, we performed a $2K$ simulation with the same box size as the $4K$ one used in Section \ref{sec:GW_from_scaling}. This setup ensures that the two simulations coincide in the IR region of the spectrum, since the lowest resolvable momentum is determined by the box size, $k_{\mathrm{IR}} = \frac{2\pi}{L}$.
The main difference lies in the UV cutoff, which depends on the grid spacing $\Delta x \equiv L/N$, given by $k_{\mathrm{UV}} = \frac{\sqrt{3}\pi}{\Delta x}$,
where the factor of $\sqrt{3}$ arises because the maximum momentum that can be represented corresponds to the diagonal of the reciprocal lattice\footnote{The reciprocal lattice corresponds to the Fourier transform of the real-space lattice.}.
By comparing the two simulations, we can precisely identify the scale at which deviations appear as we move toward the UV region. Note, however, that the $2K$ simulation due to its lower resolution cannot be evolved to very late times. Nevertheless, it suffices to study the scaling regime, and we can evolve to $m\tau \sim 20$ while this is already well established for $m\tau \gtrsim 15$.

\begin{figure}
    \centering
    \includegraphics[scale=0.52]{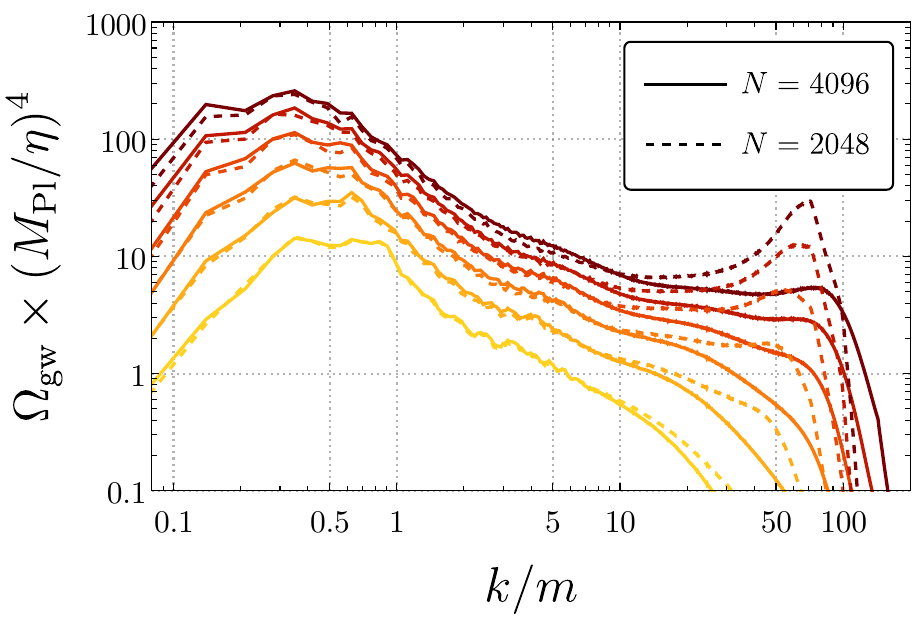}
    ~~
    \includegraphics[scale=0.48]{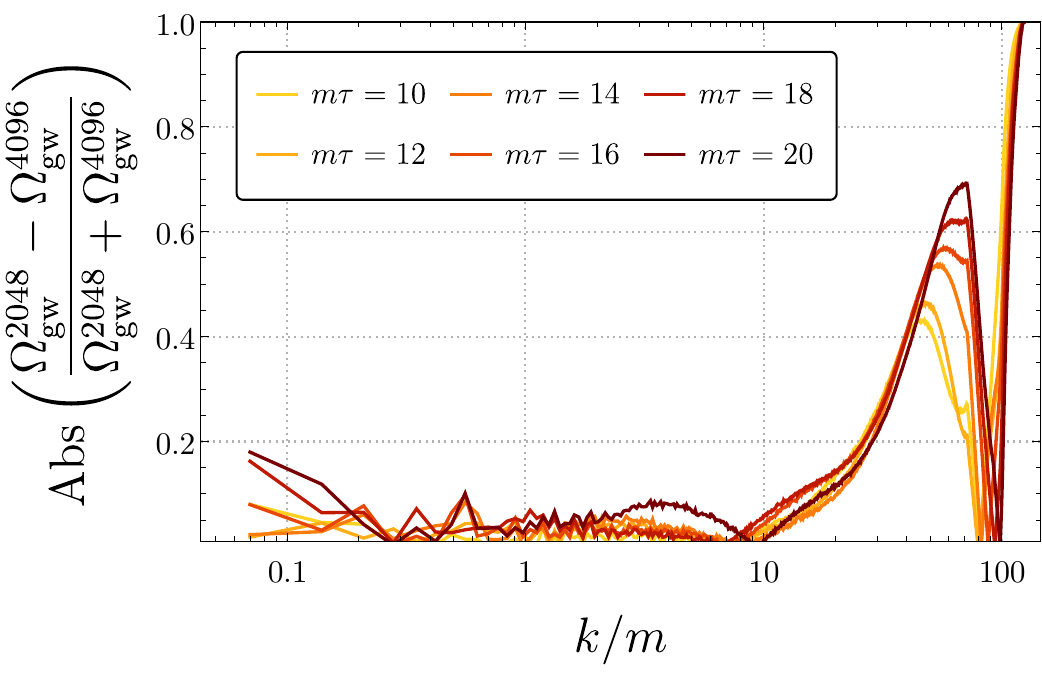}
    \caption{\textbf{Left}: GW spectra obtained from simulations using $N^3 = 4096^3$ (solid) and $N^3 = 2048^3$ (dashed) grid points. The spectra are shown from $m\tau = 10$ up to $m\tau = 20$ in steps of $2$ from bottom to top. Both simulations used the same box size, namely $mL = 90$. \textbf{Right}: Relative difference between the GW spectra from the left plot.}
    \label{fig:resolutionGW}    
\end{figure}

In the left panel of Fig. \ref{fig:resolutionGW} we show the GW spectra obtained with different resolutions, for times ranging from $m 
\tau =10$ to $m
\tau=20$.
It is manifest that the two spectra deviate for $k m \gtrsim 5-10$.
To make the comparison clearer, in the right panel of Fig. \ref{fig:resolutionGW}
we show the relative difference between the two spectra. It can be seen that, besides fluctuations, the two GW spectra start departing from each other with a clear trend from $k/m \gtrsim 10$,
and the plateau-like feature of the GW spectrum should be considered an artifact of the grid resolution.
In order to minimize the effect of this feature in the fit of the GW spectrum, we conservatively restrict to momenta $k/m \lesssim 5 $ in the main body of the paper.

\section{Rescaling arguments for the stress tensor correlator}
\label{app:rescaling}

In this Appendix, we argue that the parameterization of the energy momentum tensor leading to the expression for $\Pi^2$ in Eq.\,\eqref{eq: Pi to CT} is dictated by Weyl invariance and dimensional analysis for a scaling network of defects.
To show this, we consider the action of the underlying scalar field theory:
\beq
S = \int d^4 x \sqrt{-g} \left(
\frac{1}{2} g^{\mu \nu} 
\partial_{\mu} \phi \partial_{\nu} \phi 
-\frac{\lambda}{4} \left(\phi^2 -\eta^2 \right)^2
\right) \, ,
\eeq
where the background metric is assumed to be a
FLRW metric of the form
\beq
ds^2 = a(\tau)^2  (d \tau^2 - d x_i d x^i) \, ,
\eeq
where the type of cosmology is encoded in the $\tau$ dependence of the scale factor.

The action above is invariant at the classical level under the following global \emph{Weyl} rescaling with constant parameter $\alpha$:
\beq
\label{eq:weylscaling}
g_{\mu \nu} \to  e^{2\alpha} g_{\mu \nu} \, , \qquad \phi \to e^{- \alpha} \phi \, , \quad 
\eta^2 \to e^{-2\alpha} \eta^2 \, ,
\eeq
where we have taken the parameter $\eta$ to be a spurion that transforms non-trivially.
The rescaling does not involve the coordinates, and acts on the scale factor as 
\bea 
a(\tau)^2 \to e^{2\alpha} a(\tau)^2. 
\eea

We would like to use this property to deduce the classical scaling of the two point function of the stress tensor with the scale factor. One can show that:
\bea
T_{\mu \nu} \to e^{-2\alpha} T_{\mu \nu}.
\eea
According to the parameterization in Eq.\,\eqref{eq: T equals Pi}, it follows that $\Pi^{TT}_{ij}$ has weight zero, as the combination $a^2 \rho_S$ with $\rho_S = T^0_0$ has in fact weight $+2-4=-2$, and the TT projection does not affect the Weyl transformation properties. This implies that $C^T$ as defined in \eqref{eq: Pi to CT} has weight zero as well and is dimensionless. Therefore, it can only depend on Weyl--invariant combinations: 
$k(\delta_{\rm dw}/a(\tau))$ and $k\tau$, where $\delta_{\rm dw} \sim \eta^{-1}$.
If we now assume that the network is in the scaling regime, we can remove the dependence on $\delta_{\rm dw}$ (for scales that do not resolve the domain wall width), and conclude that $C^T$ can only depend on $k \tau$, as considered in the main text.
This motivates the general parameterization for $k^{-1} \gg \delta_{\rm dw}/(2 \pi a(\tau))$ that we used in Eq.\,\eqref{eq: Pi to CT}.

\section{UTC and GW scaling for networks with constant fraction of the energy density}\label{app: UTC CS}

In this Appendix, we apply the scaling arguments to obtain the momentum dependence of the UTC as in Ref.\,\cite{Figueroa:2012kw}.
Let us then consider Eq.\,\eqref{eq: rho_gw Caprini},
\begin{equation}
\frac{\mathrm{d}\rho^{S}_\text{gw}}{\mathrm{d}\ln k}(k,\tau) = \frac{2G}{\pi}\frac{k^3}{a^4(\tau)}\int_{\tau_i}^\tau\mathrm{d}\tau_1\int_{\tau_i}^\tau
\mathrm{d}\tau_2\ a^3(\tau_1)a^3(\tau_2)\rho_{S}(\tau_1)\rho_{S}(\tau_2)\cos\left[k(\tau_1 - \tau_2)\right]\Pi^2_{S}(k,\tau_1,\tau_2)\,,
\end{equation}
and take a source for which the energy density remains a constant fraction of the critical density, namely $\rho_S \propto H^2(\tau)$ as assumed in Ref.\,\cite{Figueroa:2012kw}:
\begin{equation}
\label{eq:rhoSapp}
\Pi^2_{S}(k,\tau_1,\tau_2) = (\tau_1\tau_2)^{3/2}C^T_{S}(x_1,x_2) \qquad a(\tau) = \sqrt{\Omega^0_\text{rad}}H_0\tau \qquad \rho_{S}(\tau) = \mu\, H^2(\tau)  \, ,
\end{equation}
where $\mu$ has the dimension of a string tension. One then obtains:
\begin{equation}
\frac{\mathrm{d}\rho_\text{gw}^{S}}{\mathrm{d}\ln k}(k,\tau) = \frac{2 \mu^2 G H_0^2\Omega^0_{\rm rad}}{\pi}\frac{k^3}{a^4(\tau)}\int_{\tau_i}^\tau\mathrm{d}\tau_1\int_{\tau_i}^\tau
\mathrm{d}\tau_2 \sqrt{\tau_1  \tau_2} \cos\left[k(\tau_1 - \tau_2)\right] C^T_{S}(x_1, x_2)\,,
\end{equation}
which leads to
\begin{align}
\Omega_{\rm gw}^{S}(\tau, k) \equiv \frac{1}{\rho_c}\frac{\mathrm{d}\rho_\text{gw}^{S}}{\mathrm{d}\ln k} & = \frac{16}{3}(G\mu)^2\Omega_{\rm rad}(\tau)  F^T_{S}(x)\,, 
 \nn F_{S}^T(x)&\equiv \int_{x_i}^x\mathrm{d}x_1\int_{x_i}^x
\mathrm{d}x_2 \sqrt{x_1  x_2} \cos\left(x_1-x_2\right) C^T_{S}(x_1, x_2) \,.
\end{align}
In \cite{Figueroa:2012kw}, the authors argued from analytical considerations that the function $F^T_{S}(x)$ should converge to a constant at late times. This can be seen by noticing that  $C^T_{S}(x_1, x_2)$ is peaked at $x_1 \approx x_2$ and decays away from the diagonal. Assuming a power law decay at large $x$, $C^T_{S}(x, x) \propto x^{-q}$, one obtains that any value of $q>2$ implies a constant $F_S^T(x)$ for $x \to \infty$\,\footnote{It should be noted that the scale invariant spectrum for this type of sources only holds in a RD Universe, while different spectral behaviors are found in other cosmologies, see discussion in Section\,\ref{sec: other cosmologies}.}. 
One then simply has:
\begin{align}
    \Omega_{\rm gw}^{S}(\tau, k)= \frac{16}{3}(G\mu)^2\Omega_{\rm rad}(\tau)F^T_{S}(x \to \infty) = \text{const.} \,
\end{align}
This result should be contrasted with the case of domain walls, where $F_{\rm dw}^T(x)$ may be constant only if very special values of $q$ were to be realized (e.g. $q\approx2$ for a totally incoherent UTC, or $q\approx1$ for the coherent case, as discussed in Sec.\,\ref{sec: GW from ETC}). The origin of this discrepancy boils down to the different functional form of $F_S^T(x)$ compared to Eq.\,\eqref{eq:main_result_for_ETC}, which stems from the different powers of $H(\tau)$ in the expression for the energy density in Eq.\,\eqref{eq:rhoSapp} compared to Eq.\,\eqref{eq:scalingenergy}. Therefore, we conclude that obtaining a scale--invariant GW spectrum requires extra assumptions on the redshift of the source energy density in addition to its self--similarity in the scaling regime.

\section{Modification of \CL  ~for the ETC}
\label{app: Implementation in CosmoLattice}

In this Appendix, we outline the algorithm employed to compute the two-point correlator of the energy–momentum tensor using the publicly available code \CL. We begin with a brief overview of how power spectra are computed within \CL. This provides the foundation for describing our modifications to the code that enable the evaluation of the correlators of the energy–momentum tensor.

\subsection{Power spectrum}

Let us first examine the computation of the power spectrum for an arbitrary real function $f$ in \CL ~(for a comprehensive review, see Ref.~\cite{PSTechNote}). In the continuum, the ensemble average of a real function $f$ is given by  
\begin{equation}\label{eq: power spectrum continuum}
    \langle f^2(\mathbf{x})\rangle = \int \frac{d^3\mathbf{k}}{(2\pi)^3}\int \frac{d^3\mathbf{q}}{(2\pi)^3}\langle f(\mathbf{k})f^*(\mathbf{q})\rangle e^{i \mathbf{x}\cdot(\mathbf{k}-\mathbf{q})} = \int \Delta_{f}(k)\, d(\ln k)\,,
\end{equation}
where  
\begin{equation}\label{eq: Delta to P}
    \Delta_f(k) \equiv \frac{k^3}{2\pi^2}\mathcal{P}_{f}(k)\,,
\end{equation}
is the dimensionless power spectrum and $\mathcal{P}_f(k)$ denotes the power spectrum, defined by  
\begin{equation}
    \langle f(\mathbf{k})f^\star(\mathbf{q})\rangle = (2\pi)^3 \mathcal{P}_f(k)\,\delta^{(3)}(\mathbf{k}-\mathbf{q})\,.
\end{equation}

To obtain a discrete representation of the power spectrum on the lattice, we first express the ensemble average as a volume average,  
\begin{equation}\label{eq: lattice volume average}
    \langle f^2\rangle_V = \frac{1}{N^3}\sum_{\mathbf{n}}f^2(\mathbf{n}) = \frac{1}{N^6}\sum_{\tilde{\mathbf{n}}}\left|f(\tilde{\mathbf{n}})\right|^2\,,
\end{equation}
where $N$ is the total number of grid points per spatial dimension, and the summations run over lattice sites $\mathbf{n}$ in real space and $\tilde{\mathbf{n}}$ in Fourier space, defined as
\begin{align}
    \mathbf{n} &= (n_x, n_y, n_z)\,, \quad n_i = 0, 1, \ldots, N-1\,,\\
    \tilde{\mathbf{n}} &= (\tilde{n}_x, \tilde{n}_y, \tilde{n}_z)\,, \quad \tilde{n}_i = -\frac{N}{2}+1, -\frac{N}{2}+2, \ldots, -1, 0, 1, \ldots, \frac{N}{2}-1, \frac{N}{2}\,.
\end{align}
The second equality in Eq.~\eqref{eq: lattice volume average} follows from the discrete Fourier transform,
\begin{equation}
    f(\mathbf{n}) = \frac{1}{N^3}\sum_{\tilde{\mathbf{n}}} e^{\frac{2\pi i}{N} \mathbf{n} \tilde{\mathbf{n}}} f(\tilde{\mathbf{n}})\,, \quad 
    f(\tilde{\mathbf{n}}) = \sum_\mathbf{n} e^{-\frac{2\pi i}{N}\mathbf{n} \tilde{\mathbf{n}}} f(\mathbf{n})\,, \quad 
    \text{with} \quad \sum_\mathbf{n} e^{\frac{2\pi i}{N}\mathbf{n} \tilde{\mathbf{n}}} = N^3\delta_{\mathbf{0}\tilde{\mathbf{n}}}\,.
\end{equation}
We can further decompose the summation into spherical bins,
\begin{equation}
     \langle f^2\rangle_V = \frac{1}{N^6}\sum_{\left|\tilde{\mathbf{n}}\right|}\sum_{\tilde{\mathbf{n}}\in R(\left|\tilde{\mathbf{n}}\right|)}\left|f(\tilde{\mathbf{n}})\right|^2\,.
\end{equation}
Here, $R(\left|\tilde{\mathbf{n}}\right|) = \left[\left|\tilde{\mathbf{n}}\right| - \tfrac{1}{2}, \left|\tilde{\mathbf{n}}\right| + \tfrac{1}{2}\right)$ denotes a spherical shell (or bin) of unit width. We thus sum over all modes $\tilde{\mathbf{n}}$ lying within the shell $R(\left|\tilde{\mathbf{n}}\right|)$, repeating this procedure for all possible radii\footnote{One may, in principle, adopt a more general binning scheme with unequal bin widths, $R(l) = [\,l - \Delta \tilde{n}_l^-,\, l + \Delta \tilde{n}_l^+\,)$, where $l = 1, 2, \ldots$ labels the bins and $\Delta \tilde{n}_l^{\pm}$ specify asymmetric radial widths. Such general binning is implemented in \CL, but for clarity we restrict ourselves to the \emph{canonical binning} with equal widths $\Delta\tilde{n}_l^{\pm} = 1/2$, which is also the default in \CL.} $\left|\tilde{\mathbf{n}}\right|$. For each bin, we define the multiplicity $\#_{\left|\tilde{\mathbf{n}}\right|}$ as the number of modes contained within the corresponding spherical shell\footnote{As discussed in Ref.~\cite{PSTechNote}, there are two possible ways to count the number of modes in each spherical shell: Type~I, which uses the exact mode count per bin, and Type~II, which approximates it as $\#_{\left|\tilde{\mathbf{n}}\right|} \approx 4\pi\left|\tilde{\mathbf{n}}\right|^2$. In this appendix, we adopt the Type~I definition.}. The angular average for a given bin is then defined as 
\begin{equation}
\langle\left|f(\tilde{\mathbf{n}})\right|^2\rangle_{R(\left|\tilde{\mathbf{n}}\right|)} \equiv \frac{1}{\#_{\left|\tilde{\mathbf{n}}\right|}}\sum_{\tilde{\mathbf{n}}\in R(\left|\tilde{\mathbf{n}}\right|)}\left|f(\tilde{\mathbf{n}})\right|^2\,.
\end{equation}

The unit displacement in the reciprocal lattice corresponds to the smallest momentum mode that can be resolved, $k_{\rm IR} = 2\pi/L$, where $L$ is the lattice box size. Each position in the reciprocal lattice therefore corresponds to a momentum $\mathbf{k} = k_{\rm IR}\tilde{{\mathbf{n}}}$. Since the bins have a width $\Delta \tilde{n} = 1$, they are associated with momentum bins of width $\Delta k = k_{\rm IR}$ centered at $k(\tilde{\mathbf{n}}) = k_{\rm IR}\left|\tilde{\mathbf{n}}\right|$. Hence, we can write $\Delta (\ln k) = k_{\rm IR}/k$ and express the volume average as
\begin{equation}\label{eq: PS lattice}
    \langle f^2\rangle_V = \sum_{\left|\tilde{\mathbf{n}}\right|}\Delta(\ln k) \frac{k(\tilde{\mathbf{n}})}{2\pi}\frac{\Delta x}{N^5}\#_{\left|\tilde{\mathbf{n}}\right|}\langle\left|f(\tilde{\mathbf{n}})\right|^2\rangle_{R(\left|\tilde{\mathbf{n}}\right|)}\,,
\end{equation}
where $\Delta x \equiv L/N$.  
Comparing with the continuum limit in Eq.~\eqref{eq: power spectrum continuum}, one identifies the lattice power spectrum as\footnote{Different versions of Eq.~\eqref{eq: PS lattice} exist depending on the definition of the momentum modulus $k(\tilde{\mathbf{n}})$. In the appendix, we chose $k(\tilde{\mathbf{n}})$ as the central momentum of each bin, while alternative definitions are discussed in Ref.~\cite{PSTechNote}.}
\begin{equation}
    \Delta_f(\left|\tilde{\mathbf{n}}\right|) = \frac{k(\tilde{\mathbf{n}})}{2\pi}\frac{\Delta x}{N^5}\#_{\left|\tilde{\mathbf{n}}\right|}\langle\left|f(\tilde{\mathbf{n}})\right|^2\rangle_{R(\left|\tilde{\mathbf{n}}\right|)}\,.
\end{equation}
\CL ~provides built-in routines to compute the angular average of any function $f$, which we employ to determine the power spectrum of the energy–momentum tensor and, consequently, the Equal Time Correlator.

\subsection{Towards the ETC}

We now proceed to the computation of the ETC. Our goal is to extract the TT component of the energy–momentum tensor, since this component allows us to compute $C^T_{\rm dw}(x,x)$ through Eq.~\eqref{eq: T2}. For a real scalar field $\phi$, this is given by $T_{ij}^{TT} = (\partial_i \phi\, \partial_j \phi)^{TT}$, 
which is most conveniently evaluated in Fourier space as  
\begin{equation}\label{eq: TTT to T}
    T^{TT}_{ij}(\mathbf{k},\tau) = \Lambda_{ij,kl}(\hat{\mathbf{k}})\, T_{kl}(\mathbf{k},\tau)\,,
\end{equation}
where $T_{kl}(\mathbf{k},\tau)$ denotes the Fourier transform\footnote{The full energy–momentum tensor is
\begin{equation}
    T_{ij}(\mathbf{x},\tau) = \partial_i\phi\,\partial_j\phi - g_{ij}\!\left[\frac{1}{2}g^{\mu\nu}\partial_\mu\phi\,\partial_\nu\phi + V(\phi)\right]\,,
\end{equation}
but only the first term contributes under the TT projection. Hence, we retain only this term in what follows.}  
of $\partial_k\phi(\mathbf{x},\tau)\,\partial_l\phi(\mathbf{x},\tau)$. The operator $\Lambda_{ij,kl}$ is the standard TT projection operator, defined as  
\begin{equation}
    \Lambda_{ij,kl} \equiv P_{il}P_{jk} - \frac{1}{2}P_{ij}P_{kl}\,, \quad \text{with} \quad P_{ij}(\hat{\mathbf{k}}) \equiv \delta_{ij} - \hat{k}_i\hat{k}_j\,,
\end{equation}
where $\hat{k}_i = k_i / k$. The implementation of these operators in \CL ~is discussed in Ref.~\cite{GWTechNote}.

We then compute the power spectrum of the energy-momentum tensor through
\begin{equation}\label{eq: PS T}
    \Delta_T(\left|\tilde{\mathbf{n}}\right|) = \frac{k(\tilde{\mathbf{n}})}{2\pi}\frac{\Delta x}{N^5}\#_{\left|\tilde{\mathbf{n}}\right|}\,
    \langle T^{TT}_{ij}(\tilde{\mathbf{n}},\tau)\,T^{TT\star}_{ij}(\tilde{\mathbf{n}},\tau)\rangle_{R(\left|\tilde{\mathbf{n}}\right|)}\,.
\end{equation}
To evaluate this, we must compute the bilinear product $T^{TT}_{ij}T^{TT\star}_{ij}$, which can be expressed as  
\begin{equation}\label{eq: bilinear}
    T^{TT}_{ij}(\tilde{\mathbf{n}},\tau)\,T^{TT\star}_{ij}(\tilde{\mathbf{n}},\tau)
    = {\rm Tr}(\mathbf{P}\mathbf{T}\mathbf{P}\mathbf{T}^\star) - \frac{1}{2}\,{\rm Tr}(\mathbf{P}\mathbf{T})\,{\rm Tr}(\mathbf{P}\mathbf{T}^\star)\,.
\end{equation}
Here, the matrices $\mathbf{T}$ and $\mathbf{P}$ have components $(\mathbf{T})_{ij} = T_{ij}(\tilde{\mathbf{n}},\tau)$ and $(\mathbf{P})_{ij} = P_{ij}(\tilde{\mathbf{n}})$, respectively. Once the bilinear $T^{TT}_{ij}\,T^{TT\star}_{ij}$ is obtained, the built-in routines of \CL ~can be used to evaluate Eq.~\eqref{eq: PS T}, which corresponds in the continuum limit to  
\begin{equation}
    \Delta_T(k) = \frac{k^3}{2\pi^2}\,T^2_{\rm dw}(k,\tau,\tau)\,,
\end{equation}
analogous to Eq.~\eqref{eq: Delta to P}. Finally, using Eq.~\eqref{eq: T2}, we can directly compute $C^T_{\rm dw}(x,x)$ via  
\begin{equation}
    T^2_{\rm dw}(k,\tau,\tau) = a^4(\tau)\,\rho_{\rm dw}^2(\tau)\,\tau^3\,C^T_{\rm dw}(x,x)\,.
\end{equation}

The entire procedure follows the same steps as those used for computing the GW spectrum, as detailed in Ref.~\cite{GWTechNote}. Accordingly, we modified the \CL\ code by introducing new fields $T_{ij}(\mathbf{x}, \tau)$ and treating them analogously to the GW perturbations\footnote{In \CL, the GW perturbations are represented as $h_{ij}(\mathbf{k},\tau) = \Lambda_{ij,kl}(\hat{\mathbf{k}})\,u_{kl}(\mathbf{k},\tau)$, where $u_{ij}(\mathbf{k},\tau)$ denotes the Fourier transform of the real-space field $u_{ij}(\mathbf{x},\tau)$. The latter evolves according to
\begin{equation}
    u_{ij}'' + 2\mathcal{H}u_{ij}' - \nabla^2u_{ij} = 16\pi G\,T_{ij}\,,
\end{equation}
with $T_{ij} = \partial_i\phi\,\partial_j\phi$. One can thus see from Eq.\eqref{eq: TTT to T} that $T^{TT}_{ij}$ ($T_{ij}$) can be treated as $h_{ij}$ ($u_{ij}$) in this formulation.}
$h_{ij}$. There are, however, three main differences that arise. The first concerns the EoMs: while $h_{ij}$ must evolve according to its own EoM, the tensor $T_{ij}$ is directly defined as $T_{ij} = \partial_i \phi \, \partial_j \phi$. The second difference is related to Eq.~\eqref{eq: bilinear}: the GW spectrum depends on the bilinear $h_{ij}' h_{ij}'^{\star}$, where a prime denotes a derivative with respect to conformal time. For the energy–momentum tensor, however, no time derivative is involved. Finally, the GW spectrum is expressed in terms of $\Omega_{\rm gw}$, which is related to $\Delta_{h'}(|\tilde{\mathbf{n}}|)$ by
\begin{equation}
    \Omega_{\rm gw}(\tilde{\mathbf{n}}, \tau) = \frac{1}{\rho_c} \frac{M_{\rm Pl}^2}{4 a^2} \, \Delta_{h'}(|\tilde{\mathbf{n}}|) \, ,
\end{equation}
whereas, in light of the ETC, we directly compute $\Delta_T(|\tilde{\mathbf{n}}|)$.

\bibliographystyle{JHEP}
{\footnotesize
\bibliography{biblio}}
\end{document}